\documentclass{article}

\makeatletter
\newcommand\ackname{Acknowledgements}
\if@titlepage
  \newenvironment{acknowledgements}{%
      \titlepage
      \null\vfil
      \@beginparpenalty\@lowpenalty
      \begin{center}%
        \bfseries \ackname
        \@endparpenalty\@M
      \end{center}}%
     {\par\vfil\null\endtitlepage}
\else
  \newenvironment{acknowledgements}{%
      \if@twocolumn
        \section*{\abstractname}%
      \else
        \small
        \begin{center}%
          {\bfseries \ackname\vspace{-.5em}\vspace{\z@}}%
        \end{center}%
        \quotation
      \fi}
      {\if@twocolumn\else\endquotation\fi}
\fi
\makeatother

\usepackage{arxiv}

\usepackage[utf8]{inputenc} 
\usepackage[T1]{fontenc}    
\usepackage{hyperref}       
\usepackage{url}  
\usepackage{xcolor}
\usepackage{booktabs}       
\usepackage{amsfonts}       
\usepackage{nicefrac}       
\usepackage{microtype}      
\usepackage{lipsum}		
\usepackage{graphicx}
\usepackage{amsmath}
\usepackage{mathtools}
\usepackage{subcaption}
\usepackage{multirow} 

\usepackage[english]{babel} 
\usepackage{blindtext}
\usepackage{ragged2e}
\usepackage{bm}
\usepackage{dsfont}
\usepackage{biblatex}
\usepackage{biblatex2bibitem}
\bibliography{references.bib}

\DeclareMathOperator*{\argmin}{arg\,min}
\title{Robust financial calibration: a Bayesian approach for neural SDEs}
\newtheorem{theorem}{Theorem}[section]

\newtheorem{remark}[theorem]{Remark}

\newtheorem{proof}{Proof} 
\date{September 9, 2024}	
\date{} 					

\author{\href{https://orcid.org/0000-0003-1308-5341}{\includegraphics[scale=0.06]{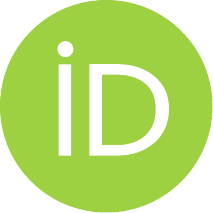}\hspace{1mm}Christa~Cuchiero\thanks{Christa Cuchiero gratefully acknowledges financial support 
through grant Y 1235 of the START-program.}}\\
	University of Vienna,\\ Department of Statistics and Operations Research, \\Kolingasse 14, 1090 Vienna, Austria\\
	\texttt{christa.cuchiero@univie.ac.at} \\
	\And
 \href{https://orcid.org/0009-0009-1494-8007}{\includegraphics[scale=0.06]{orcid.pdf}\hspace{1mm}Eva~Flonner\thanks{Eva Flonner acknowledges support from the University of Applied Sciences Wiener Neustadt where parts of this work were carried out.}} \\
	Vienna University of Economics and Business, \\
	 Welthandelsplatz 1,\\ 1020 Vienna, Austria\\
	\texttt{eva.flonner@wu.ac.at} \\
	 \And
	 \href{https://orcid.org/0000-0002-8824-1882}{\includegraphics[scale=0.06]{orcid.pdf}\hspace{1mm}Kevin Kurt\thanks{Opinions expressed in this paper are those of the author, and do not necessarily reflect the view of IQAM Invest.}} \\
	 IQAM Invest, \\
	Wollzeile 36-38,
1010 Vienna, Austria \\
	 \texttt{kevin.kurt@iqam.com} \\
}

\renewcommand{\shorttitle}{Robust financial calibration: a Bayesian approach for neural SDEs}

\hypersetup{
pdftitle={Robust financial calibration: a Bayesian approach for neural SDEs},
pdfsubject={q-fin},
pdfauthor={Christa~Cuchiero, Eva~Flonner, Kevin~Kurt},
pdfkeywords={model calibration, Bayesian, Neural SDE, option pricing, uncertainty quantification, model averaging},
}
\begin{document}
\maketitle
\begin{abstract}
	The paper presents a Bayesian framework for the calibration of financial models using neural stochastic differential equations (neural SDEs)\textcolor{black}{, for which we also formulate a global universal approximation theorem based on Barron-type estimates}. The method is based on the specification of a prior distribution on the neural network weights and an adequately chosen likelihood function. The resulting posterior distribution can be seen as a mixture of different classical neural SDE models yielding robust bounds on the implied volatility surface. Both, historical financial time series data and option price data are taken into consideration, which necessitates a methodology to learn the change of measure between the risk-neutral and the historical measure. The key ingredient for a robust numerical optimization of the neural networks is to apply a Langevin-type algorithm, commonly used in the Bayesian approaches to draw posterior samples.
\end{abstract}

\keywords{model calibration \and Bayesian \and neural SDE \and option pricing \and uncertainty quantification}

\section{Introduction}\label{Introduction}
\justifying
In financial institutions, the task of choosing one model from a pool of asset price models, given current market and historical data, is called calibration. Machine learning techniques started to offer new perspectives on this crucial task since they are computationally efficient and manage to capture empirically observed market characteristics. For instance, \cite{gierjatowicz2020robust} or \cite{Cuchiero_2020} and more recently \cite{cohen2021arbitrage} and \cite{guyon2022neural} use so-called \emph{neural stochastic differential equations (neural SDEs)} for solving the calibration problem in a setting where one is looking for a model which \textit{generates} given (derivatives') market prices. One key feature of neural SDEs is that by the universal approximation property of neural networks (see~\cite{Cybenko1989ApproximationBS},~\cite{HORNIK1991251},~\cite{SONTAG1997177}
), all continuous characteristics of any SDE can be approximated on compacts. \textcolor{black}{We aim here to go further and exploit  the approximation capacity of neural SDEs not only on the level of characteristics, but rather on the level
of the SDE solutions, e.g., in
 a (probabilistic) $L^2$-sense.
Recent progress towards such an approximation theorem can be found in~\cite{kwossek2025universal}. Another result  in this direction is provided in
\cite{neural_projection_filter}, which gives estimates based on  \cite{barron93}, however only locally up to a hitting time of certain compact sets.
Building on these developments, we present a new theorem (see Theorem~\ref{UAT_NSDE_THM}) that combines the techniques of \cite{kwossek2025universal} with the quantitative bounds of \cite{barron93}. In particular, our $L^2$-approximation error depends explicitly on the number of neurons in the hidden layer.} 

There are two drawbacks of the standard calibration approach using neural SDEs mentioned above: firstly, possible prior knowledge on appropriate distributions of the neural network weights is disregarded. Secondly, only point estimates of neural network weights are provided and thus there is no built-in uncertainty quantification on the resulting option prices. The latter is important to get robust price bounds for the asset pricing model at hand. In the following we attempt to tackle these aspects by proposing a Bayesian calibration algorithm for neural SDEs. 
Traditional calibration techniques for SDEs often aim to find a model that best replicates market data, in the sense of finding the 'true' optimum with respect to a certain optimization criterion, which can be problematic given the inherently ill-posed nature of the calibration problem. Instead of seeking a singular solution, our Bayesian framework focuses on determining a distribution of plausible solutions.

The core concept  is to start with a prior distribution for the unknown parameters of  the neural SDE model and then update this prior using the observed market data to derive a posterior distribution.
We focus on constructing practical prior and likelihood functions and assessing the sensitivity of the Bayesian approach, with respect to the hyperparameter choices of the prior and likelihood functions. Our contribution also addresses the specific challenges of jointly calibrating to time-series data under the historical measure and derivatives' price data under the risk neutral measure. This  induces the need for measure changes, which we parameterize again by neural networks.

One goal of this work is  to serve as a proof of concept for applying Bayesian techniques to the highly flexible and dynamic framework of neural SDEs, highlighting the advantages of this approach. 
By fully integrating the power of neural networks with the stochastic nature of SDEs and addressing the complexities of including time series data as well as data on options, our approach opens new avenues for modeling in finance.

\subsection{Relation to the literature}

More broadly speaking and relating the current work to the literature, the application of machine learning techniques in mathematical finance has evolved rapidly over the last decades.  Indeed, since the 1990s there
is a big and steadily growing number of papers
related to this topic, as outlined e.g., in \cite{ruf2020neural}. We do not attempt to summarize all of them, but rather
 refer to some major milestones in view of model calibration and
give a flavour of how to categorize earlier developments within this research area. 

If we focus on neural networks as a machine learning tool the first research category is utilising neural networks merely as a computational utility or as a method to store information. This is in order to avoid for instance slow and computationally heavy Monte Carlo sampling e.g. for pricing derivatives like VIX options. 
An example in this direction is the work  by \cite{bayer2018deep}. Another category builds on neural SDEs mentioned above, where the drift and diffusion function of classical SDEs are modeled by neural networks. Neural SDEs are a promising tool not only for (robust) pricing, hedging and calibration purposes (see \cite{gierjatowicz2020robust}) but also for risk management or portfolio optimization.
We also refer to \cite{kidger2020neural}, \cite{choudhary2023funvol}, \cite{cohen2021arbitrage}, \cite{cohen2022estimating} and the references therein for learning features of neural SDEs and their employment for financial applications. In particular,  in \cite{cohen2021arbitrage} an arbitrage-free neural SDE market model for European options is  learnt by  using discrete time series data as well as option price data.

Let us mention in this context also
so-called 'Sig-SDEs'  introduced by \cite{arribas2020sigsdes}, where the volatility of the asset price process is modeled by means of a linear function of the (extended) signature of the driving Brownian motion.  Signature here refers to the sequence of iterated Stratonovich integrals of the  given Brownian motion. For other papers in the realm of calibration with signature methods we refer to \cite{cuchiero2022signaturebased, cuchiero2023joint}, where the extended signature of some primary process, which does not have to be the (marked-inferred) Brownian motion, is used. 
The essence is the linearity of the model, which allows for fast and accurate calibrations results, using both time-series and implied volatility surface data, of the S\&P 500 as well as the VIX index.

Other approaches, also based on 
 machine learning tools, model directly the volatility surface without a dynamical arbitrage-free model, see e.g.,
~\cite{chataigner2020deep} and the references therein. This article presents a deep learning method to produce the whole surface of local volatilities, enforcing no-arbitrage rules by applying  Dupire's formula to impose bounds on the local volatility related to the option prices.

As far as Bayesian modeling in mathematical finance is concerned, one key reference for the current work is \cite{Gupta2014RobustCO}. The paper introduces a Bayesian framework to calibrate local volatility models under the risk neutral measure based on the market's derivatives prices. This 
leads to a posterior distribution of the model parameters. \color{black} Other papers in the area of local volatility estimation are \cite{tegner2021probabilistic} or \cite{animoku2018modeling}. In the latter the focus lies on the avoidance of negative local variances, that can occur in the more traditional ways of estimating the local volatility. To this end, a suitable prior in a Bayesian framework is suggested.
\color{black}
Another Bayesian approach is adopted in 
the paper 'Estimate nothing' by \cite{duembgen2014estimate}. 
There a mixture of different classical models is considered and a posterior distribution of the weights of the different models is obtained. To this end, also historical financial time series data is taken into consideration.
More recent work employing Bayesian methods in the area of mathematical finance has been done by \cite{Tegner2021ABT}, where a local volatility model is equipped with a Gaussian process prior. This is done by defining a prior distribution over local volatility functions through a positive transformation of a Gaussian process with zero mean function and squared exponential covariance function, ensuring a smooth and positive volatility surface. Additionally, a hyperprior is placed on the parameters of the covariance function. The paper~\cite{Luca_Bayesian} exploits Sequential Monte Carlo methods with a tempered likelihood to estimate option pricing models, allowing for two stochastic volatility factors, co-jumps between price and volatility as well as a stochastic jump intensity. The latent states in the Sequential Monte Carlo scheme are marginalized using risk-neutral cumulants which are given in closed form.
\color{black}
We would also like to acknowledge work in the area of stochastic volatility in \cite{kastner2017efficient}, \cite{gunawan2017efficient} or \cite{hosszejni2021modeling}. In the latter, a discrete-time stochastic volatility model is considered and a method to efficiently estimate such a model in a Bayesian context is proposed. More precisely, the suggested Markov Chain Monte Carlo sampler makes use of an approximate mixture representation of the latent volatility factor, and thus a conditionally Gaussian state space model is obtained. The method is suitable for multivariate stochastic volatility.
\color{black}

\color{black}
As outlined in the introduction the goal of the current paper is to unite the efforts described above and provide a Bayesian calibration framework based on neural SDEs. The paper is organized as follows: In Section \ref{Stochastic volatility models as neural SDEs} the model is introduced, in Section \ref{Change of measure for neural SDEs} we propose a change of measure for neural SDEs, section \ref{Variance reduction for Monte Carlo pricing via hedging strategies} recalls a variance reduction technique, that is an essential ingredient for the proposed algorithm, Section \ref{options} and \ref{Including time series characteristics} are the core of the paper, as they specify the Bayesian methodology to calibrate the neural SDE model under the historical and risk neutral measure, Section \ref{Sensitivity analysis using the Black-Scholes model} provides a sensitivity analysis with respect to hyperparameter values, followed by an empirical study in Section~\ref{Results for empirical data}, possible technical modifications in Section~\ref{Possible technical modifications} and a conclusion inside \ref{Conlusion}.

\section{Stochastic volatility models as neural SDEs}\label{Stochastic volatility models as neural SDEs} Similarly as in \cite{gierjatowicz2020robust}, fix $T > 0$ and assume a constant interest rate $r \in \mathbb{R}$. Furthermore consider a parameter space $\Theta=\Theta^{b^V} \times \Theta^{\sigma} \times \Theta^{\sigma^V}\subset \mathbb{R}^p$ and functions $b^V:[0,T] \times \mathbb{R} \times \Theta^{b^V} \to \mathbb{R}$, $\sigma: [0,T] \times \mathbb{R}^2 \times \Theta^{\sigma} \to \mathbb{R}$ and $\sigma^{V}: [0,T] \times \mathbb{R} \times \Theta^{\sigma^V} \to \mathbb{R}$. Let $(W_t^1)_{t \in[0,T]}$ be a standard  Brownian motion supported on $(\Omega, \mathcal{F},(\mathcal{F}_t)_{t\in [0,T]}, \mathbb{Q})$. Consider the parametric SDE
\begin{align} \mathrm{d} S_t^{\theta}&=rS_t^{\theta}\mathrm{d}t+\sigma(t,S_t^{\theta},V_t^{\theta},\theta)S_t^{\theta} \mathrm{d}W_t^1,\label{priceprocess}\\
    \mathrm{d}V_t^{\theta}&= b^V(t,V_t^{\theta},\theta)\mathrm{d}t+\sigma^{V}(t,V_t^{\theta},\theta)\mathrm{d}W_t^2, \label{varianceprocess}\\
    \color{black}X_t^{\theta}&=\left(S_t^{\theta},V_t^{\theta}\right), \label{X_neural}
\end{align} 
where $\mathrm{d} W_t^2 = \rho W_t^1+\sqrt{1-\rho^2} \color{black}U_t$, with $U_t$ a Brownian motion independent from $W_t^1$, $\rho \in [-1,1]$ \textcolor{black}{a constant parameter that will be estimated,} and where $(V_t^{\theta})_{t\geq 0}$ controls the spot variance process in our stochastic volatility model. Note that the model is free of arbitrage by construction, since $(e^{-{rt}}S_t^{\theta})_{t \in [0,T]}$ is a local martingale. From now on we will refer to the risk-neutral measure, corresponding to the neural network parameters $\theta$, as $\mathbb{Q}(\theta)$.
Furthermore define for all $t\in [0,T]$ and $\theta \in \mathbb{R}^p$, the functions $b$ and $\sigma^x$ by $$b(t,(s,v),\theta):=(rs,b^{V}(t,v,\theta)) \in \mathbb{R}^2 \text{ and } \sigma^x(t,(s,v),\theta):=(s \sigma(t,s,v,\theta),\sigma^{V}(t,v,\theta)) \in \mathbb{R}^{2},$$ 
where the first component of these functions refers to the $S^{\theta}$-part and the second to the $V^{\theta}$-part.

In the case where $(\sigma,b^V,\sigma^V)$ are \textcolor{black}{artificial neural networks} we call \eqref{priceprocess},\eqref{varianceprocess} a system of neural SDEs. \textcolor{black}{To guarantee the existence and uniqueness of \eqref{X_neural}, it is sufficient that the functions $b$ and $\sigma^x$ are locally Lipschitz continuous and satisfy a linear growth condition. Both properties are naturally inherited from the activation functions of the neural networks, in particular when they are Lipschitz continuous and bounded, as it is for instance the case for sigmoidal activation functions which we shall consider in Section \ref{UAT} below. }
 In the sequel we use $S^{\theta}$ and $V^{\theta}$ whenever we refer to the model quantities and $S$ and $V$ whenever we refer to quantities based on empirical data.
\color{black}
\subsection{Approximation capacity of neural SDEs}\label{UAT}
Understanding the approximation capacity of the above neural network model in the set of solutions to general It\^o-SDEs is crucial in view of universality properties of this model class.

Results in this direction are provided in \cite{neural_projection_filter} and the recent paper
\cite{kwossek2025universal}. 
We here provide a theorem, which can be seen as a combination of 
Theorem 2 of \cite{neural_projection_filter} and some techniques used in Proposition 3.6 in \cite{kwossek2025universal}. 
The difference to the latter is that we use the specific one-hidden neural networks considered in \cite{barron93}, and that we rely on the respective bounds obtained there for certain function classes introduced below.
Indeed, the quantitative
universal approximation result in \cite{kwossek2025universal} does not use the Barron-type bound applied in~\cite{neural_projection_filter}. Thus, our result is distinct in the sense that it yields an explicit bound involving the number of neurons in the hidden layer.

To be in line with \eqref{X_neural}, let us consider an SDE of the form
\begin{align} \mathrm{d} S_t&=rS_t \mathrm{d} t+\sigma(t,S_t,V_t)S_t \mathrm{d} W_t^1,\label{priceprocess_real}\\
    \mathrm{d} V_t&= b^V(t,V_t)\mathrm{d} t+\sigma^{V}(t,V_t)\mathrm{d} W_t^2, \label{varianceprocess_real}\\X_t&=\left(S_t,V_t\right), \label{X_real}
\end{align} 
driven by a $2$-dimensional \textcolor{black}{correlated Brownian motion} $W=(W^1, W^2)$.

The functions\[b\left(t, (s,v)\right):=\left(rs,b^V(t,v)\right) \in \mathbb{R}^2,\text{ and }\sigma^x \left(t,(s,v)\right):=\left(s \sigma(t,s,v), \sigma^V(t,v)\right) \in \mathbb{R}^2\] are \textcolor{black}{assumed to satisfy the usual Lipschitz continuity and linear growth conditions, such that a unique strong solution of $X$ exists.} To derive an estimation error between  the solution of \eqref{X_real} and the one of an approximative neural SDE given by \eqref{X_neural},  we shall consider the following class of functions introduced in \cite{barron93}.

For some constant $C >0$ denote by $\Gamma_{C}$ the set of functions $f$ for which the following conditions hold:
\\
\begin{enumerate}
\item  $f$ admits a Fourier representation of the following form
$$
f(x)=\int_{\mathbb{R}^2} e^{i \omega^{\top} x} \tilde{f}(\omega) d \omega,
$$
where $\tilde{f}(\omega)$ is the Fourier transform of $f$ defined by
$
\tilde{f}(\omega)=(2 \pi)^2 \int_{\mathbb{R}^2} f(x) e^{-i \omega^{\top} x} d x.
$
\item 
Assume that $\omega \tilde{f}(\omega)$ is integrable and that 
$$
\int_{\mathbb{R}^2}   \|\omega\| |\tilde{f}(\omega)| d \omega \leq C.
$$ 
\end{enumerate}

The neural networks that we shall consider to approximate the components of $b$ and $\sigma^x$ are of the following form
\begin{align}\label{eq:NNbasicmain}
f_p(x)= \sum_{k=1}^p c_k \phi(\langle a_k, x \rangle +b_k)+ c_0, 
\end{align}
 where $a_k \in \mathbb{R}^2$, $b_k, c_k, c_0 \in \mathbb{R}$ for $k=1, \ldots, p $, and $\phi $  a sigmoidal activation function, that is a bounded measurable function with $\lim_{z \to \infty} \phi(z)=1 $ and $\lim_{z \to -\infty} \phi(z)=0 $. 
 We shall now state the theorem which gives a precise bound on the number of neurons in the hidden layer, denoted by $p$, in terms of the approximation error measured in an $L^2$ sense.

In the following  $B_r(\mathbb{R}^2)$ denotes the ball of radius $r >0$.

\begin{theorem}\label{UAT_NSDE_THM}
\textbf{(Universal Approximation Theorem for neural SDEs):} Consider \eqref{X_real}  and $b: \mathbb{R} \times \mathbb{R}^2 \rightarrow \mathbb{R}^2$, $\sigma^x: \mathbb{R} \times \mathbb{R}^{2} \rightarrow \mathbb{R}^{2}$ be continuous functions, and $\left(W_t\right)_{t \geq 0}$  a $2$-dimensional Brownian motion.  
Assume that  the following conditions hold:
\begin{enumerate}
\item[1)] $b$ and $\sigma^x$ are locally Lipschitz continuous, meaning that for every $r >0$ there are constants  $k_{1,r}, k_{2,r}>0$, such that $\forall x, x^{\prime} \in B_r(\mathbb{R}^2)$ 
$$
\begin{aligned}
& \left\|b(t, x)-b\left(t, x^{\prime}\right)\right\|^2 \leq k_{1,r}\left\|x-x^{\prime}\right\|^2 \\
& \left\|\sigma^x(t, x)-\sigma^x\left(t, x^{\prime}\right)\right\|^2 \leq k_{2,r}\left\|x-x^{\prime}\right\|^2.
\end{aligned}
$$
\item[2)] $b$ and $\sigma^x$ satisfy the following linear growth condition
\[
\|b(t,x)\| \vee \|\sigma^x(t,x)\| \leq L(1 +\| x\|), \quad t \in [0,T],
\]
for some $L >0$. 
\item[3)] There exists some $C>0$, such that $\sup_{t \in [0,T]} b(t,\cdot) \in$ $\Gamma_{C}$ and $\sup_{t \in [0,T]} \sigma^x(t,\cdot) \in \Gamma_{C}$. 
\end{enumerate}
Then, for every $0 <\varepsilon \leq 1$ there exist 
 neural networks $b\left(t,x,\theta_p\right)$ and $\sigma^x\left(t,x,\theta_p\right)$ of form \eqref{eq:NNbasicmain} with (arbitrary) Lipschitz continuous sigmoidal activation function and $p$ 
given by
\begin{align}\label{eq:p}
p \geq \left(\frac{r^2 C^2 (256 T  +64 T^2 )}{\varepsilon}\right) 
 \exp\left(16 k_{2,r} T+4 k_{1,r} T^2\right),
\end{align}
where
\begin{align}\label{eq:radius}
r^2 \geq \frac{8(1+ 3^3 \| X_0\| ^4) e^{\beta T}}{\varepsilon}, 
\end{align} 
with $
\beta=54 K^2 T (T^2+ (32/3)^2)
$
and  $K$ a constant, which depends on the  bound on the activation function as well as as on $L, C$ and $T$,
such that the solution $X$ of \eqref{X_real}  can be approximated by the solution $X_t^{\theta_p}$  of \eqref{X_neural}, started  at the same initial value  $X_0^{\theta_p}= X_0$,  in the following sense,
\begin{equation}
E\left[\sup _{t \in \mid 0, T]}\left\|X_{t }^{\theta_p}-X_{t }\right\|^2\right] \leq  \varepsilon \label{UAT_NSDE}.
\end{equation}

\end{theorem}

\color{black}

\color{black}
\begin{remark}
Note that $p$ is of order 
\[
p\geq \frac{A}{\varepsilon^2},
\]
where $A$ is a constant that depends 
on the bound on activation function as well as $L, C, T, k_{1,r}, k_{2,r}$
Let us also remark  that the above result translates of course to higher dimensions (with changing constants since the results in \cite{barron93} depend on the dimension).
\end{remark}
\color{black}

\begin{proof}

The proof is an adaptation of the techniques used in Proposition 3.6 in \cite{kwossek2025universal}
and Theorem 2 in \cite{neural_projection_filter}.

Applying Proposition 1 in \cite{barron93}, as stated in Theorem \ref{th:barron} below, we get that
\begin{align}\label{eq_54}
& \sup _{t \in[0, T], x \in B_r(\mathbb{R}^2)}\left\|b(t, x, \theta_p)-b(t, x)\right\|^2 \leq \frac{8\left( r C\right)^2 }{p}, \\
& \sup _{t \in[0, T], x \in B_r(\mathbb{R}^2)}\left\|\sigma^x(t, x, \theta_p)-\sigma^x(t, x)\right\|^2 \leq \frac{8\left( r C\right)^2 }{p}.
\end{align}
Since the coefficients $b, \sigma^x$ of \eqref{X_real} are supposed to be of linear growth with some constant $L$, we obtain for every $x \in B_{r}(\mathbb{R}^2)$
\[
\sup _{t \in[0, T]}\left\|b(t, x, \theta_p)\right\| \leq  \sup _{t \in[0, T]}\left\|b(t, x)\right\| + \frac{\sqrt{8} r C}{\sqrt{p}} \leq L (1 + \|x \|) + \frac{\sqrt{8} r C }{\sqrt{p}}
\]
and similarly for $\sigma^x(\cdot, \cdot, \theta_p)$. Moreover, outside $B_r(\mathbb{R}^2)$, we have the following estimate 
\[
\sup _{t \in[0, T]}\left\|b(t, x, \theta_p)\right\| \leq M(1+\|x\|),
\]
where  $M$ depends on the bound of the activation function $\phi$, $C$ and $\sup_{t \in[0, T] } |b(t,0)|$ ($\sup_{t \in[0, T] } |\sigma^x(t,0)|$ respectively). This is due to the fact that $c_k^i$ in the specification of the neural networks in  \eqref{eq:NNbasic} can be chosen to satisfy $\sum_{i=1}^p |c^i_k|  < 2 rC$ (see Theorem \ref{th:barron} below).  We therefore get the following linear growth condition for
\[
\sup _{t \in[0, T]}\left\|b(t, x, \theta_p)\right\| \leq K_{p,r}(1 +\|x\|),
\]
where $K_{p,r}=\max(M, L +\frac{\sqrt{8} r C }{\sqrt{p}}) $, and the same holds for $\sigma^x(t, x, \theta_p)$ with the same constant.
By the choice of $p$ given by \eqref{eq:p}, we obtain that $\frac{r}{\sqrt{p}}=A_{T,C} \sqrt{\varepsilon}$ where $A_{T,C}$ denotes  a constant depending on $T$ and $C$. Choosing as maximal value $\varepsilon=1$ we can replace $K_{p,r}$ by a constant $K$ that is independent of $p,r$ (but still depends on $T,L,C$ and the bound on the activation function).

Note now that for any stopping time $\tau \leq T$
\begin{align*}
E\left[\sup _{t \in \mid 0, T]}\left\|X_{t }^{\theta_p}-X_{t}\right\|^2\right] &\leq E\left[\sup _{t \in \mid 0, T]}\left\|X_{t }^{\theta_p}-X_{t}\right\|^2 \mathbf{1}_{\{\tau \leq T\}} \right ]+ E\left[\sup _{t \in \mid 0, T]}\left\|X_{t \wedge \tau }^{\theta_p}-X_{t \wedge \tau}\right\|^2\right ]\\
&\leq\underbrace{\left (E\left[\sup _{t \in \mid 0, T]}\left\|X_{t }^{\theta_p}-X_{t}\right\|^4  \right ]\right)^{\frac{1}{2}}(P[\tau \leq T])^{\frac{1}{2}}}_{(I)}\\
&\quad +\underbrace{ E\left[\sup _{t \in \mid 0, T]}\left\|X_{t \wedge \tau }^{\theta_p}-X_{t \wedge \tau}\right\|^2\right ]}_{(II)}.
\end{align*}
The goal is  to estimate  each of the two summands (I) and (II) by $\frac{\varepsilon}{2}$ using the the specific stopping time 
$$\tau_r:=\inf \{t \geq 0 \,|\, \| X_t \| \geq r\}
$$
for $r$ given in \eqref{eq:radius}.
By the estimate given in Theorem 4.4 \textcolor{black}{in Chapter 2 of}
~\cite{mao2007stochastic} (applied with $p=4$ in the notation of this theorem), we can bound
\[
E\left[\sup_{t \in \mid 0, T]}\left\|X_{t }^{\theta_p}-X_{t}\right\|^4\right] \leq 8\left( E\left[\sup _{t \in \mid 0, T]}\left\|X_{t }^{\theta_p}\right\|^4\right]+ E\left[\sup _{t \in \mid 0, T]}\left\|X_{t}\right\|^4\right]\right)\leq 16(1+ 3^3 \| X_0\| ^4) e^{\beta T},
\]
where
$
\beta=54 K^2 T (T^2+ (32/3)^2).
$
Note here that by construction $K$ also serves as linear growth bound for the functions $b, \sigma^x$ and is the linear growth constant of the neural networks.

For the stopping time $\tau_r$ we obtain the following estimate by the Markov inequality and e.g. Lemma 3.2 in \textcolor{black}{Chapter 2 of}~\cite{mao2007stochastic}
\[
P[\tau_r \leq T]= P[\sup_{t \in \mid 0, T]} \|X_{t}\| \geq r] \leq \frac{E[\sup_{t \in \mid 0, T]} \|X_{t}\|^4]}{r^4} \leq \frac{(1 + 3^3 \| X_0\|^4) e^{\beta T}}{r^4}.
\]
Hence,
\begin{align*}
\left (E\left[\sup _{t \in \mid 0, T]}\left\|X_{t }^{\theta_p}-X_{t}\right\|^4  \right ]\right)^{\frac{1}{2}}(P[\tau_r \leq T])^{\frac{1}{2}} &\leq \frac{4(1+ 3^3 \| X_0\| ^4) e^{\beta T}}{r^2} \leq \frac{\varepsilon}{2},
\end{align*}
where the last inequality follows from definition of $r$. 

The rest of the proof is now dedicated to bound the term (II).

Let us define
\begin{align}
&e_r(s):=X_{s \wedge \tau_r}^{\theta}-X_{s \wedge \tau_r}\\
&\psi_r^1(s):=\left(b\left(s, X_{s \wedge \tau_r}^{\theta},\theta\right)-b\left(s, X_{s \wedge \tau_r}^{\theta}\right)\right) \mathbf{1}_{\left[\left[0, \tau_r\right]\right]}\\
&\psi_r^2(s)=\left(b\left(s, X_{s \wedge \tau_r}^{\theta}\right)-b\left(s, X_{s \wedge \tau_r}\right)\right) \mathbf{1}_{\left[\left[ 0, \tau_r \right]\right]}\\
&\gamma_r^1(s):=\left(\sigma^x\left(s, X_{s \wedge r_r}^{\theta}, \theta\right)-\sigma^x\left(s, X_{s \wedge r_r}^{\theta}\right)\right) \mathbf{1}_{\left[\left[ 0, \tau_r \right]\right]}\\
&\gamma_r^2(s):=\left(\sigma^x\left(s, X_{s \wedge \tau_r}^{\theta}\right)-\sigma^x\left(s, X_{s \wedge \tau_r}\right)\right) \mathbf{1}_{\left[\left[0, \tau_r \right]\right]}.
\end{align}

Then, using the elementary inequality

\begin{align}\label{eq_56}
&\left\|e_r(t)\right\|^2 \leq 4\left\|\int_0^t \psi_r^1(s) d s\right\|^2+4\left\|\int_0^t \psi_r^2(s) d s\right\|^2 \\
&+4\left\|\int_0^t \gamma_r^1(s) d W_s\right\|^2+4\left\|\int_0^t \gamma_r^2(s) d W_s\right\|^2.
\end{align}

Moreover,

\begin{align}
& E\left[\sup _{s \in [0, t \mid}\left\|\int_0^s \gamma_r^1(\tau) d W_{\mathrm{\tau}}\right\|^2\right]+E\left[\sup _{s \in[0, t]}\left\|\int_0^s \gamma_r^2(\tau) d W_\tau\right\|^2\right] \\
&\leq 4 E\left[\int_0^t\left\|\gamma_r^1(\tau)\right\|^2 d \tau\right]+4 E\left[\int_0^t\left\|\gamma_r^2(\tau)\right\|^2 d \tau\right],
\end{align}

where we have used Theorem 5.12 in ~\cite{mao2007stochastic}.
Now, using that $b$ and $\sigma^x$ are locally Lipschitz continuous and \eqref{eq_54}, and we obtain
$$
\mathrm{E}\left[\sup _{s \in[0, t]} \int_0^s\left\|\gamma_r^2(\tau)\right\|^2 d \tau\right] \leq k_{2, r} \int_0^t E\left[\sup _{\tau \in[0, s]}\left\|e_r(\tau)\right\|^2\right] d s.
$$
and
$$
E\left[\sup _{s \in[0, t]} \int_0^s\left\|\gamma_r^1(\tau)\right\|^2 d \tau\right] \leq \frac{8\left( r C\right)^2  T}{p}.
$$

From this it follows that

\begin{align} \label{eq_60}
& E\left[\sup _{s \in[0, t]}\left\|\int_0^s \gamma_r^1(\tau) d W_\tau\right\|^2\right]+E\left[\sup _{s \in[0, t ]}\left\|\int_0^s \gamma_r^2(\tau) d W_{\tau}\right\|^2\right] \\
& \leq \frac{32 T\left( r C\right)^2 }{p}+4 k_{2, r} \int_0^t E\left[\sup _{\tau \in [ 0, s]}\left\|e_r(\tau)\right\|^2\right] d s.
\end{align}

Similarly, by using H\"older's inequality
$$
\begin{aligned}
\left\|\int_0^t \psi_r^1(s) d s\right\|^2 
& \leq T \int_0^t\left\|\psi_r^1(s)\right\|^2 d s .
\end{aligned}
$$

Thus

\begin{align}\label{eq_62}
E\left[\sup _{s \in[0, t]}\left\|\int_0^s \psi_r^1(\tau) d \tau\right\|^2\right] & \leq T \mathrm{E}\left[\int_0^t \sup _{\tau \in [ 0, s]}\left\|\psi_r^1(\tau)\right\|^2 d s\right] \\
& \leq \frac{T^2 8\left(r C\right)^2 }{p}
\end{align}

and
\begin{align}\label{eq_63}
E\left[\sup _{s \in[0, t]}\left\|\int_0^s \psi_r^2(\tau) d \tau\right\|^2\right] \leq k_{1, r} T \int_0^t E\left[\sup _{\tau \in[0, s]}\left\|e_r(\tau)\right\|^2\right] d s.
\end{align}

Therefore, in view of \eqref{eq_56}, \eqref{eq_60}, \eqref{eq_62}, and \eqref{eq_63}, we have
$$
\begin{aligned}
& E\left[\sup _{s \in [0, t]}\left\|e_r(s)\right\|^2\right]  \leq\left(16 k_{2,r}+4 k_{1,r} T\right) \int_0^t E\left[\sup _{\tau \in [0, s]}\left\|e_r(\tau)\right\|^2\right] d s +\frac{128 T r^2  C^2 +32 T^2 r^2 C^2 }{p}.
\end{aligned}
$$

Then, Gr\"onwalls lemma yields
$$
\begin{aligned}
& \mathrm{E}\left[\begin{array}{l}
\sup _{t \in[0, T \mid}\left\|e_r(t)\right\|^2
\end{array}\right] \leq\left(\frac{r^2 C^2(128 T   +32 T^2 ) }{p}\right) 
 \exp\left(16 k_{2,r} T+4 k_{1, r} T^2\right) \leq \frac{\varepsilon}{2},
\end{aligned}
$$
where the last inequality follows from the choice of $p$.
\end{proof}

We here state the approximation result by Barron, see Propostion~1 in \cite{barron93} that we use in the above proof.

\begin{theorem}\label{th:barron}
Let $r >0$ and $C>0$ and consider the class of functions $\Gamma_{C}$. Then 
for every function $f: \mathbb{R}^2 \to \mathbb{R}^2$ with $f^i \in \Gamma_{C}$, for $i=1,2$, there exists a one-hidden layer neural network of the form $f_p=(f^1_p, f^2_p)$
\begin{align}\label{eq:NNbasic}
f^i_p(x)= \sum_{k=1}^p c^i_k \phi(\langle a^i_k, x \rangle+b^i_k)+ c^i_0, \quad i=1, 2,
\end{align}
where $a^i_k \in \mathbb{R}^2$, $b^i_k, c^i_k, c^i_0 \in \mathbb{R}$ for $k=1, \ldots, p $, $i=1, 2$, and $\phi $  a sigmoidal activation function such that for every probability measure on $B_r(\mathbb{R}^2) \subseteq \mathbb{R}^2 $
\[
\int_{B_r(\mathbb{R}^2)} \| f(x)-f_p(x)\|^2 \mu(dx) \leq \frac{8(r C)^2 }{p}.
\]
Moreover,  the coefficients $c_k^i$ may be restricted such that $\sum_{j=1}^p |c_k^i|\leq 2rC$ and $c^i_0=f^i (0)$.
\end{theorem}

Having  established this universality result, we shall now exploit the  advantages of neural SDEs. That is, they allow to combine the appealing flexibility of neural networks, in particular in view of optimization techniques via stochastic gradient descent methods, with well-known theory from mathematical finance based on continuous semimartingales.
In view of model calibration, one main goal of  this paper is to present a robust methodology to learn the parameters of the neural networks $(\sigma, b^V, \sigma^V)$ from market data. To this end we take both, derivatives' prices and time series data into account.

\color{black}
\section{Change of measure for neural SDEs}\label{Change of measure for neural SDEs}

The calibration of neural SDEs can in principle be extended to the real world measure by considering an additional parametric function $\zeta: [0,T] \times \mathbb{R}^2\times \mathbb{R}^p \to \mathbb{R}^2$ that could be another neural network and by extending the parameter space to $\Theta = \Theta^b \times \Theta^{\sigma}\times \Theta^{\zeta}\subset \mathbb{R}^p$\footnote{For notational convenience we still denote the dimension of the parameter space by $p$}. Let $Y_t = \log S_t$ and define the historical measure $\mathbb{P}(\theta)$ via the Radon-Nikodym derivative
\begin{align}
    \frac{\mathrm{d} \mathbb{P}(\theta)}{\mathrm{d} \mathbb{Q}(\theta)} \coloneqq \exp \Big( \int_{0}^T \begin{pmatrix} \zeta^1(t, Y^\theta_t, V^\theta_t, \theta) \\ \zeta^2 (t, Y^\theta_t, V^\theta_t, \theta) \end{pmatrix}^{\top} \begin{pmatrix} \sigma(t, S^\theta_t, V^\theta_t, \theta) & 0 \\ \rho \sigma^V(t, V^\theta_t, \theta) & \sqrt{1-\rho^2}\sigma^V(t, S^\theta_t, V^\theta_t, \theta) \end{pmatrix} \begin{pmatrix} \mathrm{d} W_t^{1} \\ \mathrm{d} W_t^{2} \end{pmatrix}\Big).
\end{align}

The drift parts under this measure are now defined via
\begin{align}
    b^{Y,\mathbb{P}}(t,Y_t^{\theta},V_t^{\theta},\theta) &\coloneqq r- \frac{1}{2}\sigma (t,\exp(Y_t^{\theta}),V_t^{\theta},\theta)^2+(\sigma(t,\exp(Y_t^{\theta}),V_t^{\theta},\theta))^2\zeta^1(t,Y_t^{\theta},V_t^{\theta},\theta)\\
&\quad +\rho \sigma(t,\exp(Y_t^{\theta}),V_t^{\theta},\theta) \sigma^V(t,V_t^{\theta},\theta)\zeta^2(t,Y_t^{\theta},V_t^{\theta},\theta), 
\label{real_world_drift1}\
\end{align}
\begin{align}
   b^{V,\mathbb{P}}(t,Y_t^{\theta},V_t^{\theta},\theta) &\coloneqq b^V(t,V_t^{\theta},\theta)+\rho \sigma(t,\exp(Y_t^{\theta}),V_t^{\theta},\theta)  \sigma^{V}(t,V_t^{\theta},\theta)\zeta^1(t,Y_t^{\theta},V_t^{\theta},\theta)\\&+(\sigma^{V}(t,V_t^{\theta},\theta))^2 \zeta^2(t,Y_t^{\theta},V_t^{\theta},\theta),\label{real_world_drift2}\end{align}
   where $\zeta$ plays the role of the market price of risk.
Assuming that $\zeta$ is such that Novikov's condition is satisfied, Girsanov's theorem yields a Brownian motion $(W_t^{\mathbb{P}(\theta)})_{t\in [0,T]}$ such that\\ 
\[
W_t^{\mathbb{P}(\theta)}=W_t-\int_0^t \begin{pmatrix}\zeta^1(s, Y^\theta_s, V^\theta_s, \theta) \sigma(s, S^\theta_s, V^\theta_s, \theta) + \zeta^2 (s, Y^\theta_s, V^\theta_s, \theta) \rho \sigma^V(s, S^\theta_s, V^\theta_s, \theta)\\
\zeta^2 (s, Y^\theta_s, V^\theta_s, \theta) \sqrt{1-\rho^2} \sigma^V(s, S^\theta_s, V^\theta_s, \theta) \end{pmatrix} \mathrm{d} s,
\]
where $W=(W^1,W^2)$
and
\begin{equation}\label{eq:real_world_Y}
 \mathrm{d} Y_t^{\theta}=b^{Y,\mathbb{P}}(t,Y_t^{\theta},V_t^{\theta},\theta)\mathrm{d} t+\sigma(t,\exp(Y_t^{\theta}),V_t^{\theta},\theta)\mathrm{d} W_t^{\mathbb{P}(\theta),1},   
\end{equation}
\begin{equation}
  \mathrm{d} V_t^{\theta}=b^{V,\mathbb{P}}(t,V_t^{\theta},\theta)\mathrm{d} t+\sigma^V(t,V_t^{\theta},\theta)(\rho \mathrm{d}W_t^{\mathbb{P}(\theta),1}+\sqrt{1-\rho^2}  \mathrm{d} W_t^{\mathbb{P}(\theta),2}).
\end{equation}
This neural SDE under the historical measure $\mathbb{P}(\theta)$ can now be used to learn the market price of risk using market data such as a time series of asset prices.

Note that the proposal of~\cite{gierjatowicz2020robust} to calibrate under the historical measure as well as under the risk neutral one is actually implemented and transferred into a Bayesian framework. 
\section{Variance reduction for Monte Carlo pricing via hedging strategies}\label{Variance reduction for Monte Carlo pricing via hedging strategies}
\color{black}In the following chapters, we will propose a method to train the neural SDE introduced in Section \ref{Stochastic volatility models as neural SDEs} that relies on classical Monte Carlo option pricing. The variance of a Monte Carlo estimator decreases as the number of simulated trajectories increases. The square root of this variance, known as the standard error, can theoretically be reduced to any desired level by simulating enough paths. However, practical constraints on computational resources often make it difficult to achieve sufficiently low error. To address this, variance reduction techniques are employed to improve the efficiency of Monte Carlo simulations. Among available methods, variance reduction via control variates is a widely used approach and in the following we elaborate on a version of this technique, that relies on deep learning.
\color{black}
As outlined, e.g. in
\cite{Cuchiero_2020} and \cite{gierjatowicz2020robust} variance reduction in the computation of the Monte Carlo price of an option can be achieved using an approximate hedging portfolio with a large correlation between the option payoff and the hedge. In our setting this variance reduction will be achieved using hedging strategies which are parameterized via neural networks. More specifically, consider $p$ traded hedging instruments $(Z_t)_{t\in[0,T]}$ (including in  particular the traded assets $S^{\theta}$) which are square integrable martingales under the risk neutral measure $\mathbb{Q}(\theta)$ and take values in $\mathbb{R}^p$. Assume that the payoff of the option is a function of the terminal values of $Z$, i.e. $C=g(Z_T)$. \textcolor{black}{We choose to specify the hedging strategy as a function of time and $Z$, }\color{black} i.e., $$h: \mathbb{R}_{+} \times \mathbb{R}^{p} \to \mathbb{R}^{p}, \; h_t=h(t,z)  ,$$ where $h$ will be parameterized as a neural network. The optimal hedge determined via some convex loss function $l: \mathbb{R} \to \mathbb{R}_{+}$ 
for the claim $C$ with corresponding market price $C_{mkt}$ can now be computed by \color{black}$$\inf_{\xi} E^{\mathbb{Q}(\theta)}\left[l \left(-C+C_{mkt}+\left(h\left(\cdot,Z_{\cdot},\xi\right) \boldsymbol{\cdot} Z_{\cdot}\right)_T \right)\right],$$ 
\color{black}where  $\xi$ stands for the weights of the neural network $h$ and $(h \boldsymbol{\cdot}Z)_T$ denotes the stochastic integral with respect to $Z$. Modifying the calibration objective via the above explained variance reduction by means of hedging strategies allows a substantial reduction in the \color{black} variance \color{black}of the Monte Carlo samples. \color{black} In this context, we also refer to related work on variance reduction using neural networks, see \cite{arandjelovic2024importance}.
\color{black}
\section{Robust option calibration} \label{options}
To obtain a posterior distribution over derivatives prices and in turn price bounds for derivatives, first a likelihood function will be specified. Together with a prior distribution $p(\theta)$ on the weights $\theta$ of the neural networks $(\sigma,b^V,\sigma^V)$ appearing in the neural SDE system \eqref{priceprocess},\eqref{varianceprocess}, the desired posterior distribution will then be obtained explicitly. The calibration to derivatives 
will be with respect to call options only. However, the loss function specified in the following subchapter can be easily modified to allow for different terminal value payoffs.

\subsection{Option calibration}
The likelihood is defined via assumptions on the
deviations of the options' market prices to the 'true' model prices as specified below. The observed market prices are specified as mid prices of bid and ask prices, i.e.~$C_{mkt}^{(i)}=\frac{1}{2}\big (C_{mkt}^{(i)bid}+C_{mkt}^{(i)ask} \big) $, $i=1,...,J$. 
The options are written on the underlying $S$ and their payoffs are denoted by $\{\Phi_i\}_{i=1}^J$, where $J$  denotes  the number of options considered. From now on we assume for notational simplicity that $r=0$, unless specified otherwise. We write $C_{mod}^{(i)}(\theta)=E^{\mathbb{Q}(\theta)}[\Phi_i]$ for the corresponding
model prices with model parameters $\theta$.  Similarly as in~\cite{Gupta2014RobustCO}, let $\delta_i=\frac{10^4}{S_0} \mid C_{mkt}^{(i) ask}-C_{mkt}^{(i) bid} \mid$ denote the basis point bid-ask spread of the $i^{th}$ option at time zero. Then the basis point calibration error for the $i^{th}$ option \textcolor{black}{for the true, underlying $\theta$} is supposed to be normally distributed as 

\begin{equation}
   \frac{10^4}{S_0}( C_{mod}^{(i)}(\theta)-C_{mkt}^{(i)})\sim \mathcal{N}(0,\delta_{i}^{2})\label{calibrationmodel}.
\end{equation}
The natural interpretation of $\eqref{calibrationmodel}$ is that in theory there exists a 'true' model yielding an arbitrage free market and that the market price of the option equals its model price plus some
normally distributed random noise. 
In order to fit the implied volatility surface the basis point error function for calibration will be modified using a weighting scheme (compare~\cite{vegaweights}). Indeed, 
define $\delta^2=\frac{1}{\sum_{i=1}^J \frac{1}{\delta_i^2} }$ and $w_i=\frac{\delta^2}{\delta_i^2}$ so that

\begin{equation} \label{eq:multnormal}
\frac{10^4}{S_0}    \left(\sqrt{w_1} ( C_{mod}^{(1)}(\theta)-C_{mkt}^{(1)}), \ldots, \sqrt{w_J} ( C_{mod}^{(J)}(\theta)-C_{mkt}^{(J)} ) \right) \sim   \mathcal{N}(0, \delta^2 I_J ),
\end{equation}
where $I_J$ denotes the $J \times J$ identity matrix. Note that$\sum_{i=1}^J w_i = 1$, by definition.
The corresponding likelihood is then proportional to $
\exp \left(-\frac{1}{2 \delta^2}G(\theta)\right)$ where 
$G(\theta)$
is defined as
\begin{equation}
G(\theta)= \frac{10^8}{S_0^2} \sum_{i=1}^J    w_i \mid C_{mod}^{(i)}(\theta)-C_{mkt}^{(i)} \mid ^2.
\end{equation} 
With this specify the Bayesian likelihood function as \\
{\color{white}-}\\ %
\begin{equation}
     l(C_{mkt} \mid \theta) =  l\left((C_{mkt}^1,\ldots,C_{mkt}^J) \mid \theta \right)\propto  \exp \left(-\frac{1}{2 \delta^2}G(\theta)\right).
\end{equation}

\color{black}
The prior distribution for the neural network weights $\theta$ plays a crucial role in the Bayesian modeling framework.
In many applications, historical data or expert knowledge may be used for this prior decision. In this work, the prior choice is found through theoretical considerations about neural network initialization.  More precisely, it corresponds to the so called Glorot normal initialization, introduced in~\cite{Glorot2010UnderstandingTD}. In this work the authors study the forward pass equation and back-propagated gradients, inspired by the paper \cite{Bagnell2010LearningIM}, who established that the variance of the back-propagated gradients diminishes as one moves from the output layer to the input layer, even if no training has occurred yet. ~\cite{Glorot2010UnderstandingTD} consider a simplified setting to propose a heuristic for the initialization of neural network parameters, maintaining a high information flow across layers. The authors suggest to randomly initialize neural network weights drawing from a normal distribution centered at zero with standard deviation \begin{equation}\sigma_{prior}= g \cdot \sqrt{\frac{2}{\text{fan in} + \text{fan out}}}, \label{sigma_prior} \end{equation}  
where $g$ denotes an optional scaling factor, fan in stands for the number of neurons in the predecessor layer and fan out stands for the number of neurons in the successor layer, so that the initialization scheme depends on the width of layers.  Thus, the prior distribution is specified as \begin{equation}
    p(\theta) \sim \mathcal{N}(0,\sigma_{prior}^2 I_J)\label{prior}.
\end{equation} 
Since in our architecture, the hidden layers are of width 100, the value of $\sigma_{prior}$ is approximately 0.15 for most neural network weights, as the scaling factor $g$ is chosen to be 1.5. We will study in a later section how the prior variance affects the convergence of the suggested algorithm. \\ 
The posterior distribution can now be derived explicitly using the prior and likelihood formulated above as \begin{equation}
    f\left(\theta \mid C_{mkt}\right) = l\left(C_{mkt} \mid \theta\right) \cdot p\left(\theta\right) \propto \exp \left(-\frac{1}{2 \delta^2} G(\theta)\right) \cdot \exp \left(-\frac{\|\theta\|^2}{2\sigma_{prior}^2}\right)\label{posterior} .
\end{equation}
We would like to emphasize that the inclusion of the prior term corresponds to Tikhonov regularization with parameter $\delta^2/\sigma_{prior}^2$ as is explained in \cite{greenberg} for instance.

\subsection{Training the model}
Note that in practice the true model price for the call option $i$ with maturity $T_i$, strike $K_i$ and payoff $\Phi_i=(S_{T_i}^{\theta}-K_i)^+$, given by  $C^{(i)}_{mod}(\theta)=E^{\mathbb{Q}(\theta)}[\Phi_i]$, needs to be approximated. The choice of the standard Monte Carlo estimator means that after simulating $M$ price trajectories $(\omega_m)_{m=1}^M$ a Monte Carlo approximation to $C_{mod}^{(i)}(\theta)$ is given by computing $C^{(i)}_{mod}(\theta)=\frac{1}{M} \sum_{m=1}^M (S_{T_i}^{\theta}(\omega_m)-K_i)^{+}$, under the risk-neutral measure.
This then leads to an approximation of  $G(\theta)$ as $$G(\theta)= \frac{10^8}{S_0^2} \sum_{i=1}^J    w_i \left | C_{mod}^{(i)}(\theta)-C_{mkt}^{(i)} \right | ^2 \approx \frac{10^8}{S_0^2} \sum_{i=1}^J    w_i \left| \left(\frac{1}{M} \sum_{m=1}^M \left (S_{T_i}^{\theta}(\omega_m)-K_i \right)^{+}\right)-C_{mkt}^{(i)} \right| ^2.$$ Of course this approximation entails a Monte Carlo error of order $\frac{1}{\sqrt{M}}$, meaning that the number of simulated trajectories $M$ needs to be quite large in order to find a good approximation of the true model prices. The hedging control variates introduced above provide a remedy for this issue and reduce the number of sample paths that are needed for a good approximation. For this reason the final approximation used for $G(\theta)$ is of the form 

\begin{equation}
G(\theta) \approx \frac{10^8}{S_0^2} \sum_{i=1}^J    w_i \left | \left(\frac{1}{M} \sum_{m=1}^M \left(S_{T_i}^{\theta}(\omega_m)-K_i\right)^{+}\right)-C_{mkt}^{(i)} - \left(h(\cdot,Z_{\cdot}^{\theta}(\omega_m),\xi) \cdot Z_{\cdot}^{\theta}(\omega_m)\right)_{T_i} \right | ^2.\label{MC_estimator_reduced_V}\end{equation}

The model parameters will be estimated using a gradient descent scheme. More precisely, 
during the training we perform an updating step $\Delta \theta_j$, with step size $\epsilon_j$ of the neural network weights in iteration $j$ of the training algorithm utilizing the Langevin diffusion (see for example \cite{Roberts1996ExponentialCO}) via \begin{equation}
     \Delta \theta_j :=\frac{\epsilon_j}{2} \Big(\nabla  \log p(\theta_j) +  \nabla \log l(C_{mkt} \mid  \theta_j) \Big) +\eta_j,\;\; \eta_j \sim \mathcal{N}(0,\epsilon_j)\label{eq:18}
\end{equation}
where $\log l(C_{mkt} \mid  \theta)$ is given by
\begin{equation}
    \log l(C_{mkt} \mid \theta)=-\frac{J}{2} \log(2\pi)- \frac{J}{2}\log(\delta^2)-\frac{1}{2\delta^2}G(\theta).\label{eq:19}
\end{equation} 

The updated neural network weights are now taken as an input for the next iteration to compute the model prices, given $\theta_j$. This procedure results in a sequence of Monte Carlo  option prices, given the neural network weights. After an adequate burn-in phase, this chain will stabilize around the posterior distribution of option prices. \textcolor{black}{Note, that \ref{MC_estimator_reduced_V} depends on $h$, the control variate introduced in Section \ref{Variance reduction for Monte Carlo pricing via hedging strategies} as a neural network with parameters $\xi$. Thus, there is an interplay between the neural network parameters $\theta$ and $\xi$ which has to be taken into account for the training procedure. We follow the approach suggested in \cite{gierjatowicz2020robust}, where in a given iteration of the algorithm, the parameters $\xi$ are frozen, meaning that gradients are not computed and propagated, while the parameters $\theta$ are updated. Then the procedure is done vice versa.}
The samples drawn after the burn-in phase then yield approximate realizations of the desired posterior distribution, which can then be utilized to obtain bounds on derivatives prices.
The detailed algorithm, where the hedging instrument $Z$ is chosen to be the traded asset $S^{\theta}$ to compute \eqref{MC_estimator_reduced_V}, is explained below: \\
{\color{white}-}\\ %
\textbf{Algorithm 1}
\par\noindent\rule{\textwidth}{0.4pt}
\textbf{Input:}  $\{t_0,t_1,...,t_{N_{steps}}\}$ time grid for numerical scheme, $(\Phi_i)_{i=1}^{J}$ option payoffs, market option prices $C_{mkt}(\Phi_i),\;i=1,...,J$, weights $w_{i},\;i=1,...,J$\\
 initialize neural network parameters $\theta$ via $p(\theta)$ and choose step size $\epsilon$ 
\\ 
\textbf{for} epoch: 1:$N_{epochs}$ \textbf{do}
\begin{itemize}
    \item[]  Generate $M$ paths $\left(S_{t_n}^{\theta}(\omega_m),V_{t_n}^{\theta}(\omega_m)\right)_{n=0}^{N_{steps}} , \; m=1,...,M $ using e.g. an Euler scheme.
    \item[]  \textbf{During one epoch:} Freeze $\xi$, use Langevin Dynamics to update $\theta$, via $$ \Delta \theta = \frac{\epsilon}{2}  \Big[  \nabla \Big( -\frac{J}{2} \log(2 \pi)- \frac{J}{2}\log(\delta^2)-\frac{1}{2\delta^2} \tilde{G}(\theta)\Big)+ \nabla \log p(\theta)\Big] + \eta,$$
    where $\tilde{G}(\theta)= \frac{10^8}{S_0^2} \sum_{i=1}^J    w_i \left | \tilde{C}_{mod}^{(i)}(\theta)-C_{mkt}^{(i)} \right | ^2$,\\$\tilde{C}_{mod}^{(i)}(\theta)=
    \frac{1}{M} \sum_{m=1}^M \left[\left(S_{T_i}^{\theta}(\omega_m)-K_i\right)^{+}-\sum_{n=0}^{T_i}\overline{h}\left(t_n,S_{t_n}^{\theta}(\omega_m),\xi\right)\left(S^{\theta}_{t_{n+1}}(\omega_m)-S_{t_{n}}^{\theta}(\omega_m)\right)\right]$ with $\overline{h}$ being the specific control variate explained in the previous Section, $\eta \sim \mathcal{N}(0,\epsilon)$.
    \item[]  \textbf{During one epoch:} Freeze $\theta$, use ADAM (~\cite{kingma2017adam}) to update $\xi$, by optimizing the sample variance $$ \xi^{*} \in  \argmin_{\xi} \sum_{i=1}^J Var^{\mathbb{Q}(\theta)}\left[\tilde{C}_{mod}^{(i)}(\theta)\right], $$ 
where $Var^{\mathbb{Q}(\theta)}[\cdot]$ denotes the sample variance under $\mathbb{Q}(\theta)$.
\end{itemize}
\textbf{end for}\\
{\color{white}-}\\ %
\textbf{return} $\theta,\xi $ for all prices $(\Phi_i)_{i=1}^J.$ \\\\The output of this optimization procedure yields a posterior distribution of the parameters $\theta$. This posterior can in turn be used to obtain robust price bounds for options and the corresponding implied volatility surface. 
In the case where the set of derivatives  can be grouped by maturity, as in the case of call and put options, one can split the networks into one network per maturity. One can then perform a special kind of stepwise learning. More precisely let $\theta=(\theta_1,...,\theta_{N_{maturity}})$, where $N_{maturity}$ is the number of maturities and define
$b^V(t,V_t^{\theta},\theta):= \sum_{i=1}^{N_{maturity}}\mathds{1}_{t \in [T_{i-1},T_i)}(t)b^{V,i}(t,V_t^{\theta},\theta_i)$,
$\sigma(t,S_t^{\theta},V_t^{\theta},\theta):= \sum_{i=1}^{N_{maturity}}\mathds{1}_{t \in [T_{i-1},T_i)}(t)\sigma^i(t,S_t^{\theta},V_t^{\theta},\theta_i)$ and $\sigma^V(t,V_t^{\theta},\theta):= \sum_{i=1}^{N_{maturity}}\mathds{1}_{t \in [T_{i-1},T_i)}(t)\sigma^{V,i}(t,V_t^{\theta},\theta_i)$, where each $b^{V,i}$, $\sigma^i$ and $\sigma^{V,i}$ is a feedforward neural network. Note that the corresponding neural network parameters $(\theta_i)_{i=1}^{N_{maturity}}$ can be initialized using different standard deviations $\sigma_{prior,i}$ for each maturity.\\
\newline
\textbf{Algorithm 2}
\par\noindent\rule{\textwidth}{0.4pt}
\textbf{for} $k:$  $1:N_{maturity}$ \textbf{do}
\begin{itemize}
    \item[] Calibrate the neural SDE to market prices observed for maturity $k$ via the Algorithm 1
    \item[] Freeze the neural network parameters corresponding to maturity $k$
\end{itemize}
\textbf{end for}

\subsection{The rough Bergomi model}\label{Rough_Bergomi_section}
\begin{figure}[htbp]
    \centering

    \begin{subfigure}[b]{0.45\textwidth}
        \centering
        \includegraphics[width=\textwidth]{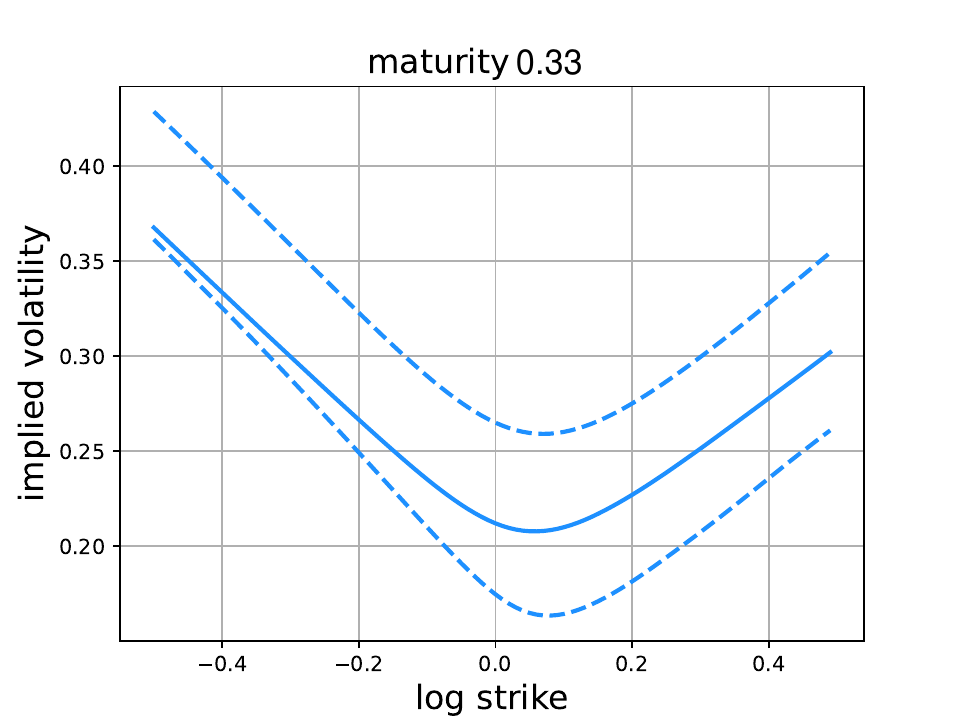}
        \label{fig:subfig1}
    \end{subfigure}
    \hfill
    \begin{subfigure}[b]{0.45\textwidth}
        \centering
        \includegraphics[width=\textwidth]{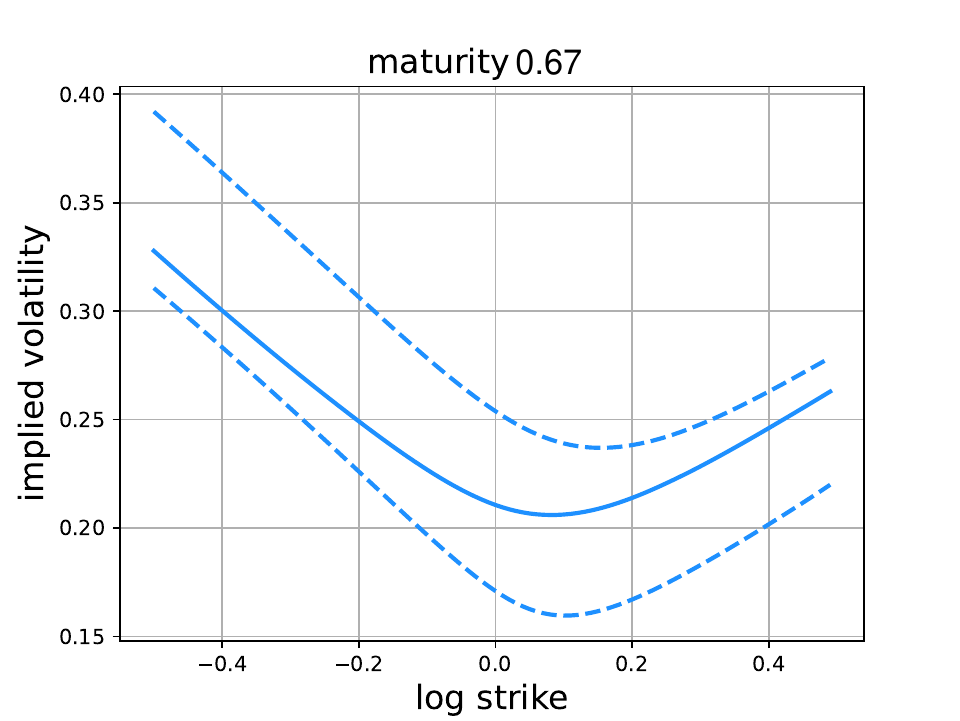}
        \label{fig:subfig2}
    \end{subfigure}

    \caption{Robust calibration bounds on the implied volatility surface in the rough Bergomi model, exemplified with 2 maturities.}
    \label{rbergomi_ivol}
\end{figure}

 \textcolor{black}{For the first simulation study, we calibrate with respect to artificial data generated by the rough Bergomi model, as introduced in~\cite{bayer2016pricing}. The original motivation behind rough volatility models is that neither local volatility nor traditional stochastic volatility models can simultaneously capture realistic market dynamics, while also accurately fitting the shape of the implied volatility surface. Thus, an alternative perspective involves moving beyond conventional Brownian motion-based stochastic volatility models and exploring models driven by fractional Brownian motion.
 One representative of this model class is the rough Bergomi model.
This model is specified as 
 $$S_t := \exp \left( \int_0^t \sqrt{ V_u } \mathrm{d} W^1_u - \frac{1}{2}\int_0^t V_u \mathrm{d} u \right),\quad W^2_u:=\rho W_t^1 + \sqrt{1 - \rho^2}U_t, $$
 $$V_t := \xi\ \exp \left( \eta Y^a_t - \frac{\eta^2}{2} t^{2a + 1}\right), \quad Y_t^a := \sqrt{2a + 1} \int_0^t (t - u)^a  \mathrm{d} W^2_u,$$
 for two independent Brownian motions $(W^1, U)$. \textcolor{black}{The correlation parameter $\rho \in [-1, 1]$ is typically negative and introduces the leverage effect, where falling asset prices are associated with increasing volatility — an effect commonly observed in equity markets. The parameter $\xi > 0$ sets the initial level of variance. It determines the scale of the volatility surface and acts as a base level around which the stochastic variance process fluctuates. The parameter $\eta > 0$ controls the volatility of volatility. 
Finally, the exponent $a \in (-\frac{1}{2}, 0)$ determines the roughness of the volatility path. Smaller  values of $a$ produce rougher trajectories. The stochastic Volterra process \( Y_t^a \) is defined by a singular kernel \( (t-u)^a \), and introduces non-Markovian memory into the volatility dynamics.} Even though this is a non-Markovian model, we illustrate via a simulation study that the neural SDE is nonetheless able to adequately capture the implied volatility surface stemming from this more complex modeling framework. The theoretical justification to mimick non-Markovian models with Markovian dynamics goes back to \cite{gyongy1986mimicking}, who showed that for a certain class of non-Markovian stochastic processes, there exists a stochastic differential equation that has the same one-dimensional marginal law as the original non-Markovian stochastic process at every time point. This idea is often referred to as \textit{Markovian projections} and a recent account on this topic is given in \cite{larsson2024markovian}, where the idea is extended to It\^o semimartingales with jumps. In the context of rough volatility, we would like to mention the work in \cite{bourgey2023local}, where the a Markovian projection is used to study the local volatility surface that is generated by rough volatility dynamics. We argue here empirically, that the neural SDE is able to learn such a Markovian projection in the case of the rough Bergomi model.\\\\
 We use the GitHub repository \href{https://github.com/ryanmccrickerd/rough_bergomi/blob/master/notebooks/rbergomi.ipynb}{github.com/ryanmccrickerd/rbergomi}, where the hybrid scheme suggested in~\cite{bennedsen2017hybrid} for efficient computation of  Volterra processes (needed in the rough Bergomi model) is implemented. The approach in~\cite{McCrickerd_2018} then yields the target implied volatility surface of the rough Bergomi model. 
 As parameters we choose $\rho=-0.5$, $a=-0.43$, $\xi=0.235^2$ and $\eta=1.9$. 
 The neural SDE is discretized via the Euler Maruyama scheme with $N_{steps}=8 \times 12$ uniform time steps for $T=1$, which corresponds to 8 steps for every month. 
 As hyperparameters for the Bayesian calibration we choose $\delta^2 = 3$ and \textcolor{black}{$\sigma_{prior}=0.15$.} 
 All numerical studies performed in this paper have been implemented using \textbf{\textsf{PyTorch}} introduced in ~\cite{NEURIPS2019_9015} and are available in form of a GitHub repository at \href {https://github.com/evaflonner/Calibration-of-Neural-SDEs-using-Bayesian-Methods}{github.com/evaflonner/Calibration-of-Neural-SDEs-using-Bayesian-Methods}. Furthermore, for all numerical studies presented the number of Monte Carlo paths used to train the neural SDE was $M=5000$ and training on a standard machine with hardware accelerated GPU availability takes less than 20 minutes, which is quite fast compared to classical Markov Chain Monte Carlo schemes. The model \eqref{priceprocess} \eqref{varianceprocess} is calibrated using the Bayesian algorithm explained above, with the goal to match vanilla call prices with bi-monthly maturities for 
 101 log-strike prices evenly distributed between -0.5 and 0.5, which are generated by the rough Bergomi model, where the initial price of the underlying is $S_0=1$. After the training procedure, it is possible to use the resulting distribution on neural network weights in order to obtain bounds on the implied volatility smile, as illustrated in Figure~\ref{rbergomi_ivol}. More specifically the lower (upper) bound corresponds to an interpolation of the $10$-$\%$-quantile ($90$-$\%$-quantile)
 of the draws obtained by the Langevin sampler after convergence, where convergence has been judged by the stabilization of the trace plots of the posterior draws.}
 
\color{black}

\color{black}
\section{Including time series characteristics}\label{Including time series characteristics}

In the following, we suggest to extend the above calibration to option price data additionally by including 
 time series data of the logarithmic asset prices $(Y_t)_{t\geq 0}$.
The result of this combined approach is a model that is  calibrated to time-series data while yielding a distribution on the implied volatility surface with robust price bounds for option prices. We would like to emphasize that we assume both, $(Y_t)_{t\geq 0}$ and $(V_t)_{t\geq 0}$ to be observable. In any empirical application, $(V_t)_{t\geq 0}$ would have to be estimated from market data. Recall that the dynamics of the neural SDE under the real world measure $\mathbb{P}(\theta)$ are given by \begin{equation}
 \mathrm{d} Y_t^{\theta}=b^{Y,\mathbb{P}}(t,Y_t^{\theta},V_t^{\theta},\theta)\mathrm{d} t+\sigma(t,\exp(Y_t^{\theta}),V_t^{\theta},\theta)\mathrm{d}W_t^{\mathbb{P}(\theta),1}, 
\end{equation}
\begin{equation}
  \mathrm{d}V_t^{\theta}=b^{V,\mathbb{P}}(t,V_t^{\theta},\theta)\mathrm{d}t+\sigma^V(t,V_t^{\theta},\theta)(\rho \mathrm{d}W_t^{\mathbb{P}(\theta),1}+\sqrt{1-\rho}  \mathrm{d}W_t^{\mathbb{P}(\theta),2}).
\end{equation}
Similarly as in~\cite{Yu} note that conditional on the neural network parameters $\theta$, $Y_{t_n}$ and $V_{t_n}$, for the timegrid $\{t_0,t_1,...,t_{N_{steps}}\}$, $Y_{t_{n+1}}-Y_{t_n}$ and $V_{t_{n+1}}-V_{t_n}$ follow a bivariate normal distribution
\[\begin{pmatrix}
Y_{t_{n+1}}-Y_{t_n} \\
V_{t_{n+1}}-V_{t_n}
\end{pmatrix}
\mid  \theta, Y_{t_n}, V_{t_n}
\sim N\left(\begin{pmatrix}
b^{Y,\mathbb{P}} \Delta t_n \\
b^{V,\mathbb{P}} \Delta t_n
\end{pmatrix},\Delta t_n\begin{pmatrix}
\sigma^2 & \rho \sigma \sigma^V \\
\rho \sigma \sigma^V & (\sigma^V)^2
\end{pmatrix}\right),
\]

where $\Delta t_n=t_{n+1}-t_n$ stands for one time step.\\
Based on that the likelihood of the log price $Y=(Y_t)_{t\geq0}$ and the volatility $V=(V_t)_{t \geq 0}$ given the neural network parameters is given by $$h(Y,V \mid \theta) = \prod_{n=0} ^{N_{steps}-1} \frac{1}{2\pi  \sqrt{1-\rho^2}}  \exp\left(-\frac{1}{2(1-\rho^2)}\left((\psi_{t_{n+1}})^2-2\rho \psi_{t_{n+1}}^V \psi_{t_{n+1}} + (\psi_{t_{n+1}}^V)^2\right)\right), $$
where $\psi_{t_{n+1}}=(Y_{t_{n+1}}-Y_{t_n} - b^{Y,\mathbb{P}} \Delta t_n)/  (\sigma \sqrt{\Delta t_n})$ and $\psi_{t_{n+1}}^V = (V_{t_{n+1}}-V_{t_n}
-b^{V,\mathbb{P}} \Delta t_n)/(\sigma^V \sqrt{\Delta t_n})$. This approach is linked to so called Approximated-Likelihood-Methods (see for example in \cite{sarkka} chapter 11.4). As the transition densities typically used in the computation of the likelihood are not tractable, they  have to be approximated. One  approach  is  to  approximate  the  SDE  with  a continuous- or discrete-time system whose transition density we can evaluate. For example, recalling that one step in the Euler-Maruyama method has a Gaussian transition density this leads to an approximation as given above, from which the corresponding log-likelihood can be derived easily. 

The joint posterior distribution of neural network parameters given time series data and option prices observed on the market is given as \begin{equation}
    g(\theta \mid Y,C_{mkt}) 
    \propto l(C_{mkt} \mid  \theta)\cdot h(Y,V \mid \theta) \cdot p(\theta).
\end{equation} This posterior distribution now serves as the new loss function in Algorithm 1.1.\\ 
\textbf{Algorithm 1.1}
\par\noindent\rule{\textwidth}{0.4pt}
\textbf{Input:}  $\{t_0,t_1,...,t_{N_{steps}}\}$ time grid for numerical scheme, observations of market data $S=(S_{t_n})_{t_n=0}^{N_{steps}}$ and $V=(V_{t_n})_{_nt=0}^{N_{steps}}$, $(\Phi_i)_{i=1}^{J}$ option payoffs, market option prices $C_{mkt}(\Phi_i),\;i=1,...,J$, weights $w_{i} ,\;i=1,...,J$\\
 initialize neural network parameters $\theta$ via $p(\theta)$ and choose step size $\epsilon$ \\ 
\textbf{for} epoch: 1:$N_{epochs}$ \textbf{do}
\begin{itemize}
    \item[]  Generate $M$ paths $(S_{t_n}^{\theta}(\omega_m),V_{t_n}^{\theta}(\omega_m))_{n=0}^{N_{steps}} , \; m=1,...,M $ using e.g. Euler scheme under $\mathbb{Q}(\theta)$.\\
    Compute log-price process $(Y_{t_n}^{\theta})_{n=0}^{N_{steps}}$ under $\mathbb{P}(\theta)$ from $(X_{t_n}^{\theta})_{n=0}^{N_{steps}}$, using the neural network $\zeta$ for the market price of risk.
    \item[]  \textbf{During one epoch:} Freeze $\xi$, sample index $k$ from integers $\{1,2,...,M\}$ for the time-series dynamics, use Langevin Dynamics to update $\theta$  via 
    \begin{align}
\Delta \theta & = \frac{\epsilon}{2} \Bigg[ 
    \nabla \left( 
        -\frac{J}{2} \log(2\pi) 
        - \frac{J}{2} \log(\delta^2) 
        - \frac{1}{2\delta^2} \tilde{G}(\theta) 
    \right) \nonumber \\
    &+ \nabla \sum_{n=0}^{N_{\text{steps}}} \left( 
        -\frac{1}{2} \log(1 - \rho^2) 
        - \frac{1}{2(1 - \rho^2)} \left( 
            \left(\psi_{\Delta t_n})^2 
            - 2\rho\, \psi_{\Delta t_n}^{\mathrm{V}} \psi_{\Delta t_n} 
            + (\psi_{\Delta t_n}^{\mathrm{V}}\right)^2 
        \right) 
    \right) \nonumber \\
    &+ \nabla \log p(\theta)
\Bigg] + \eta
\end{align}
    where $\tilde{G}(\theta)$ is as defined above, $\eta \sim \mathcal{N}(0,\epsilon)$ and where
    $$\psi_{\Delta t_n} = \Big(Y_{t_{n+1}}-Y_{t_n} - b^{Y,\mathbb{P}}(t_n,Y_{t_n}^{\theta},V_{t_n}^{\theta},\theta)(\omega_k) \Delta t_n \Big)/ \sigma(t_n,\exp(Y_{t_n}^{\theta}),V_{t_n}^{\theta},\theta)(\omega_k)\sqrt{\Delta t_n}$$ and $$\psi_{\Delta t_n}^V = \Big(V_{t_{n+1}}-V_{t_n}-b^{Y,\mathbb{P}}(t_n,Y_{t_n}^{\theta},V_{t_n}^{\theta},\theta)(\omega_k) \Delta t_n \Big)/ \sigma^V(t_n,\exp(Y_{t_n}^{\theta}),V_{t_n}^{\theta},\theta)(\omega_k)\sqrt{\Delta t_n}$$
    \item[]  \textbf{During one epoch:} Freeze $\theta$, use ADAM (~\cite{kingma2017adam}) to update $\xi$, by optimizing the sample variance $$ \xi^{*} \in  \argmin_{\xi} \sum_{i=1}^J Var^{\mathbb{Q}(\theta)}\left[\Phi_i-\sum_{n=0}^{N_{steps}-1} \overline{h}\left(t_n, S_{t_n}^{\theta},\xi \right) \left(S^{\theta}_{t_{n+1}}-S_{t_{n}}^{\theta}\right)\right] $$ 
     
\end{itemize}
{\color{white}-}\\ %
\textbf{end for}
\\ \textbf{return} $\theta,\xi $ for all prices $(\Phi_i)_{i=1}^J$. 
\subsection{\textcolor{black}{Black-Scholes model}}
As a first step the approach suggested above will be tested using the~\cite{BS_model} model. This model is of course not well suited to match the empirically observed implied volatility surface.
The reason why we consider this model is that here the market price of risk is known in closed form and thus it can be verified if the suggested method is able to learn the theoretically given change of measure. Adapting the standard notation of the Black-Scholes model, the neural SDE for the log price process $Y$ under the historical measure then takes the form 
$$\mathrm{d}Y_t^{\theta}=\mu(t, Y_t^{\theta},\theta)\mathrm{d}t - \frac{1}{2} \sigma(t,\exp(Y_t^{\theta}),\theta)^2 \mathrm{d}t +  \sigma(t,\exp(Y_t^{\theta}),\theta)\mathrm{d} W_t,$$
where $\mu$ and $\sigma$  are neural networks.
Note that in the notation of 
\eqref{eq:real_world_Y}, $b^{Y,\mathbb{P}}(t, Y_t^{\theta},\theta)$ is given by
$$b^{Y,\mathbb{P}}(t, Y_t^{\theta},\theta)=\mu(t, Y_t^{\theta},\theta) - \frac{1}{2}
\sigma(t,\exp(Y_t^{\theta}),\theta)^2 \mathrm{d}t.$$
As mentioned above, in the Black-Scholes model the market price of risk is known in closed form and so in the notation of equation \eqref{real_world_drift1} and \eqref{real_world_drift2} we specify $\zeta^1$ as $(\mu-r)/\sigma^2$ and $\zeta^2$ as 0.
Let us remark that from a learning perspective it is equivalent to parameterize $\zeta_1$ as a neural network and to specify $b^{Y,\mathbb{P}}$ (or $\mu$) according to \eqref{real_world_drift1} or the other way round as we do here.

This neural SDE is initialized via the Glorot normal prior explained above and $M$ trajectories of it are simulated to match the law of one given trajectory of the log-price process. In order to also match the Black-Scholes option prices the Monte Carlo estimator for $C_{mod}(\theta)$, the price for an option with one specific maturity and strike $K$, is computed using \eqref{MC_estimator_reduced_V}.\\
Results for the posterior distribution that is obtained based on this procedure are presented below for a trading year consisting again of 96 time steps. The input data for the historical measure is one trajectory from the logarithm of a geometric Brownian motion with drift $0.05$ and diffusion $0.3$.
$$\mathrm{d} Y_t = (0.05-0.5 \cdot 0.3^2)\mathrm{d}t+0.3 \mathrm{d}W_t,\; Y_0 = 0 $$
simulated via the Euler-Maruyama scheme. The input for the $\mathbb{Q}$-measure are call option prices obtained from the Black Scholes formula for risk-free rate $r=0.025$, 21 strikes from 0.8 to 1.2 and maturities 16, 32, 48, 64, 80 and 96. 

\begin{table}[ht]
    \centering
    \begin{tabular}{cccc}
        \includegraphics[scale=0.33]{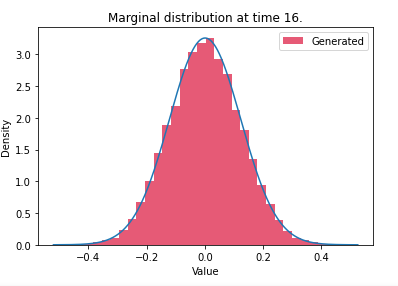}
   &   \includegraphics[scale=0.33]{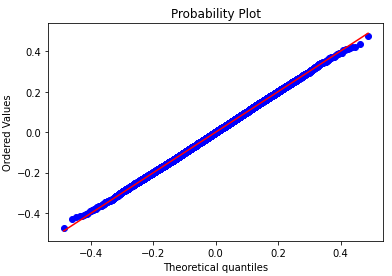}&\includegraphics[scale=0.33]{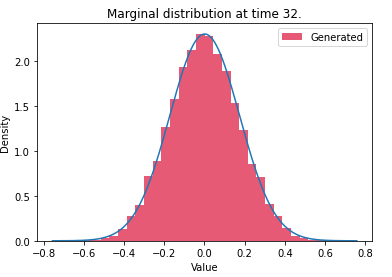} &
         \includegraphics[scale=0.33]{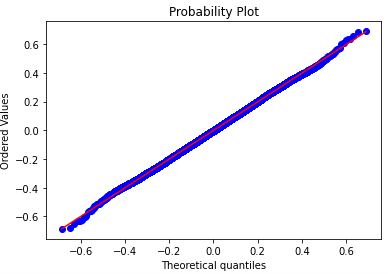}\\
         \includegraphics[scale=0.33]{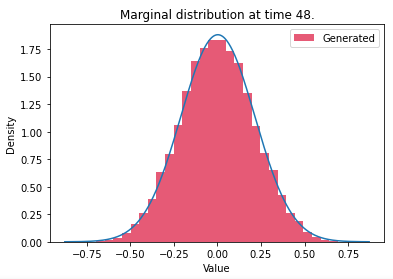} &
         \includegraphics[scale=0.33]{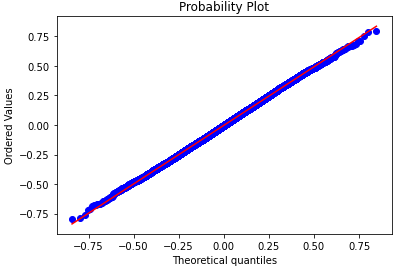}&
         \includegraphics[scale=0.33]{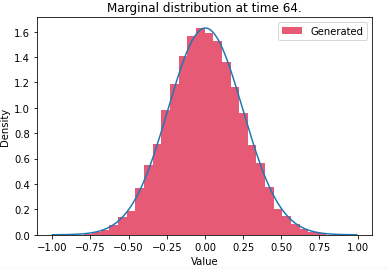} &
         \includegraphics[scale=0.33]{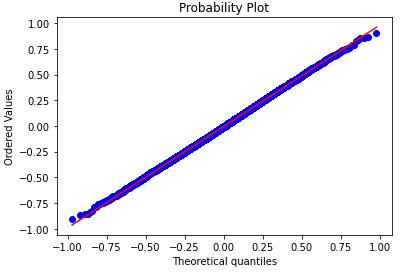}\\
         \includegraphics[scale=0.33]{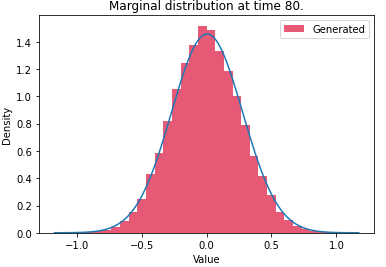}&
         \includegraphics[scale=0.33]{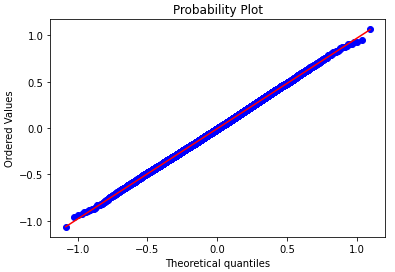}&
         \includegraphics[scale=0.33]{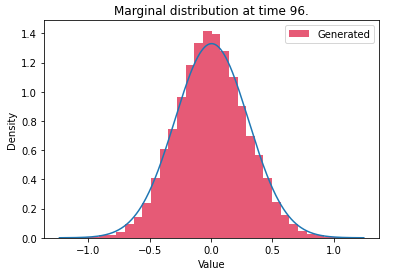}&
         \includegraphics[scale=0.33]{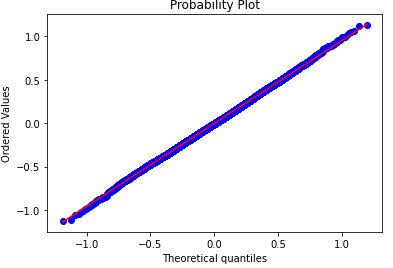}
         
    \end{tabular}
    \caption{Marginal distributions and qq-plots resulting from a calibration to simulated time series data.}
    \label{tb:table5}
\end{table}
{\color{white}-}\\ %
It is interesting to investigate the paths and the corresponding quantiles that are generated by the calibrated neural SDE in comparison to paths generated from the original Black-Scholes model, see Table \ref{tb:table5}.
\newline
{\color{white}-}\\ %
Additionally one can also compare the real Black-Scholes prices calculated via the  Black-Scholes formula for a given maturity, strike, risk-free rate and volatility and the posterior model prices that are obtained from the neural SDE by the above explained procedure in Table \ref{tb:table6}. Note that the dashed line represents the true Black-Scholes prices.
\newline

\begin{table}[!h]
    \centering
    \begin{tabular}{cccc}
         \includegraphics[scale=0.27]{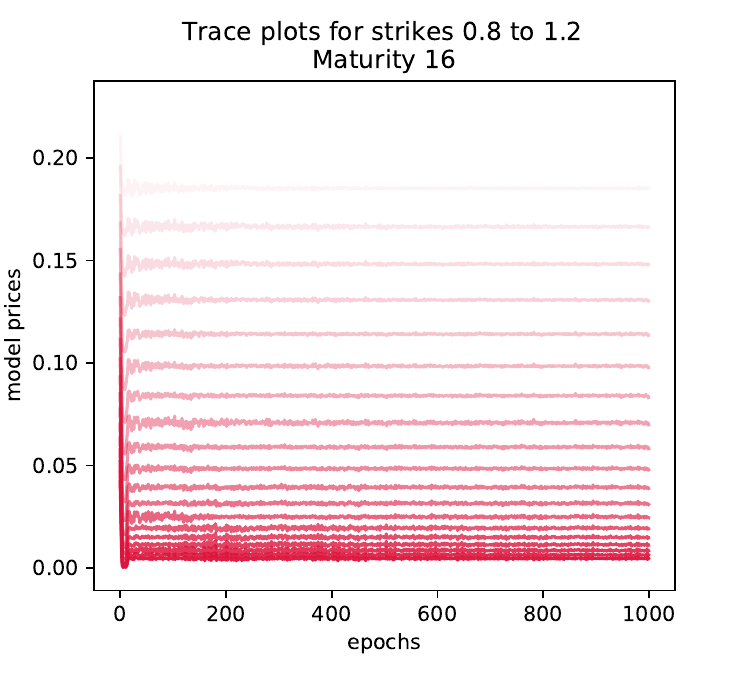}&
         \includegraphics[scale=0.25]{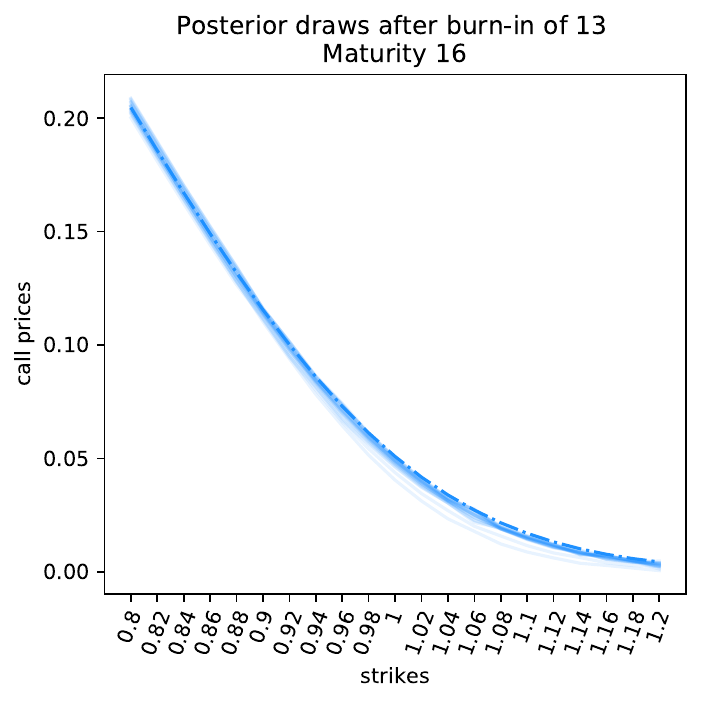} &
          \includegraphics[scale=0.27]{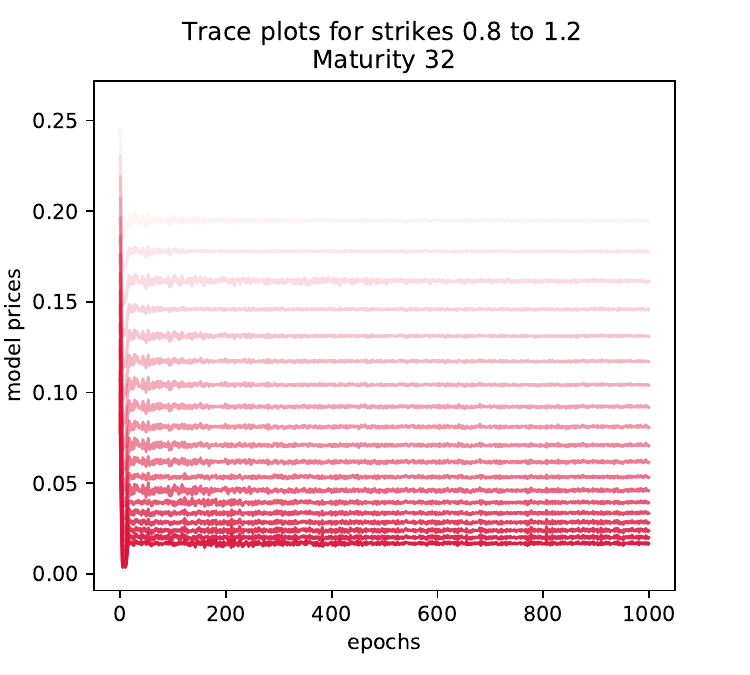}&
          \includegraphics[scale=0.25]{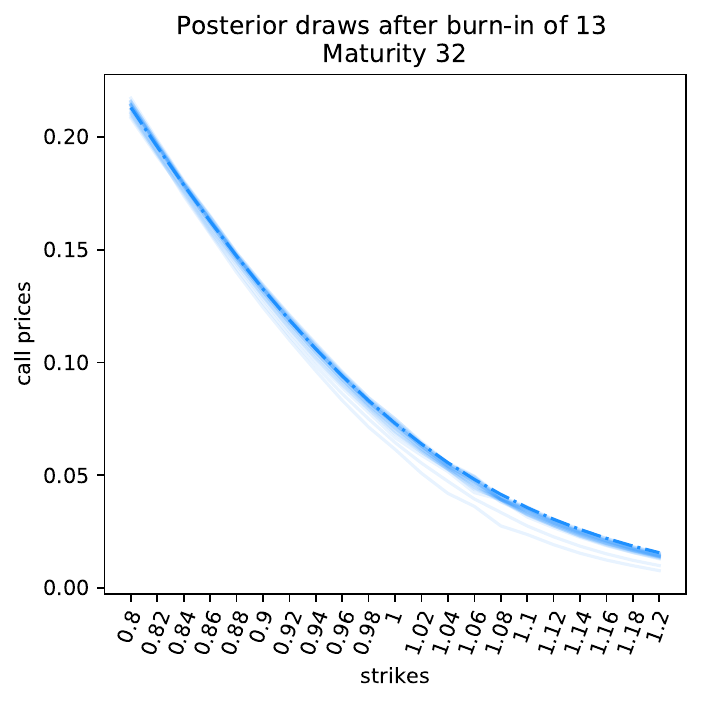}\\
          \includegraphics[scale=0.27]{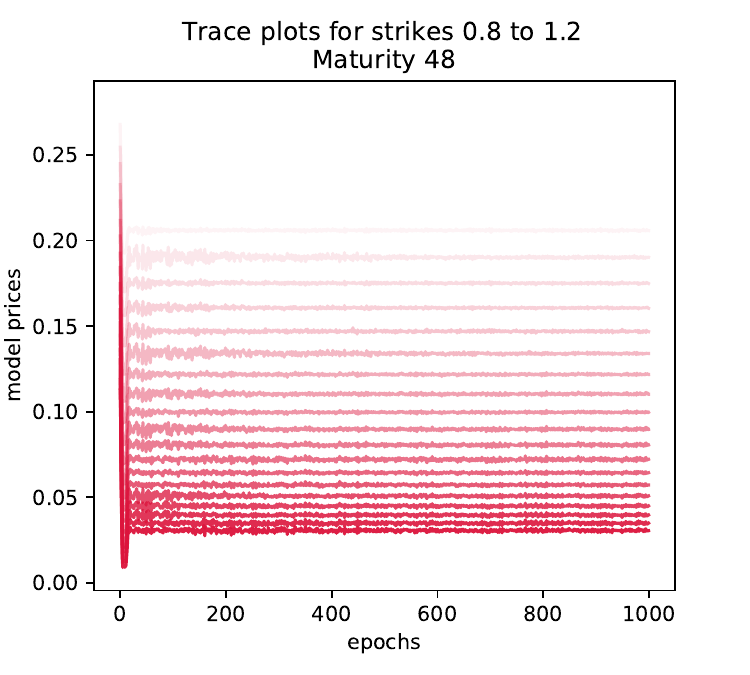}&
          \includegraphics[scale=0.25]{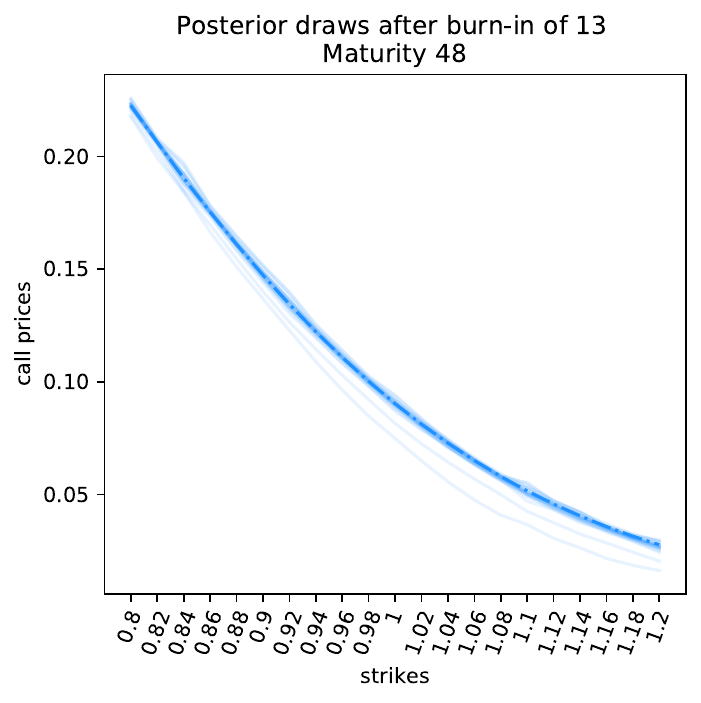}&
           \includegraphics[scale=0.27]{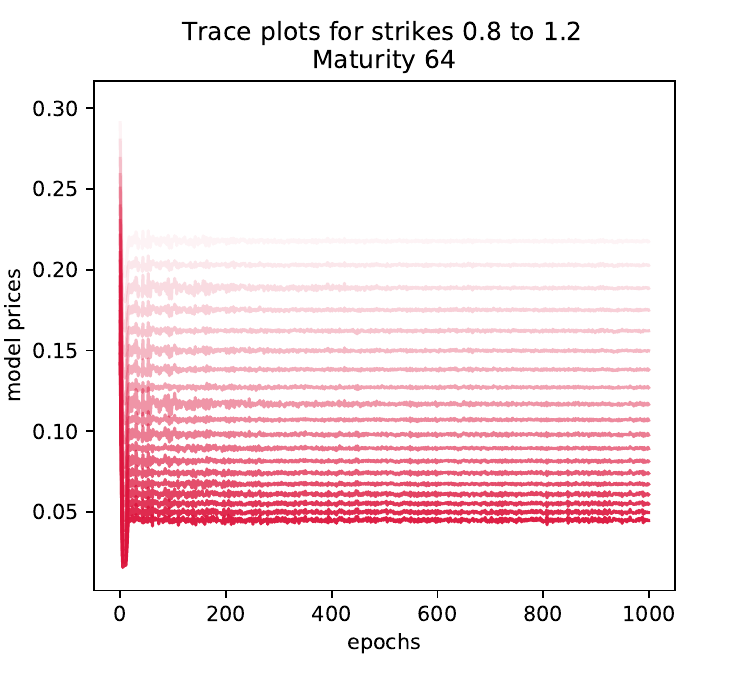}&
           \includegraphics[scale=0.25]{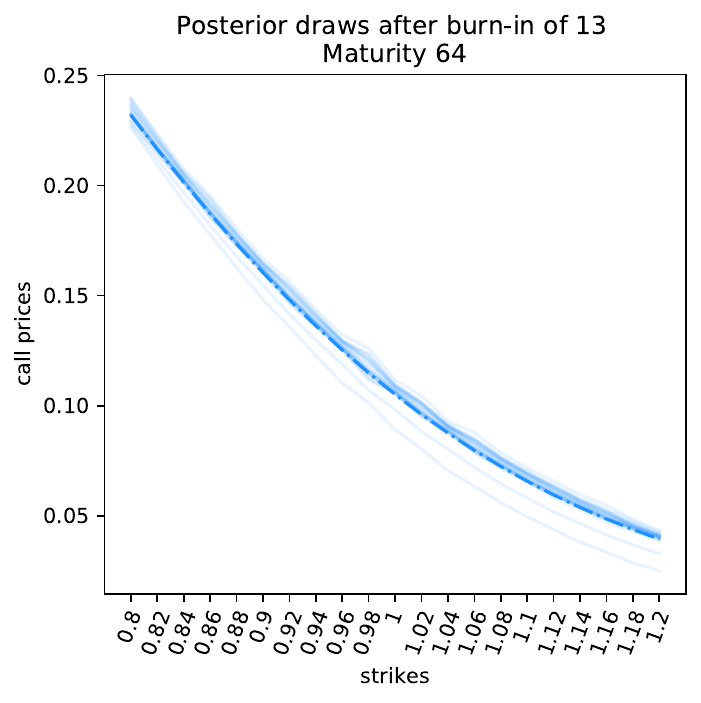}\\
          \includegraphics[scale=0.27]{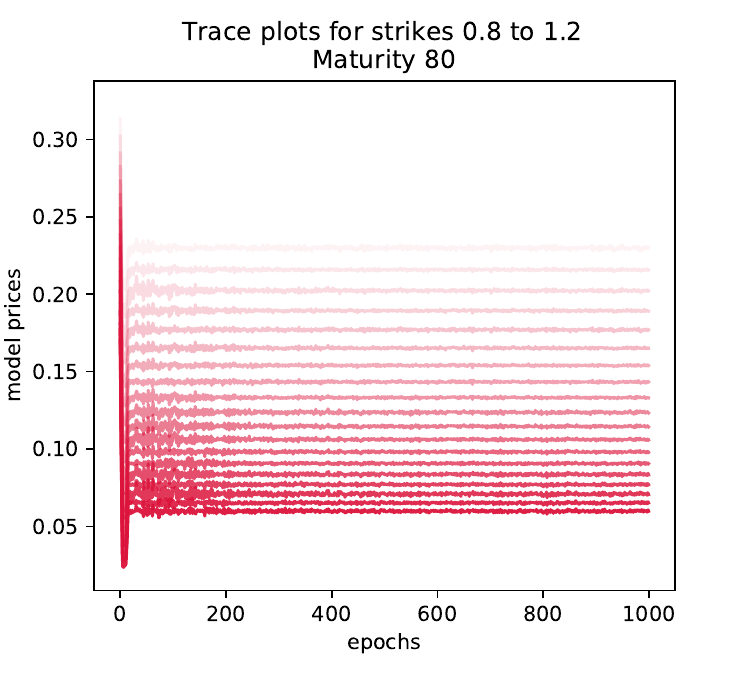}&
          \includegraphics[scale=0.25]{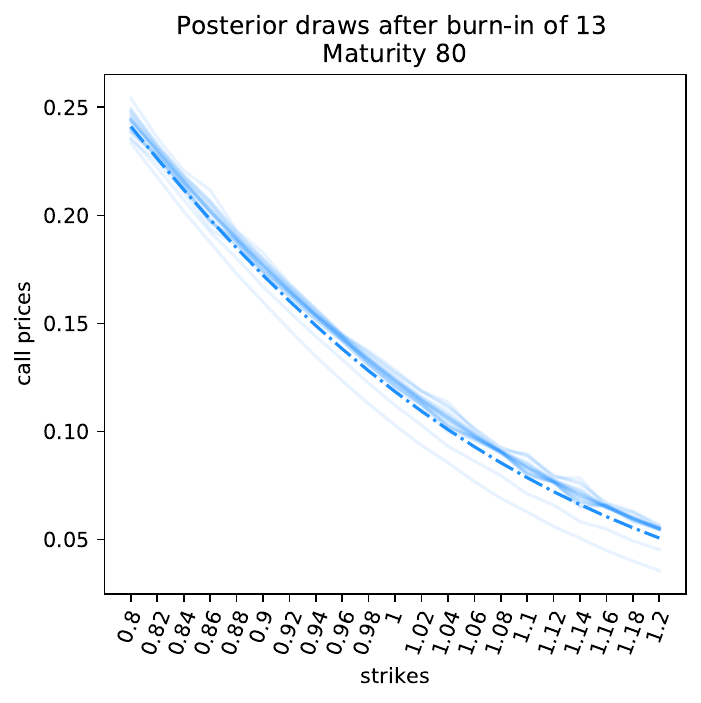}&
          \includegraphics[scale=0.27]{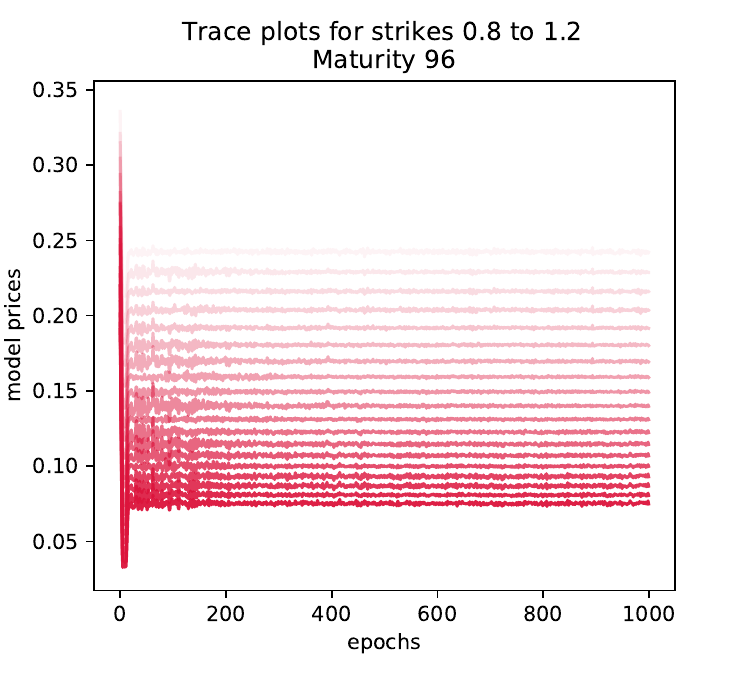}&
          \includegraphics[scale=0.25]{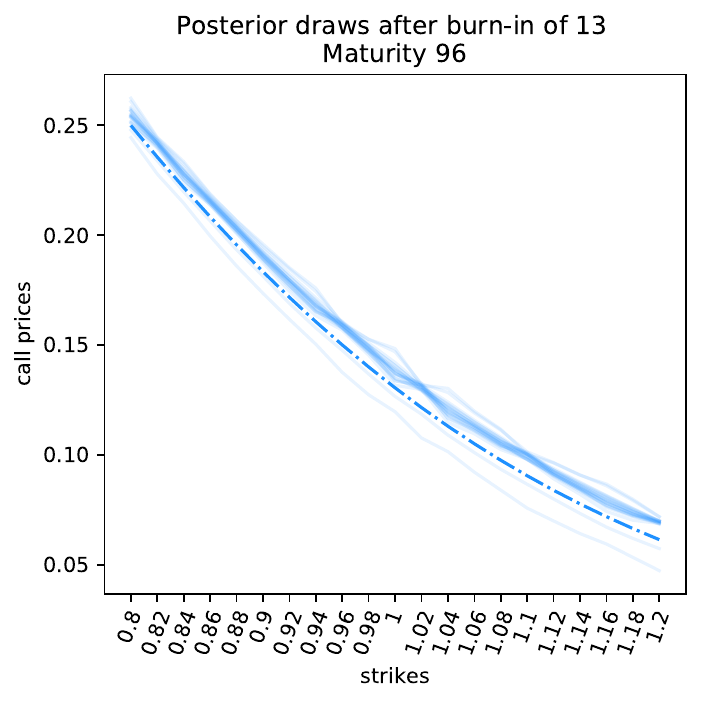}
    \end{tabular}
    \caption{Trace plots and call option prices resulting from a calibration to simulated data from a Black-Scholes model.}
    \label{tb:table6}
\end{table}
The trace plots reveal that the algorithm converged around epoch 13. This means that after this epoch we are dealing with approximate draws from the posterior distribution, which is why the plots in the second row of the above panel use model prices from the 13th epoch onwards. Recall that the model was trained to strikes between $0.08$ and $1.2$. It is very interesting to observe that at those areas where there is no market data (for example for strikes between 1 and 1.02 for maturity 64) the posterior draws are spread out in a wider range, reflecting the larger uncertainty that is present in these regions. The implied volatility surface in the Black-Scholes model is just a flat line with value equal to $\sigma$, by the definition of implied volatility. The graphs in Table \ref{tb:table7} reveal that the trained neural SDE comes rather close to this line.

\begin{table}[h!]
    \centering
    \begin{tabular}{ccc}
         \includegraphics[scale=0.3]{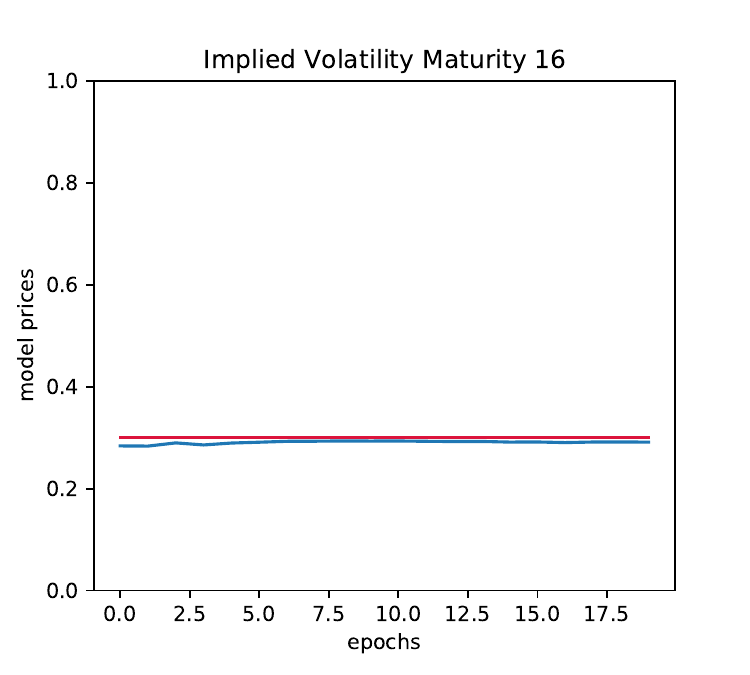}
   &
     \includegraphics[scale=0.3]{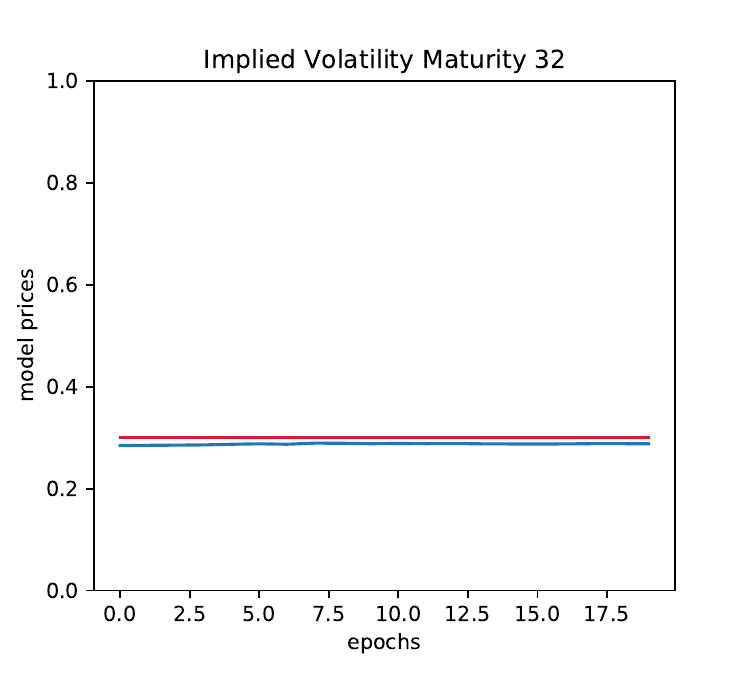}
   &
     \includegraphics[scale=0.3]{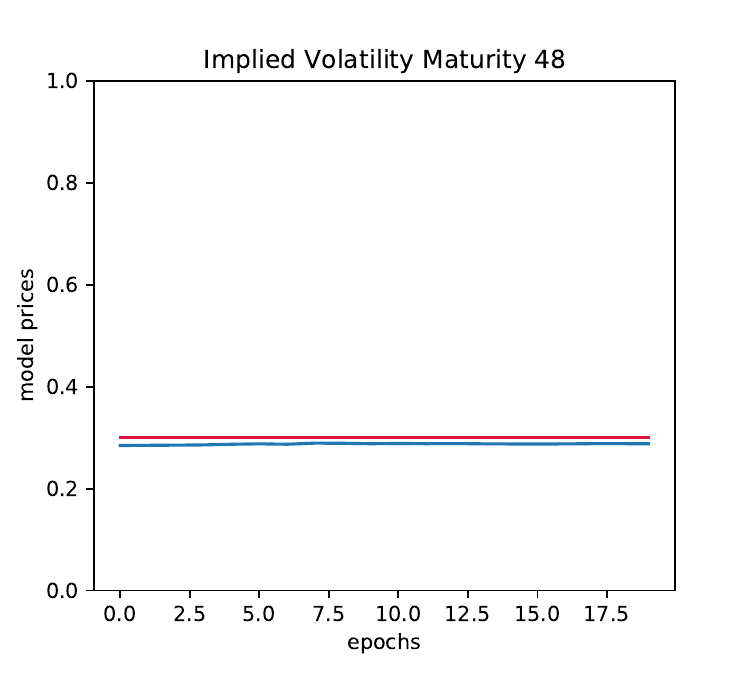}\\
     \includegraphics[scale=0.3]{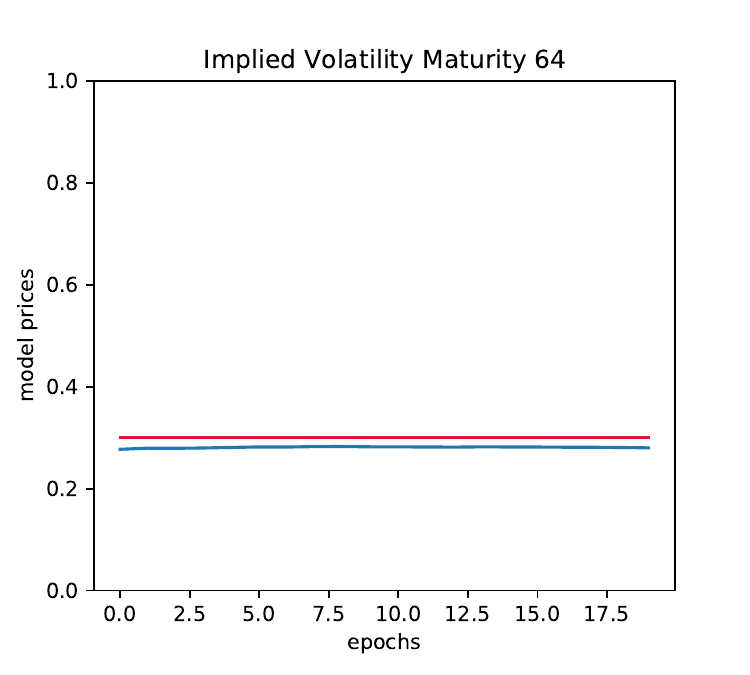}
   &
     \includegraphics[scale=0.3]{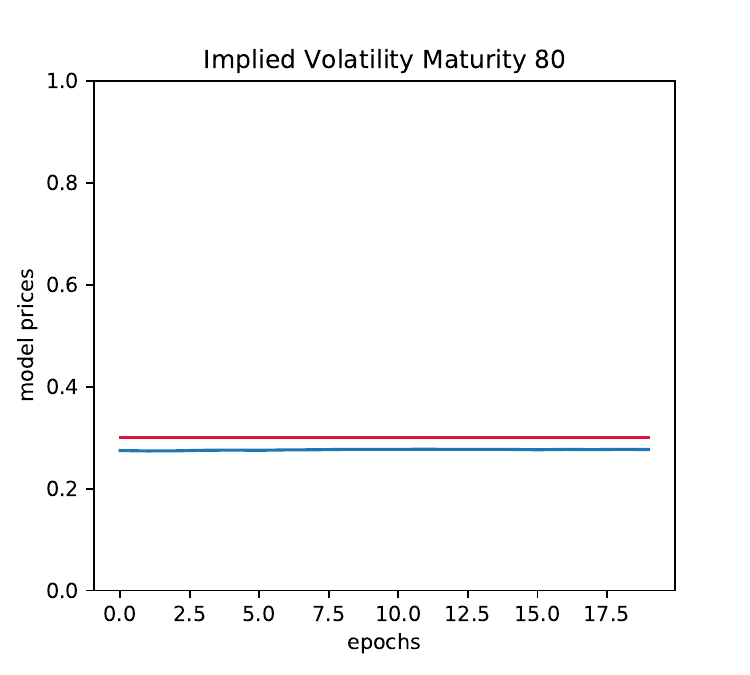}
   &
     \includegraphics[scale=0.3]{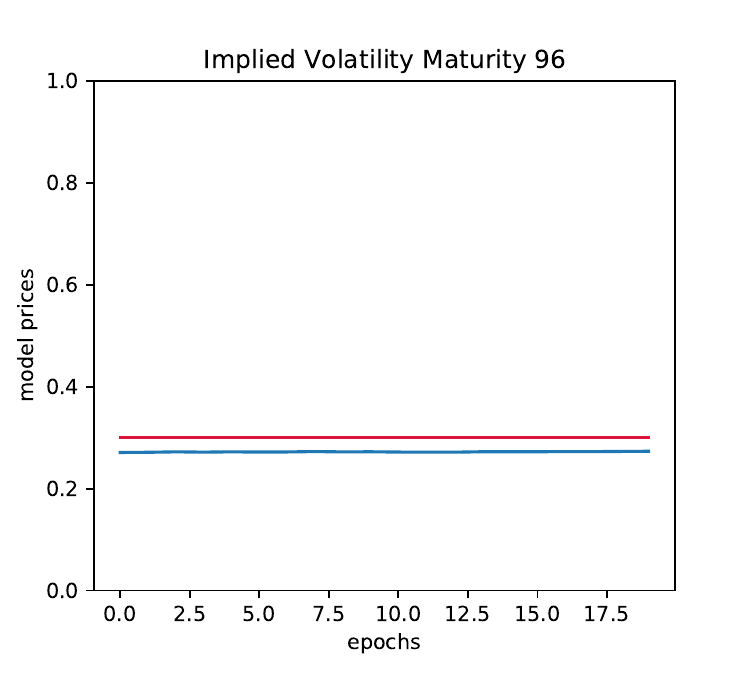}
    \end{tabular}
    \caption{Implied volatility surface resulting from a calibration to simulated data from a Black-Scholes model.}
    \label{tb:table7}
\end{table}
\newpage
\subsection{\textcolor{black}{Heston model}}\label{Heston Model}
\color{black}The Heston model~\cite{heston1993closed} is a (still) widely used stochastic volatility model in financial mathematics, capturing key market features such as mean-reverting volatility and the leverage effect. It assumes that the price $S$ of an asset and its variance $V$ are given by the following dynamics
$$
\begin{gathered}
\mathrm{d} S_t=\mu S_t \mathrm{d} t+\sqrt{V_t} S_t \mathrm{d} W_t^{\mathbb{P},1} \\
\mathrm{d} V_t=\kappa[\theta-V_t] \mathrm{d} t+\sigma \sqrt{V_t} \mathrm{d} W^{\mathbb{P},2}_t,
\end{gathered}
$$
$\mu$, $\theta \in \mathbb{R}$, $\kappa$, $\sigma \in \mathbb{R}_{+}$, where $W^{\mathbb{P},1}_t$ and $W^{\mathbb{P},2}_t$ are two correlated Brownian motions as introduced above.

This model is under the historical measure $\mathbb{P}$. Under $\mathbb{Q}$ we have the dynamics
$$
\begin{gathered}
\mathrm{d} S_t=r S_t \mathrm{d} t+\sqrt{V_t} S_t \mathrm{d} W^{1}_t \\
\mathrm{d} V_t=\bar{\kappa}[\bar{\theta}-V_t] \mathrm{d} t+\sigma \sqrt{V_t} \mathrm{d} W^{2}_t,
\end{gathered}
$$
where $\bar{\kappa}=\kappa+\lambda$ and $\bar{\theta}=\kappa \theta /(\kappa+\lambda)$ with $\lambda \in \mathbb{R}$, $W^{\mathbb{P},1}_t=W^1_t+\frac{\mu-r}{\sqrt{V_t}} t$ is the Wiener process of $S_t$ under $\mathbb{Q}$ and\\ $W^{\mathbb{P},2}_t=W^2_t+\frac{\lambda \sqrt{V_t}}{\sigma} t$ is the the Wiener process of $\sqrt{V_t}$ under $\mathbb{Q}$ and $\lambda$ can be interpreted as a volatility risk-premium. As regards the calibration with respect to the risk neutral measure, we choose exactly the same specification as in~\cite{gierjatowicz2020robust}, $r=0.025$, $\bar{\kappa}=0.78$, $\bar{\theta}=0.11$, $\sigma=0.68$, $V_0=0.04$ and $\rho=0.044$ to simulate the ground-truth option prices, to ensure comparability. The key difference in our work is that we jointly calibrate with respect to the historical measure and to this end we choose $\lambda=0.2$ and $\mu=0.25$ to simulate the paths $Y_t=\log S_t$. 

\begin{figure}[!ht]
\begin{center}
   \centering \includegraphics[scale=0.85]{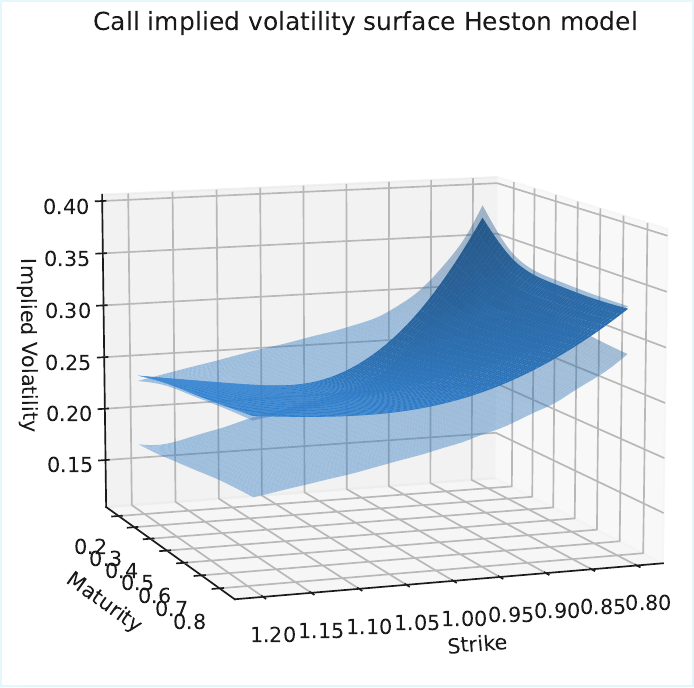}
    \caption{Robust calibration bounds on the implied volatility surface in Heston model, where the model is calibrated with respect to both measures.}
    \label{Heston_ivol}
 \end{center}
\end{figure}

\begin{figure}
    \centering
    \includegraphics[width=0.5\linewidth]{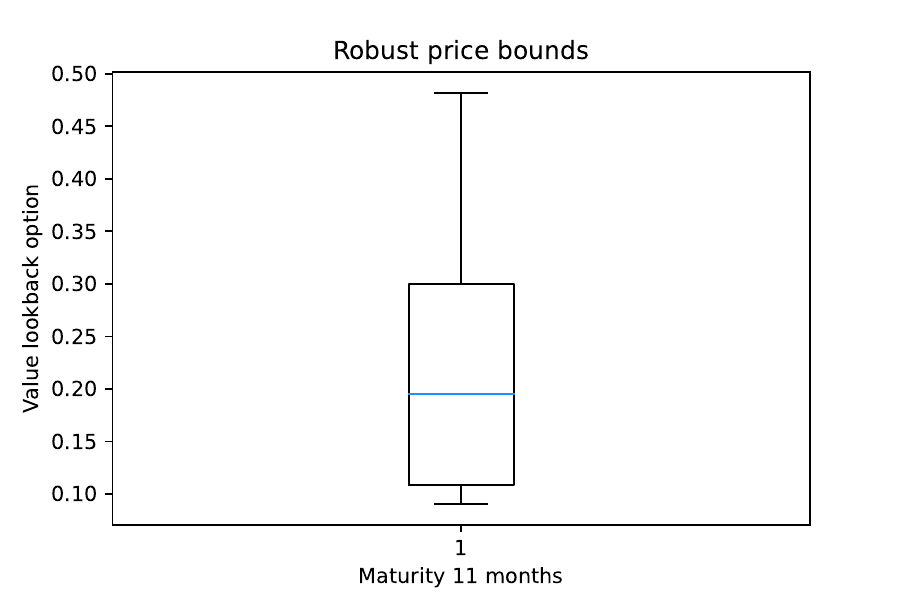}
    \caption{Robust price bounds for a floating lookback put option.}
    \label{fig:lookback}
\end{figure}

The hyperparameters are set to be $\delta= 6$ and $\sigma_{prior}=0.18$. Additionally, prices of lookback put options are computed at each iteration. \textcolor{black}{Note however, that these do not enter in the optimization procedure.} It has to be highlighted that even though we do not optimize with respect to these exotic options, price bounds are obtained naturally by means of the Bayesian setting explained above.  After the training procedure, it is possible to use the resulting distribution in order to obtain bounds on the implied volatility smile, as illustrated in Figure~\ref{Heston_ivol}. Furthermore, the model naturally produces a distribution on prices for exotic options, by computing the variance-reduced Monte Carlo prices of these exotics using the calibrated neural SDE model and Langevin sampling. In our case these bounds are illustrated by the example of lookback options. This is one of the key features of the proposed Bayesian approach which is made clear in Figure~\ref{fig:lookback}. The computed exotic option prices can be compared to the range of exotic options prices in the boxplots in~\cite{gierjatowicz2020robust} on page 21.

In Figure~\ref{tb:Heston_paths}, we present a comparative analysis between the simulated paths of the log-price process $Y$ in the Heston model under the real-world probability measure \(\mathbb{P}\), using the true characteristics and the paths generated by the neural SDE $Y^{\theta}$.  Our comparison aims to evaluate to which extent the neural SDE is capable of replicating the typical characteristics of the 
 Heston model  such as mean-reversion in the spot variance.
 
It is evident from the figure that the neural SDE effectively learns the underlying statistical properties of the Heston model, demonstrating its ability to approximate both the drift and diffusion components of the stochastic process. 
\textcolor{black}{The slightly lower calibration accuracy of the more complex model, when compared to the Black-Scholes model, is a natural outcome and does not indicate a fundamental flaw in the proposed approach. Unlike the Black-Scholes model, which assumes a fixed volatility and requires only the calibration of a single parameter to match asset price dynamics, the more advanced model must simultaneously learn both the asset price behavior and the evolution of stochastic volatility. This joint learning task inherently introduces greater estimation complexity, which can modestly affect calibration precision. Moreover, from a computational standpoint, the inversion from option prices to implied volatility becomes increasingly challenging in more sophisticated models—such as Heston—due to the need for iterative numerical methods such as Newton-Raphson or Brent's method. However, this trade-off is expected and acceptable, as the more complex model is designed to capture richer market features that the Black-Scholes framework cannot represent. Therefore, the observed reduction in accuracy should not be interpreted as a significant limitation, but rather as a consequence of the model's broader scope and the more ambitious learning objective it pursues.}
\begin{table}[ht]
    \centering
    \begin{tabular}{cccc}
        \includegraphics[scale=0.24]{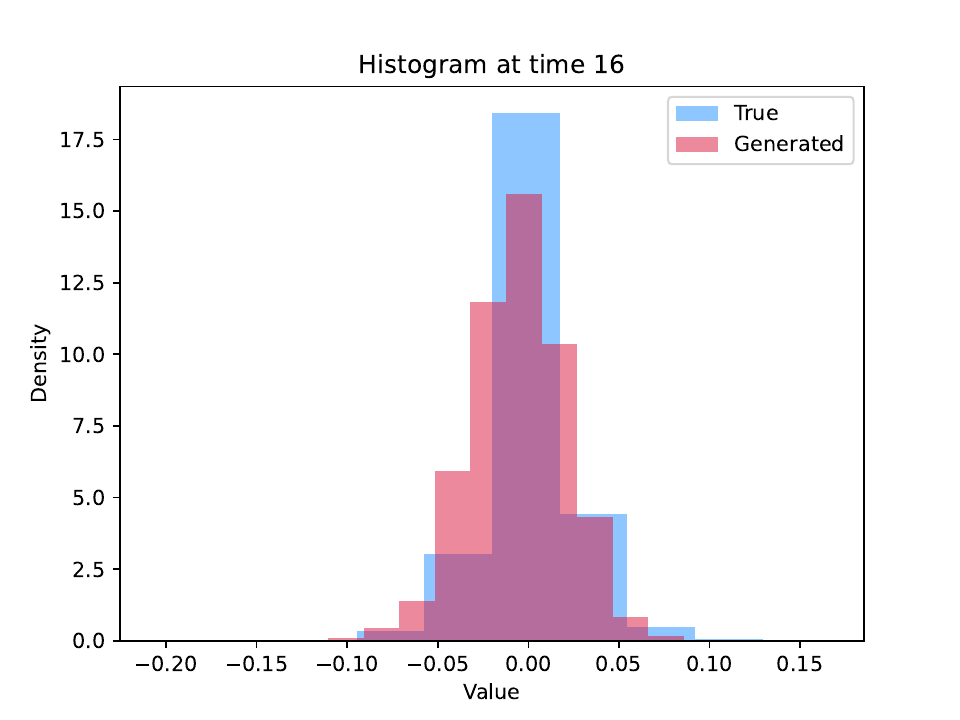}
   &   \includegraphics[scale=0.24]{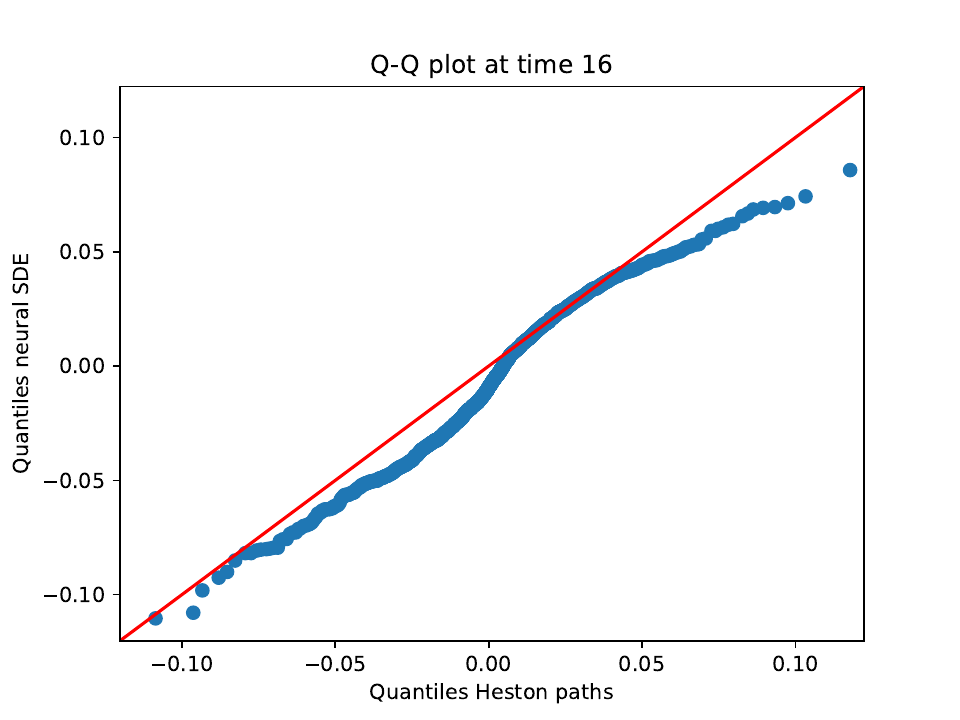}&\includegraphics[scale=0.24]{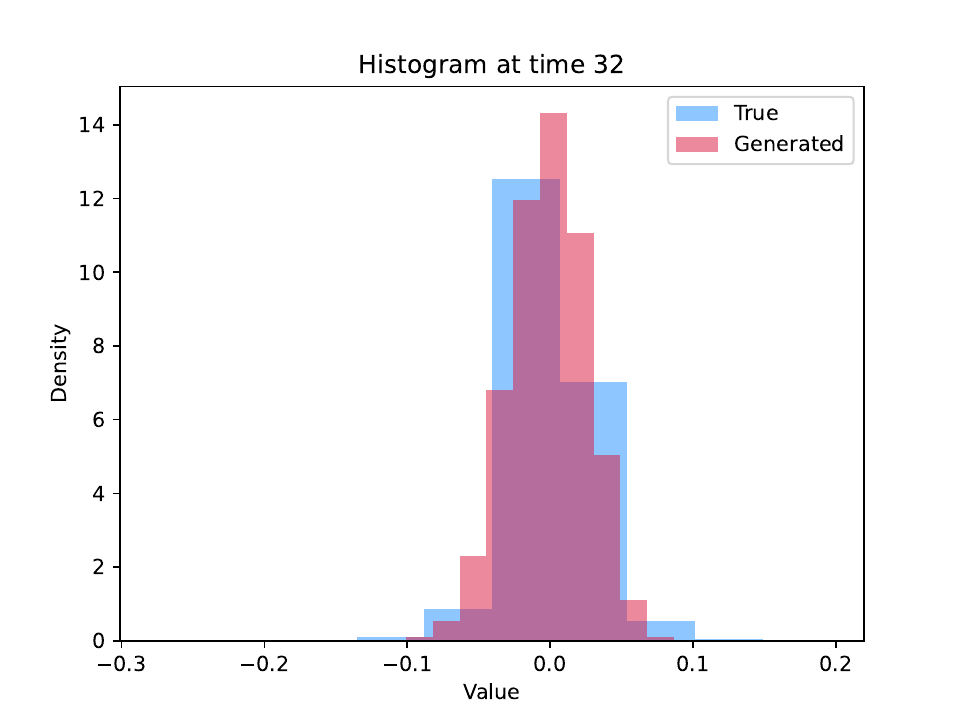} &
         \includegraphics[scale=0.24]{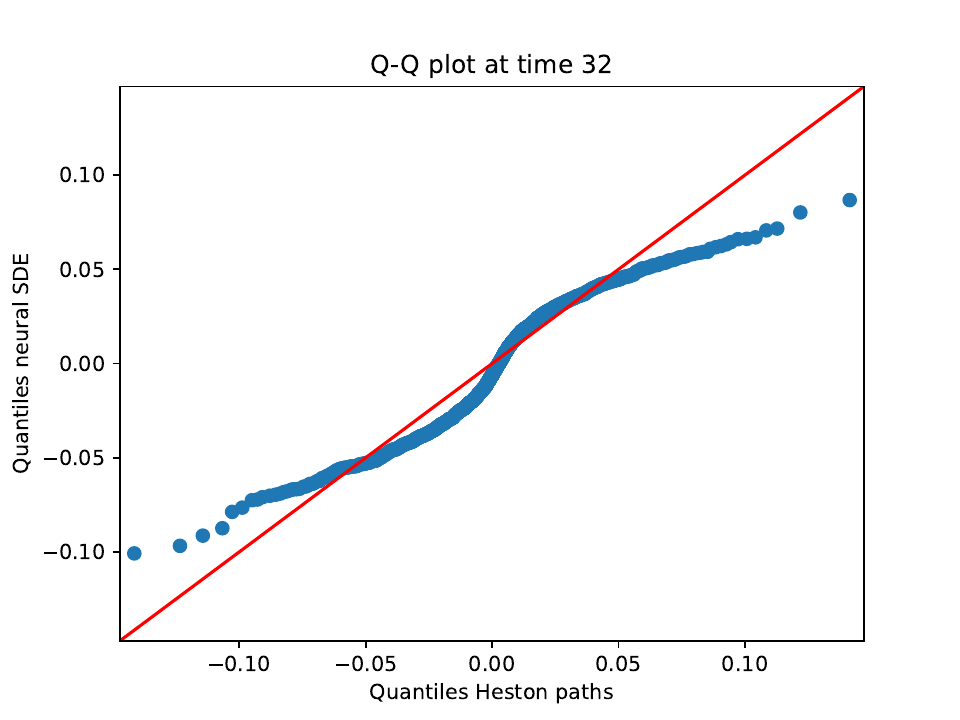}\\
         \includegraphics[scale=0.24]{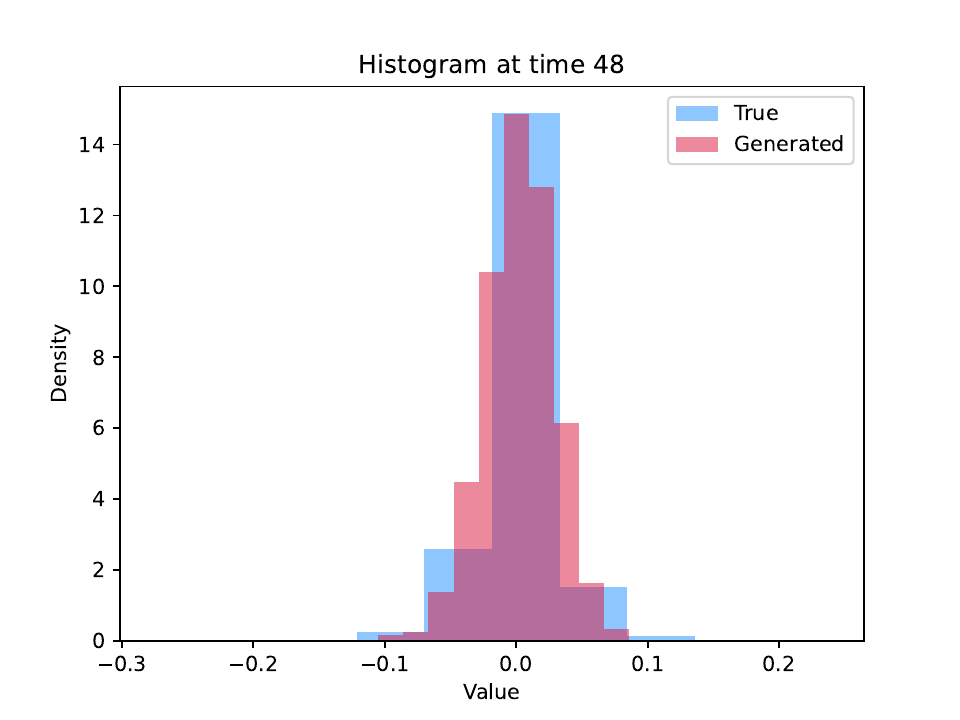} &
         \includegraphics[scale=0.24]{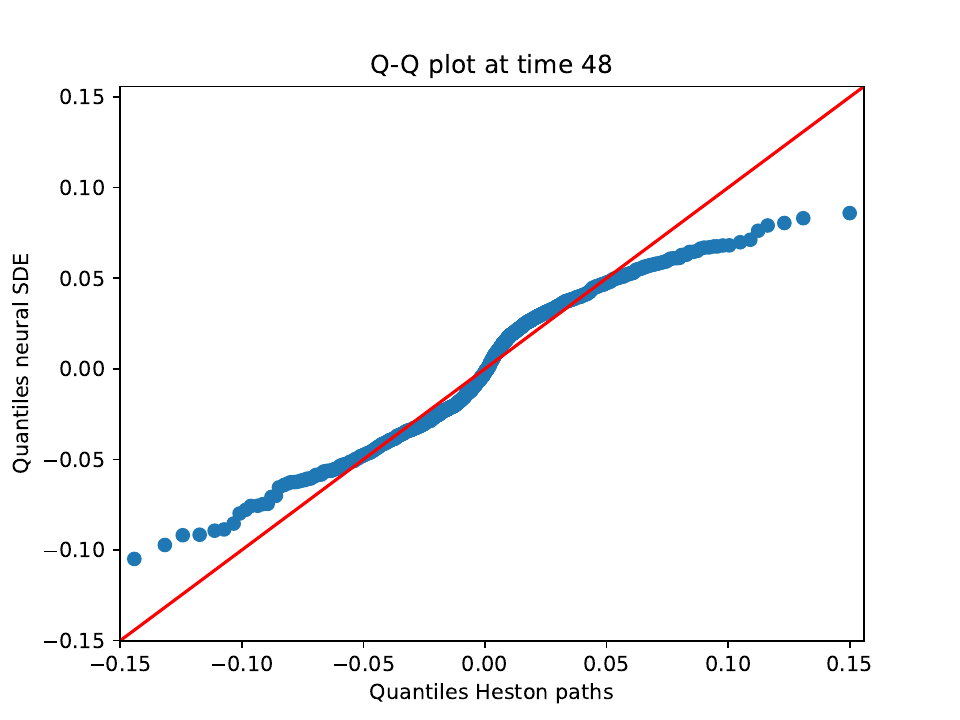}&
         \includegraphics[scale=0.24]{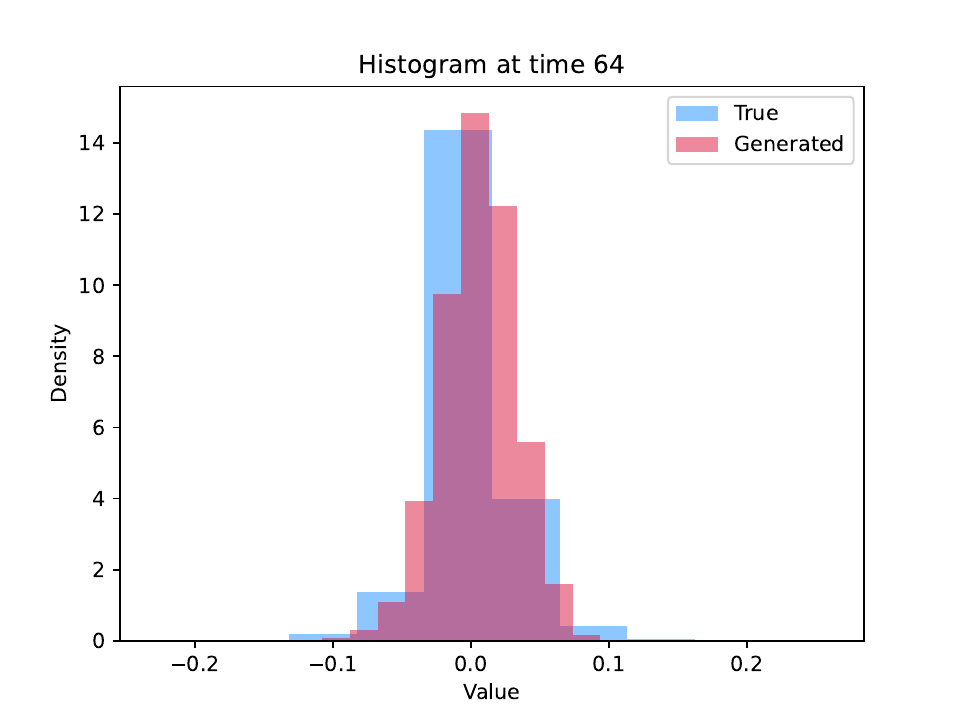} &
         \includegraphics[scale=0.24]{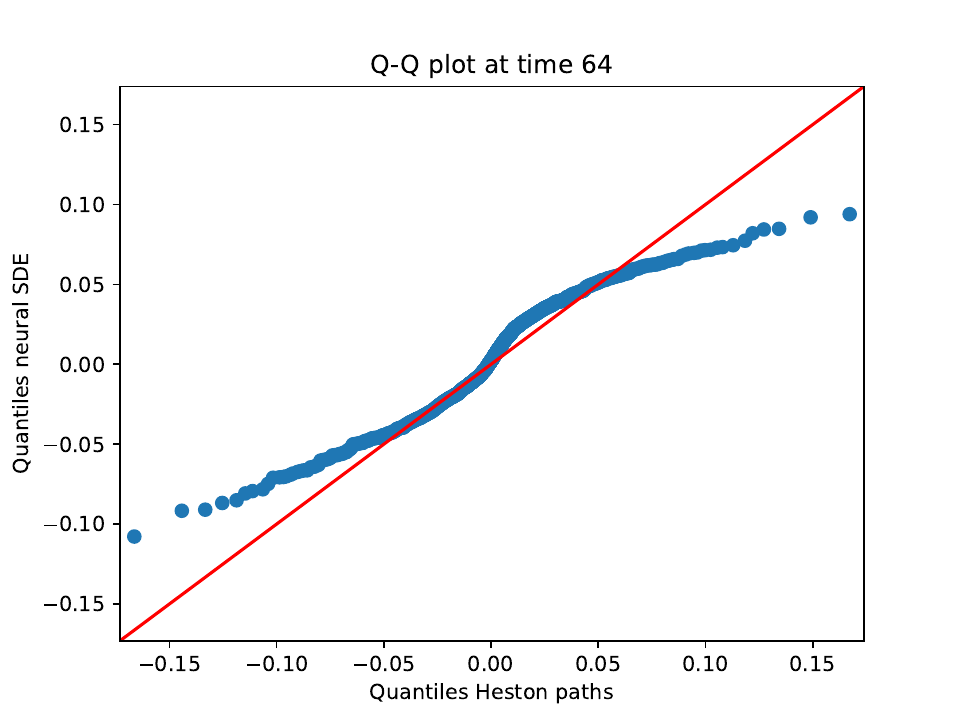}\\
         \includegraphics[scale=0.24]{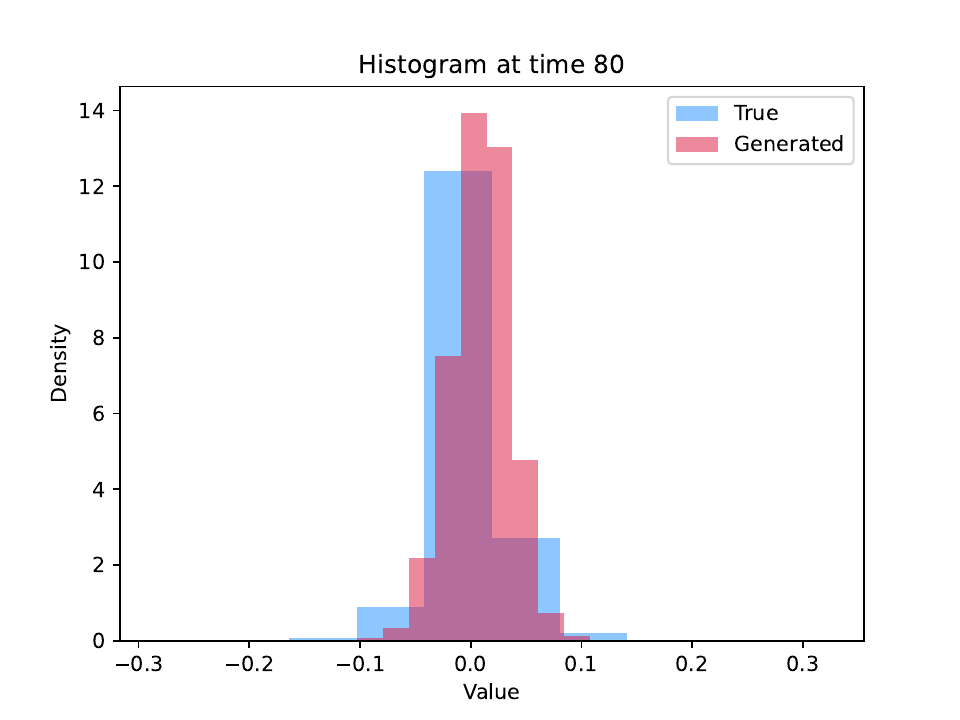}&
         \includegraphics[scale=0.24]{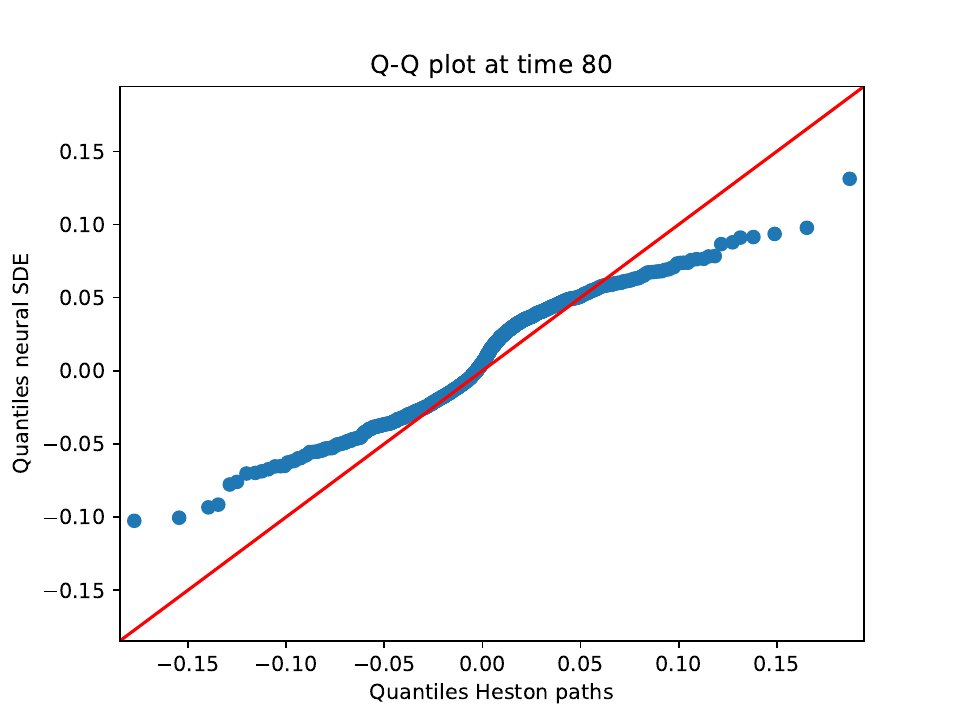}&
         \includegraphics[scale=0.24]{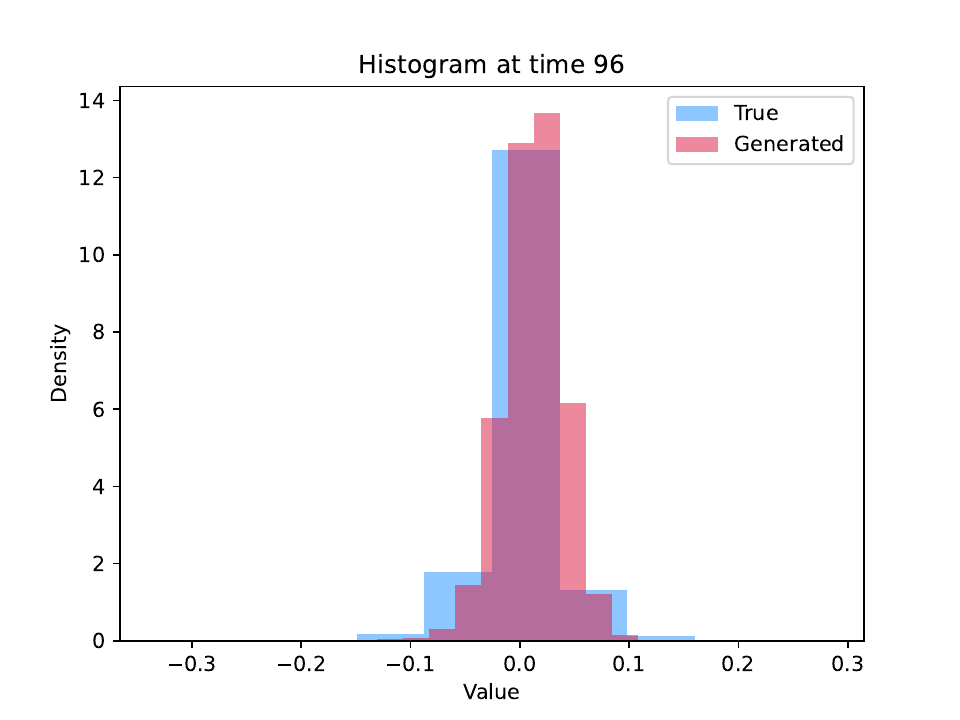}&
         \includegraphics[scale=0.24]{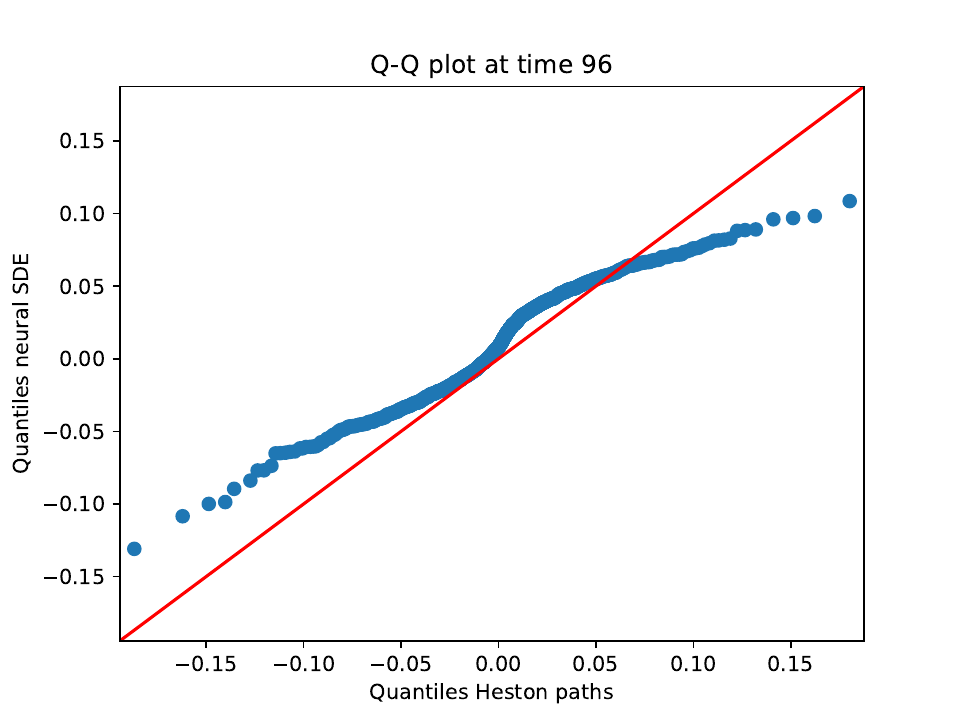}
         
    \end{tabular}
    \caption{Histograms and qq plots at different time points in the Heston model, resulting from a calibration to simulated time series data.}
    \label{tb:Heston_paths}
\end{table}
{\color{white}-}\\ %
\color{black}
\section{Sensitivity analysis} 
\subsection{Sensitivity analysis using the rough Bergomi model}
\textcolor{black}{We conduct a comprehensive sensitivity analysis of the rough Bergomi model with respect to the hyperparameters $\sigma_{prior}$ and $\delta$, both of which play a critical role in controlling the degree of regularization imposed on posterior sampling. By systematically varying these hyperparameters, we assess their impact on posterior inference, model calibration, and the stability of estimated option prices. The 'base case' is the hyperparameter choice of Section \ref{Rough_Bergomi_section}, where $\sigma_{prior}$ was set to $0.15$ and $\delta$ to $3$. We then see how the calibration changes when we consider changed values of $\sigma_{prior}$ equal to $0.25$ and $\delta$ equal to $4.5$.}

\begin{figure}
\begin{center}
\includegraphics[scale=0.2]{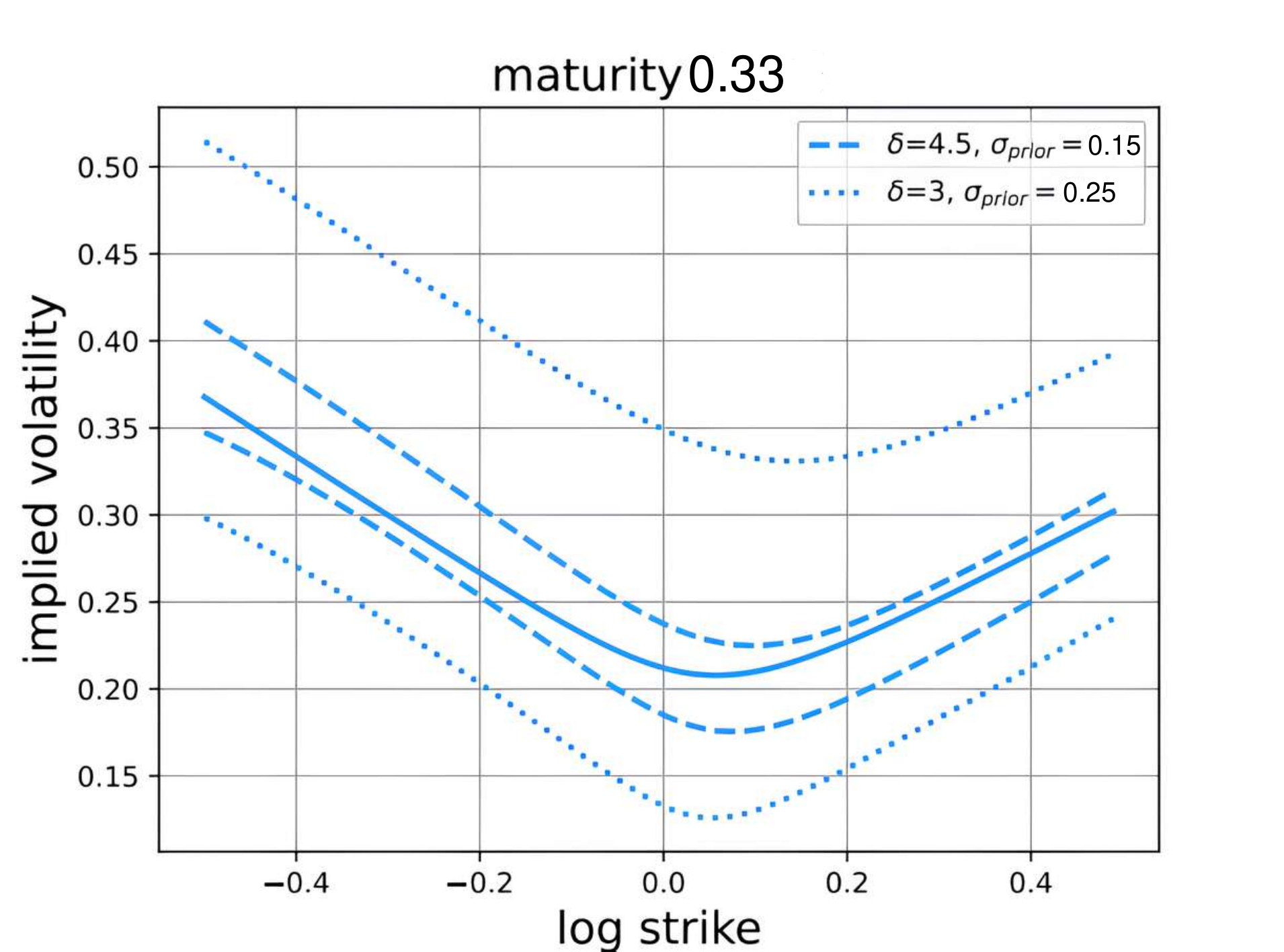}
    \caption{Sensitivity analysis of bounds on the implied volatility surface obtained in the rough Bergomi model.}
    \label{fig:rough_bergomi_sensitivity}
    \end{center}
\end{figure}

\textcolor{black}{We exemplify the sensitivity with respect to the hyperparameter choices, by considering the option with shortest maturity, and emphasize that the behavior is very similar for other options and thus no generality is lost in this representation. When conducting the numerical experiment, it was observed that when the prior standard deviation $\sigma_{prior}$ is set to a low value, the posterior samples converge more quickly to the correct model prices, indicating more stable and efficient sampling. In contrast, higher values of $\sigma_{prior}$ introduce weaker regularization, which increases the variation in posterior draws and slows down convergence. Therefore, setting $\sigma_{prior}$ too high can negatively impact inference quality, suggesting that care should be taken not to choose an excessively large value. This is also reflected in Figure \ref{fig:rough_bergomi_sensitivity}, where we consider a very high value of $\sigma_{prior}$. Due to the fact, that in this case the algorithm converges more slowly, the bounds attained are wider.} 

\textcolor{black}{The results reveal that the convergence behavior and stability of the posterior samples are not highly sensitive to the choice of $\delta$. Across both $\delta$ values, similar convergence properties and bounds are obtained, indicating that variations in this hyperparameter do not substantially alter the efficiency or behavior of the sampling process. This suggests that the impact of $\delta$ on posterior sampling is relatively limited.}

\textcolor{black}{In conclusion we can say the Figure \ref{fig:rough_bergomi_sensitivity} reveals that the bounds, computed across various configurations of the hyperparameters $\sigma_{prior}$ and $\delta$, remain rather stable, demonstrating low sensitivity to moderate changes in these hyperparameters. The consistency of the bounds across different hyperparameter choices highlights the robustness of the Bayesian calibration procedure and supports the conclusion that the method yields reliable and stable results within a reasonable range of prior assumptions.}

\subsection{Sensitivity analysis using the Black-Scholes model}\label{Sensitivity analysis using the Black-Scholes model}
In this section the sensitivity of the suggested algorithm in the Black-Scholes model is studied. We start with an analysis with respect to the parameter $\sigma_{prior}$. Recall that the prior distribution is specified as \begin{equation}
    p(\theta) \sim \mathcal{N}(0,\sigma_{prior}^2).
\end{equation} 
In our architecture, the hidden layers are of width 100, which implies that the value of $\sigma_{prior}$ is approximately 0.15 for most neural network weights, as the scaling factor $g$ used in the definition of \eqref{sigma_prior} is chosen to be 1.5. We observe in Table \ref{tb:table8} that the calibration to time series data is not highly sensitive with respect to $\sigma_{prior}$. The best fit however seems to be captured by $\sigma_{prior} \in (0.1, 0.15)$.
The results of the last iteration of the algorithm are depicted in Table \ref{tb:table9}.
The results suggest that also the calibration to option data is quite robust with respect to moderate changes in $\sigma_{prior}$. However, it is indicated that larger values of $\sigma_{prior}$ yield an inferior fit to option prices. In the Table \ref{tb:table10} we study the trace plots for each different product and varying $\sigma_{prior}$. It appears that the algorithm has converged to its stable regime by iteration 500 at latest in Table \ref{tb:table10}. 
Table \ref{tb:table11} presents the posterior surface of implied volatility for each combination of $\sigma_{prior}$ and maturity after a burn-in based on the trace plots above. It can be observed that the model adequately captures the true implied volatility surface for each small change of $\sigma_{prior}$.
Numerical experiments have shown that smaller values of $\sigma_{prior}$ yield more stable results. This seems to be natural as smaller values of $\sigma_{prior}$ lead to more shrinkage, meaning that the set of neural network parameters is sparse in the sense that they are close to zero.

We now analyze the robustness of the algorithm with respect to the parameter $\delta$. Recall that the definition of this parameter is linked to the bid-ask spread of options. Nevertheless we would like to provide an intuition of how sensitive the proposed method is with respect to the value of this parameter as the bid-ask spread may vary due to different market conditions, liquidity, trading volume, and the specific broker through which the trade is executed. Depicted in Table \ref{tb:table12} and \ref{tb:table13} are results of the last iteration of the algorithm.
The results reveal that the calibration to option data, as well as time series data, is not very sensitive with respect to changes in $\delta$. Table \ref{tb:table14} shows the trace plots for each different maturity and varying $\delta$. One can see that the algorithm stabilizes after iteration 500 at latest. 
We present the posterior surface of implied volatility for each combination of $\delta$ and maturity after a burn-in based on the trace plots above in Table \ref{tb:table15}. The plots reveal that the model is adequately calibrated to the true implied volatility surface for each choice of $\delta$.
\section{Results for empirical data}\label{Results for empirical data}
In this final Section the model consisting of equations \eqref{priceprocess} and \eqref{varianceprocess} is calibrated to real call options considering empirical S\&P 500 implied volatility data with 10 strikes and 4 maturities to determine the prices of the corresponding European call options. The spot price of the underlying at time 0 is $S_0 = 590$, $\delta=4.5$, $\sigma_{prior}=4$, the interest rate is $r = 0.060$ and dividend rate $d = 0.026$. The prices are given in Appendix 4. The reason why we take this data set as an example is that even tough the data are from 1990s, they allow for a direct comparison with the results given in~\cite{Gupta2014RobustCO}, as we choose exactly the same data and hyperparameters. This is also the reason why the $\sigma_{prior}$ is chosen relatively high compared to the values in \ref{Sensitivity analysis using the Black-Scholes model}: the Tikhonov regularization parameter $\delta^2/\sigma_{prior}^2$ depends on the scale of the input data and as the data in~\cite{Gupta2014RobustCO} is not standardized we also refrain from it in this study. The plots in Table \ref{tb:table17} reveal that in contrast to the bounds on the implied volatility surface presented in~\cite{Gupta2014RobustCO}, where the spread of implied volatilities is, according to the authors, enormous, we obtain much tighter bounds given by the minimum and maximum implied volatility for each strike and maturity that are attained after algorithmic convergence. \color{black} To put this into numbers the upper bound of implied volatilities for low strike and maturity obtained in~\cite{Gupta2014RobustCO} is above $0.6$, whereas for us it is about $0.3$.\color{black}
\newline
 \begin{table}[h!]
 \centering
  \small
  \caption{Robust bounds on implied volatility surface for calibration with respect to real option data.}
  \label{tb:table17}
     \hspace*{0cm}\begin{tabular}{ c }
      
     \includegraphics[width=\textwidth]{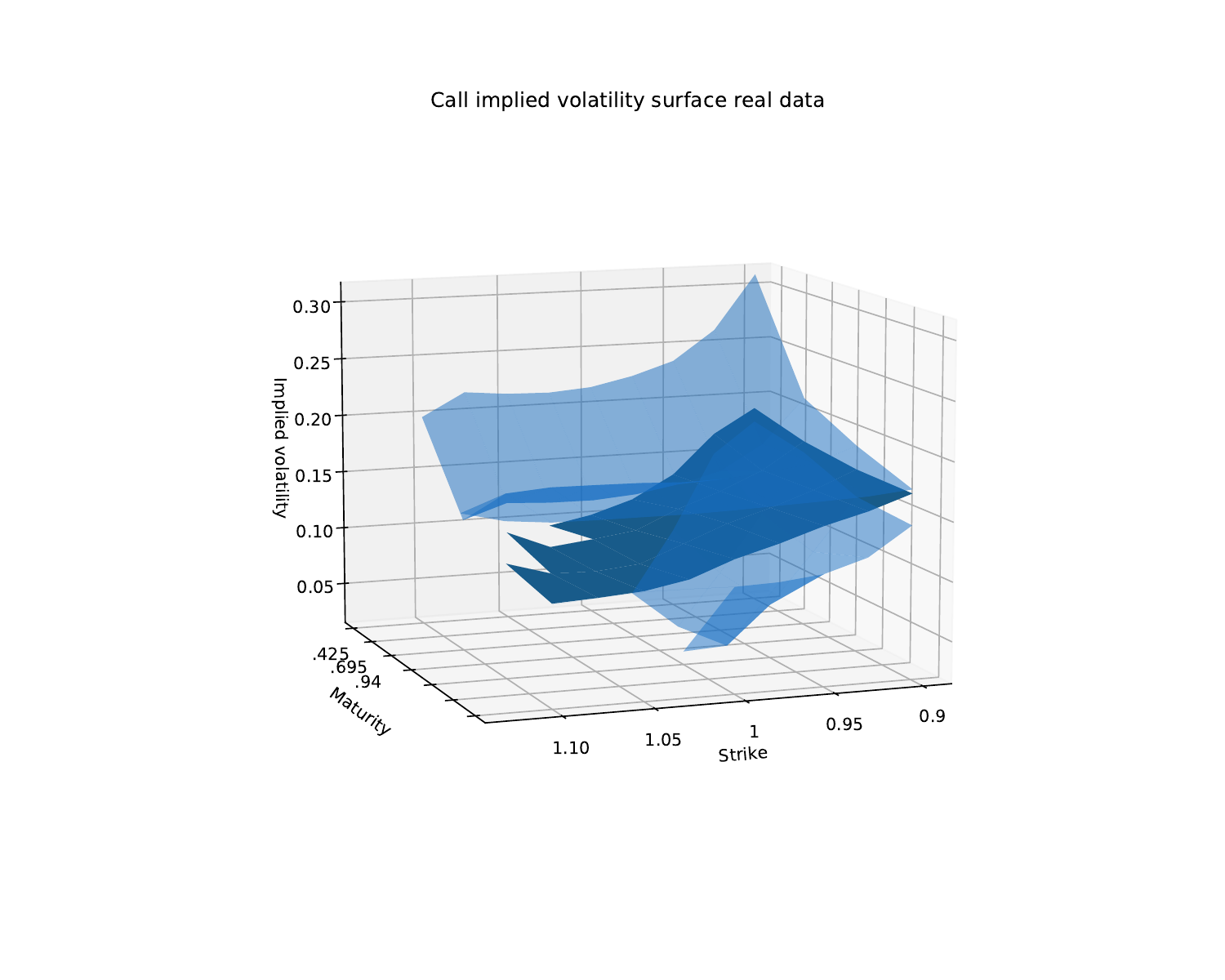}
      \label{Reisinger_plots}
      \\ 
      \end{tabular}
      \end{table}
 We do not attempt to jointly calibrate to real option data and the historical price trajectory of the S\&P 500, to ensure comparability with the results presented in ~\cite{Gupta2014RobustCO}. The plots in Table \ref{Reisinger_plots} suggest that the neural SDE has been calibrated adequately to the historical call prices. \color{black} The surface shows higher implied volatility for both high and low strikes, reflecting the presence of a volatility smile or skew, a common phenomenon in options markets where deep in-the-money and out-of-the-money options have higher implied volatilites than at-the-money options. Moreover, the term structure of implied volatility is visible. The robust bounds provided in the plot can serve as a basis for price bounds for more complex and path-dependent options as illustrated in Section \ref{Heston Model}.

\color{black}
\section{Possible technical modifications}\label{Possible technical modifications}
It might be possible to consider a Metropolis-Hastings element in the Algorithm 1.1 for the joint calibration under the historical measure and the risk neutral measure, so that an update of the neural network parameters $\theta$ is only considered to be a proposal, that is rejected with a certain probability. This is known as the Metropolis-adjusted Langevin algorithm and in the Bayesian literature this algorithm typically shows better convergence properties than the unadjusted Langevin algorithm. Similarly as in \cite{Yu}, on may also chose a more robust distribution family for the error/noise distribution which then defines the likelihood function such as the Student-t distribution. 
\section{Conclusion}\label{Conlusion}
In conclusion, this paper has presented a comprehensive Bayesian framework for the calibration of neural SDEs, addressing the inherent challenges of joint calibration with respect to option data and financial time series data, necessitating a measure change. By shifting the focus from finding a singular best-fit solution to determining a distribution of plausible solutions, our approach offers a robust framework for model calibration.

The integration of Bayesian methods allows for the incorporation of prior information and the systematic update of this information using observed market data, resulting in a posterior distribution that provides a richer concept of parameter uncertainty. This approach not only enhances the applicability of neural SDE models but also aligns with the broader goals of capturing model risk and uncertainty. Additionally, by incorporating considerations for changes of measure, our framework further extends the applicability and utility of neural SDE models in the financial domain.

Future work can build on this foundation by exploring alternative prior distributions, optimizing computational strategies, and considering a time-series of option data, rather than one single cross-section.
\appendix
\color{black}

\color{black}
\section{Details on the neural network architecture}
This section is dedicated to the simulation study aiming at matching log price trajectories as well as call option prices in the Black-Scholes model. The neural network architecture for all other simulation studies is similar to the one described  here. The trainable elements for the neural SDE in the Black-Scholes case are $\mu(.)$ and $\sigma(.)$. For each of $\mu(.)$ and $\sigma(.)$ there is one neural network for each period that takes as an input the time step as well as the current value of the neural SDE. There are four hidden layers with 100 neurons for each of these neural networks. \textcolor{black}{The activation function in the hidden layers is the tanh and in the output layer it is the softplus activation function for $\sigma(.)$, and the identity for $\mu(.)$. The initial learning rate is chosen to be $1e-4$.} As mentioned above, each of these neural networks is initialized via the so called Glorot normal prior. 

\section{Supplementary plots}
This appendix collects the plots describes in Section~\ref{Sensitivity analysis using the Black-Scholes model}
\begin{table}
\centering
  \small
  \caption{Sensitivity of calibration to time series data, with respect to $\sigma_{prior}$.}
  \label{tb:table8}
  \begin{tabular}{||c c ||} 
 \hline
   $\sigma_{prior}=0.1$&   $\sigma_{prior}=0.15$  \\ [0.1ex] 
 \hline\hline
 \includegraphics[scale=0.5]{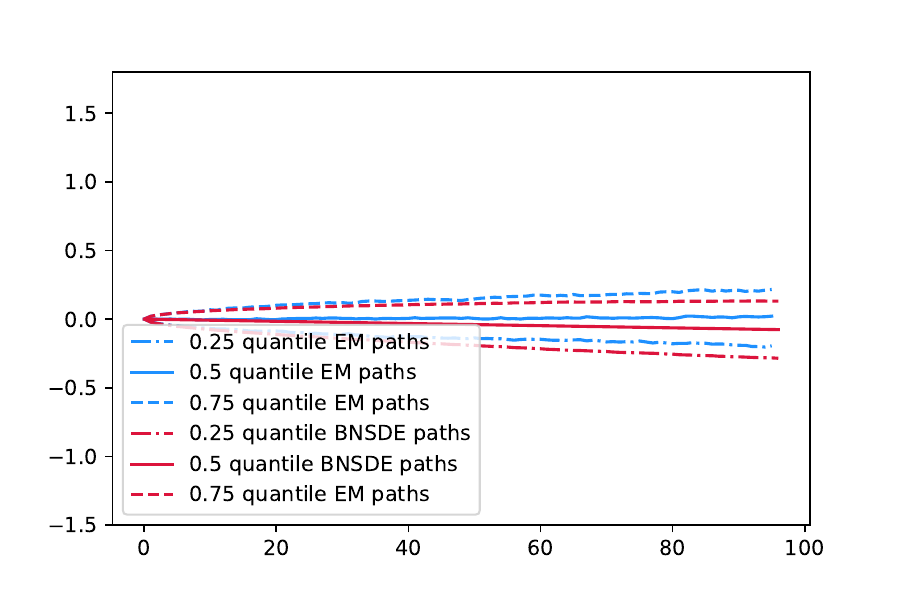}&
 \includegraphics[scale=0.5]{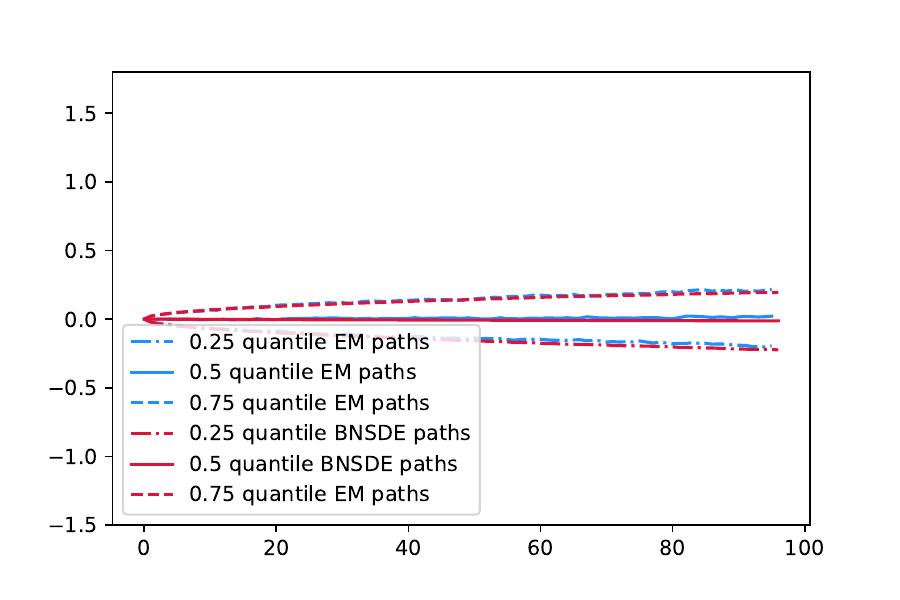} \\  [0.1ex] 
 \hline
 
  $\sigma_{prior}=0.2$&   $\sigma_{prior}=0.25$   \\ [0.1ex] 
 \hline\hline
 \includegraphics[scale=0.5]{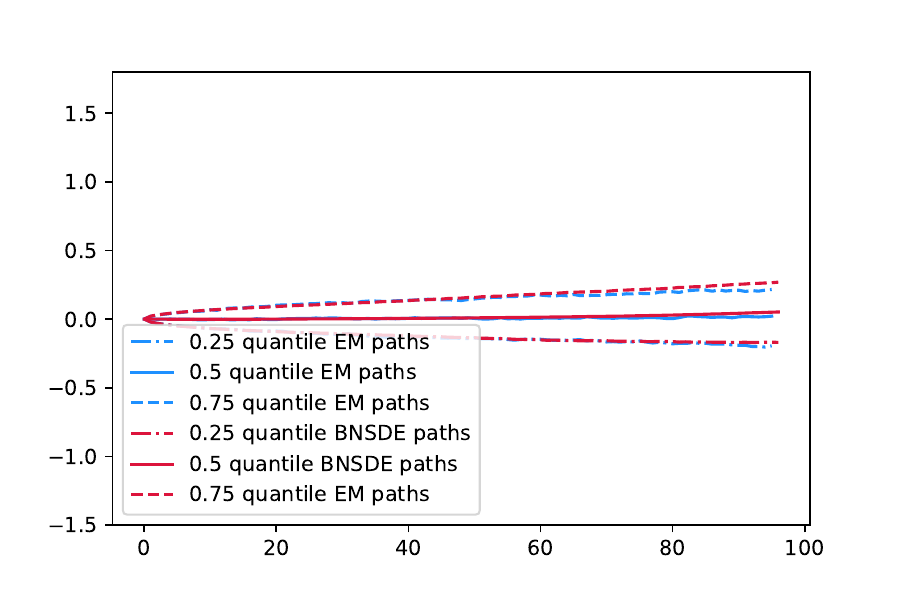}& \includegraphics[scale=0.5]{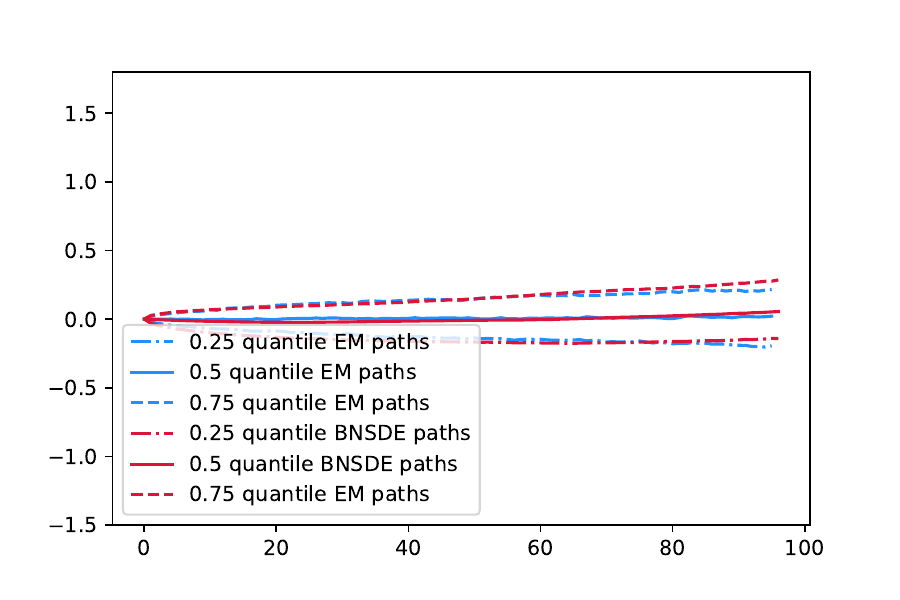}\\  [0.1ex] 
 \hline
\end{tabular}
\end{table}

\begin{table}
\centering
  \small
  \caption{Sensitivity of calibrated call option prices, with respect to $\sigma_{prior}$.}
  \label{tb:table9}
  \begin{tabular}{||c c ||} 
 \hline
   $\sigma_{prior}=0.1$&  $\sigma_{prior}=0.15$  \\ [0.1ex] 
 \hline\hline
 \includegraphics[scale=0.6]{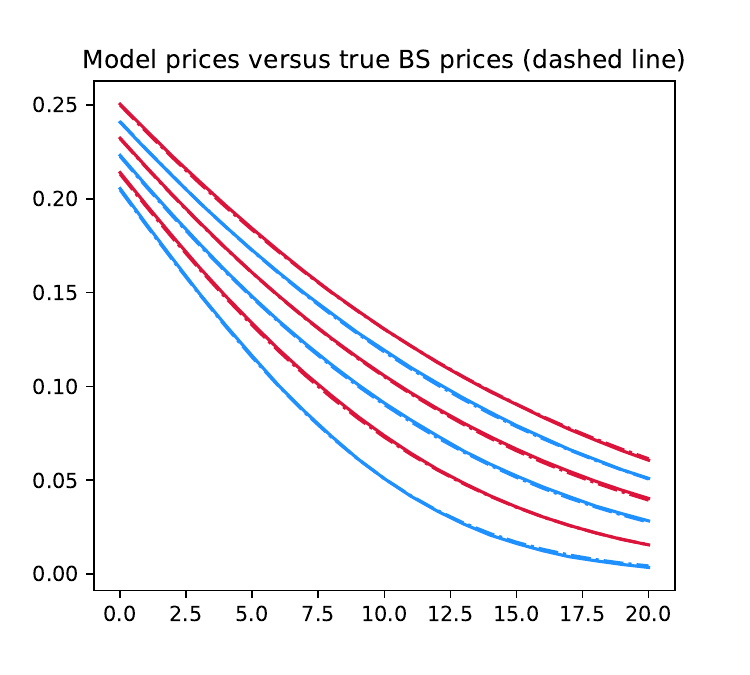} &
 \includegraphics[scale=0.6]{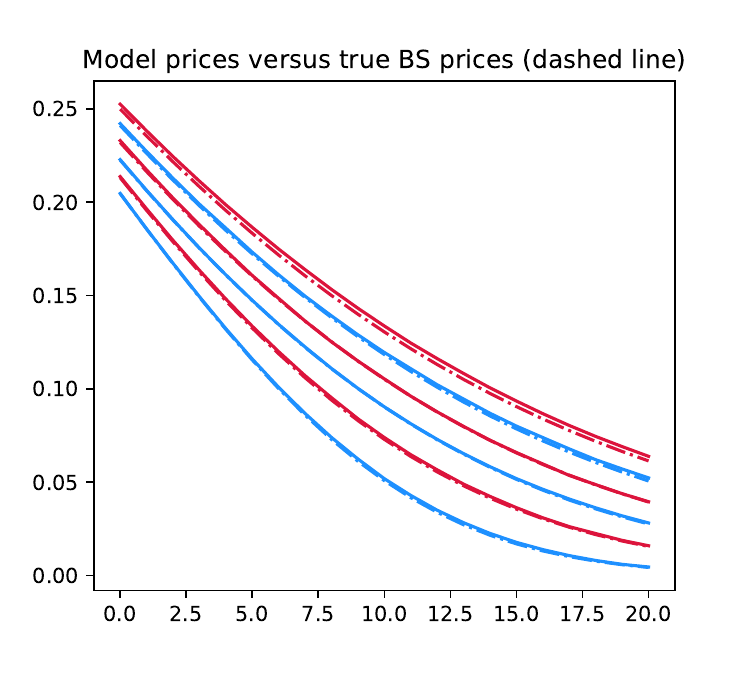} \\  [0.1ex] 
 \hline
 
 $\sigma_{prior}=0.2$ &  $\sigma_{prior}=0.25$   \\ [0.1ex] 
 \hline\hline
 \includegraphics[scale=0.6]{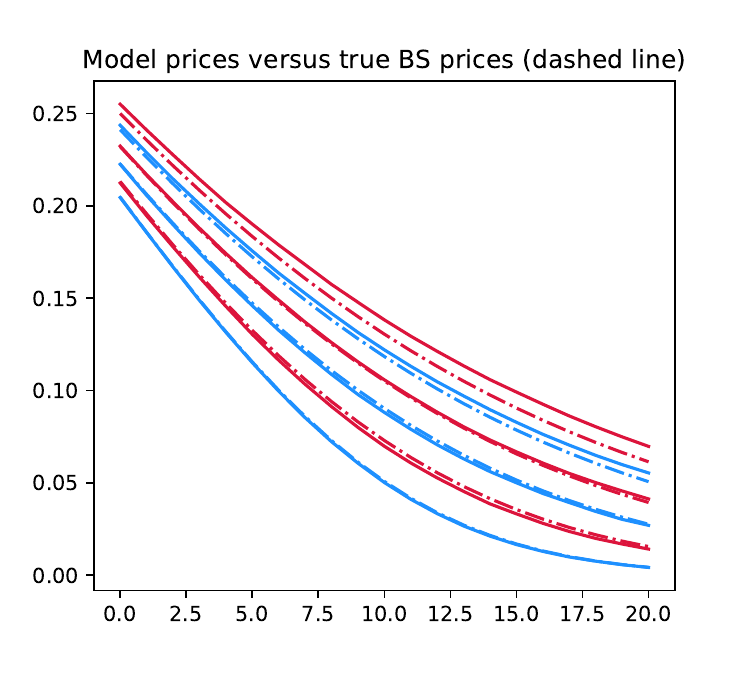}&\includegraphics[scale=0.6]{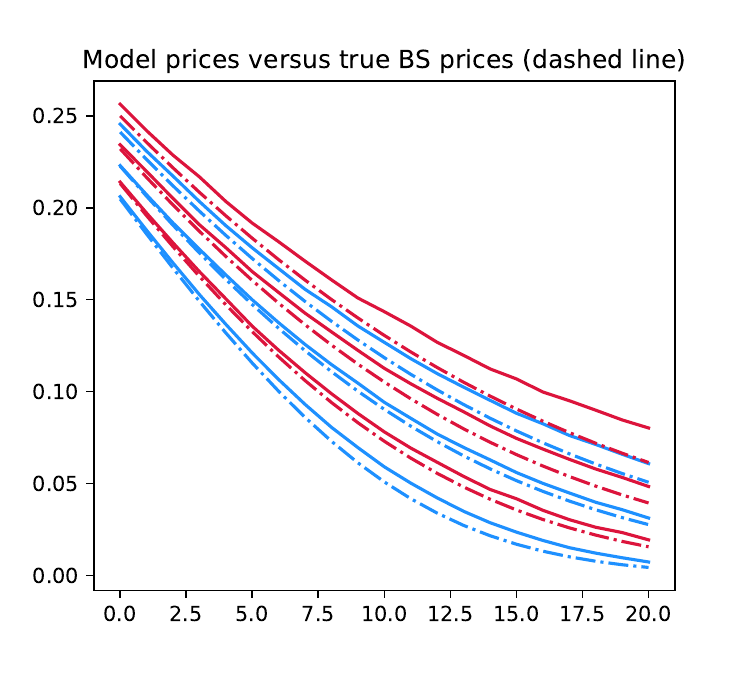}\\  [0.1ex] 
 \hline
\end{tabular}
\end{table}

\clearpage

\begin{table}
\centering
  \small
  \caption{Trace plots for calibration results with varying $\sigma_{prior}$.}
  \label{tb:table10}
  \begin{tabular}{||c | c c c c c c||} 
\hline
$\sigma_{prior}=0.1$&
\includegraphics[scale=0.14]{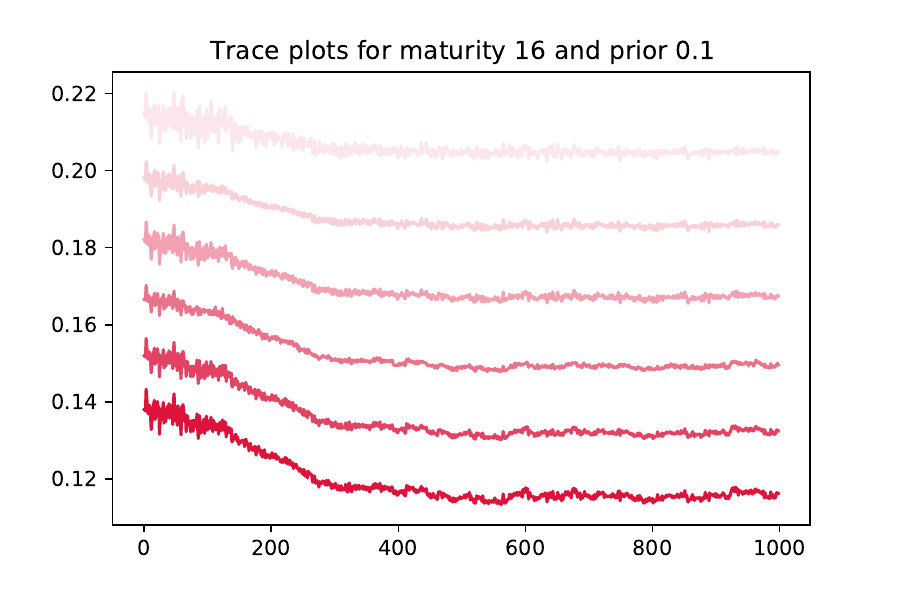}&
\includegraphics[scale=0.14]{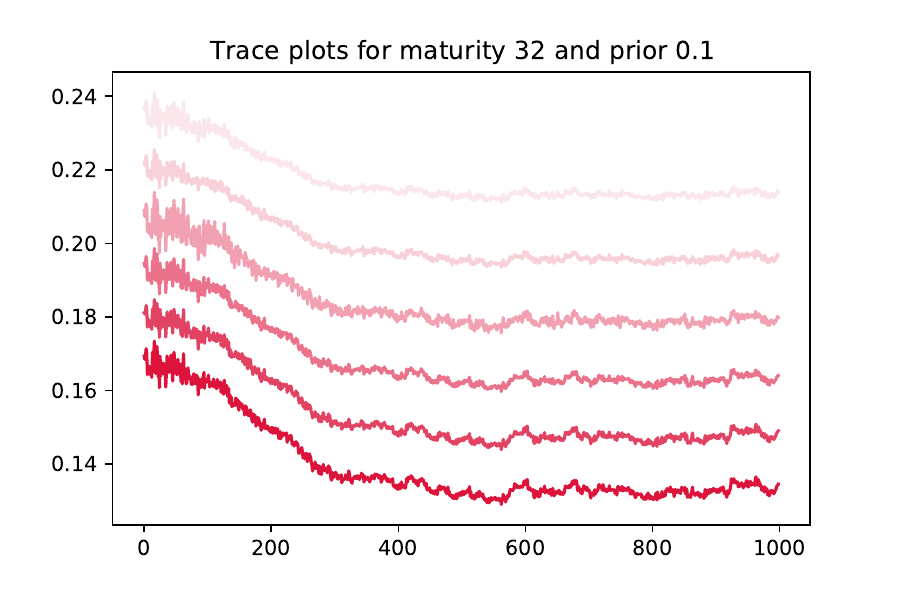}&
\includegraphics[scale=0.14]{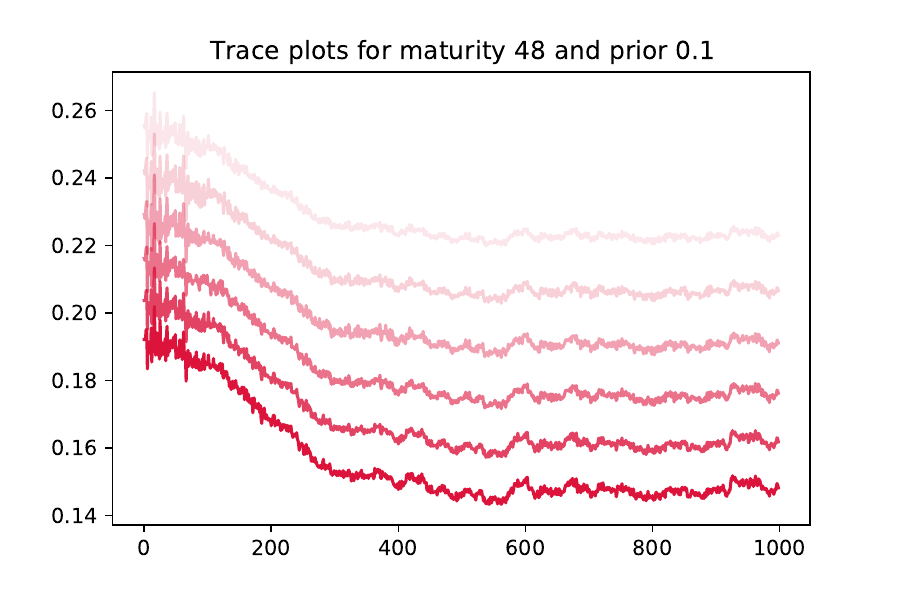}&
\includegraphics[scale=0.14]{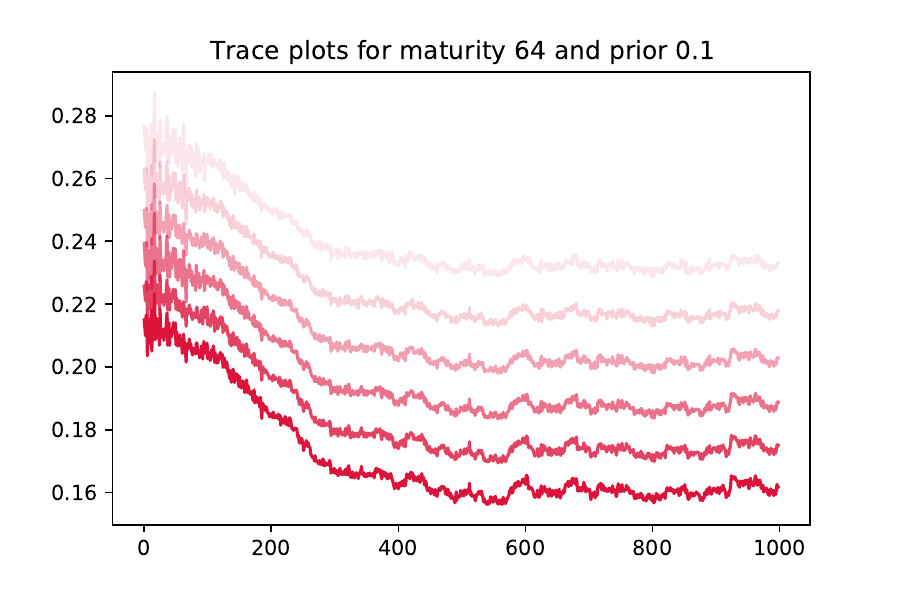}&
\includegraphics[scale=0.14]{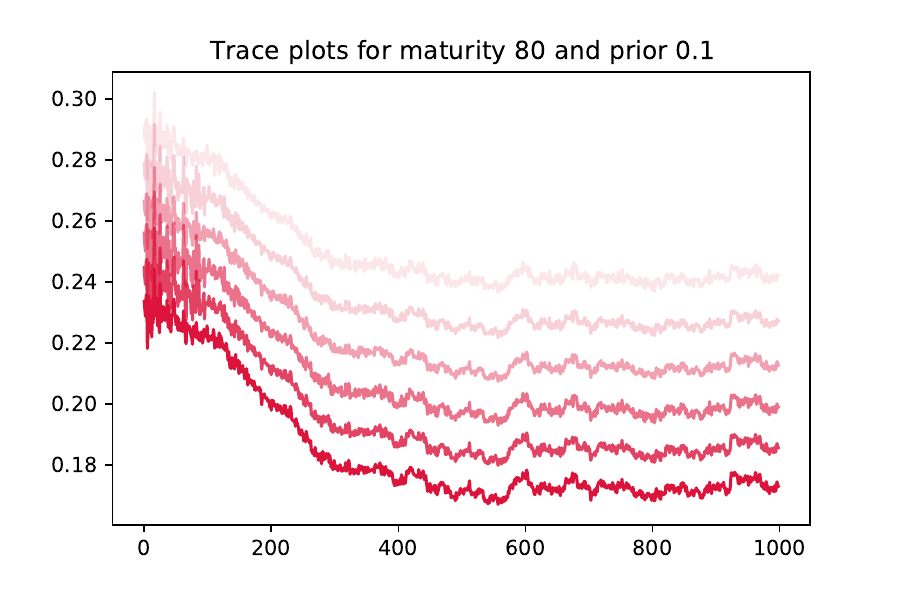}&
\includegraphics[scale=0.14]{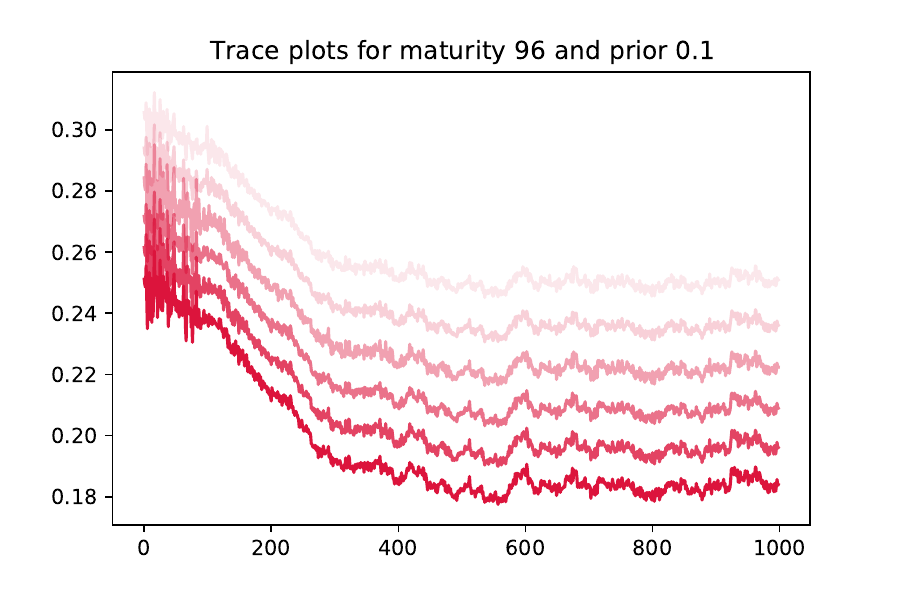}\\  [0.1ex] 
 \hline
$\sigma_{prior}=0.15$&\includegraphics[scale=0.14]{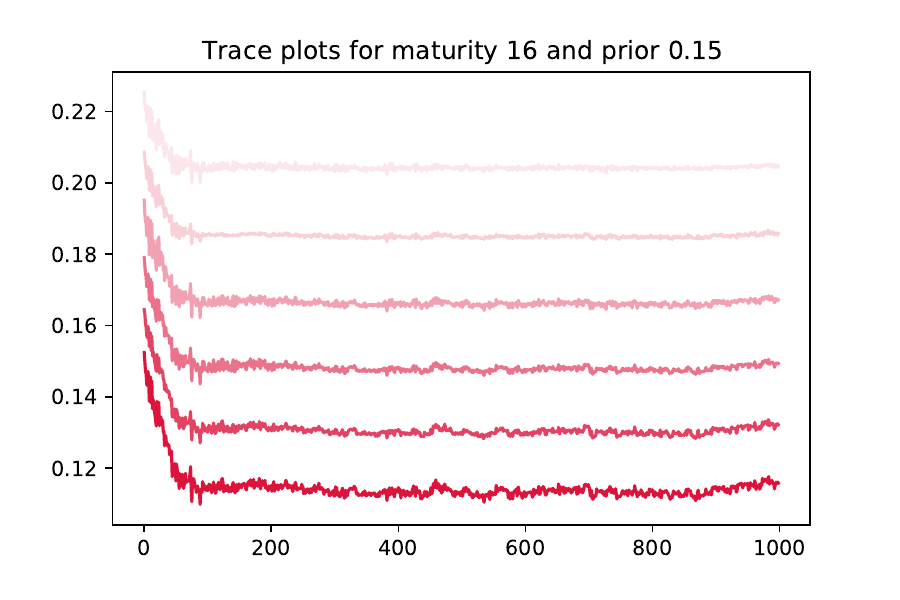}&
\includegraphics[scale=0.14]{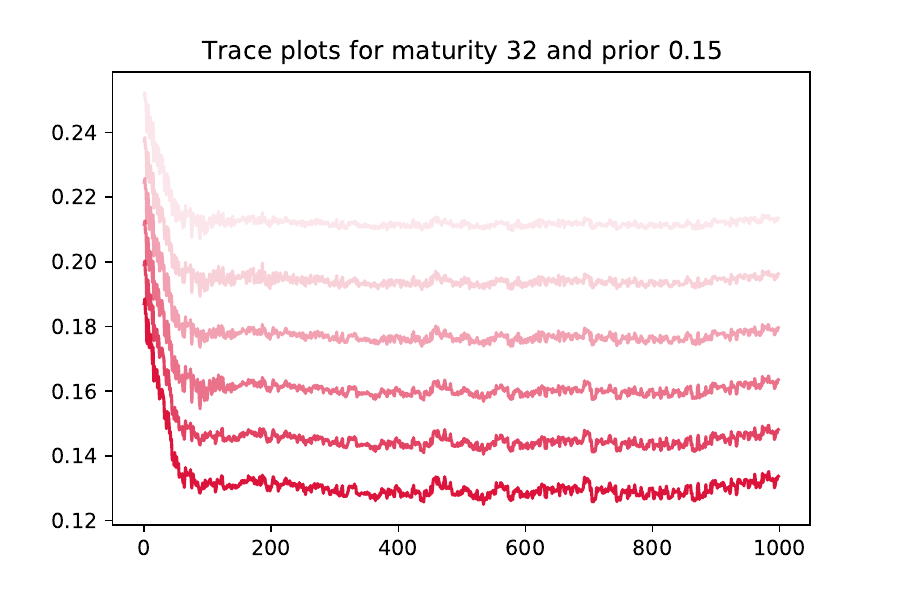}&
\includegraphics[scale=0.14]{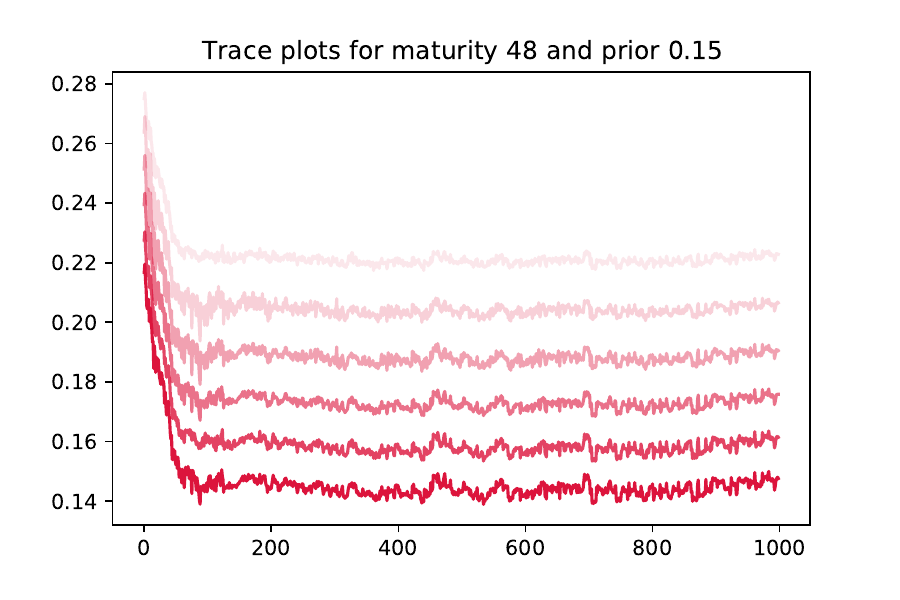}&
\includegraphics[scale=0.14]{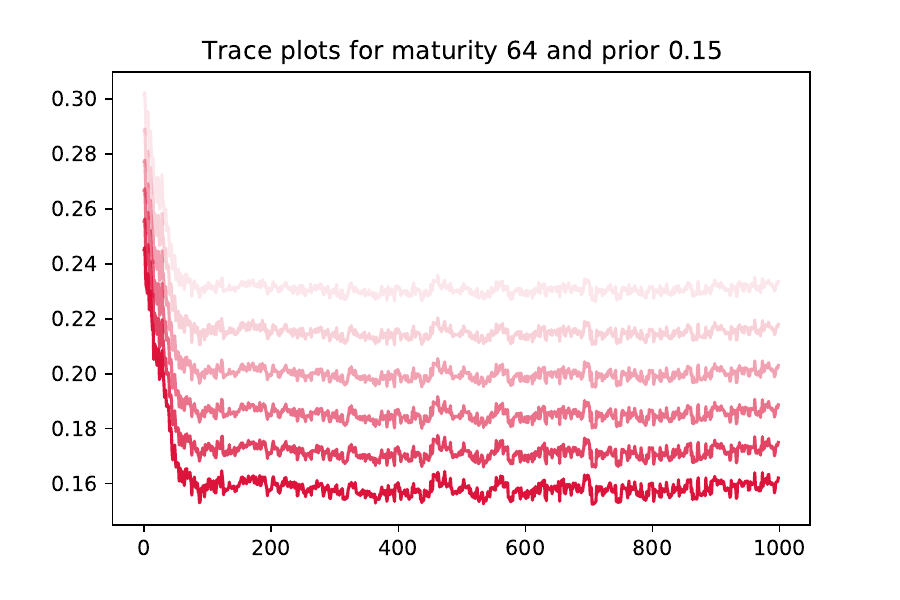}&
\includegraphics[scale=0.14]{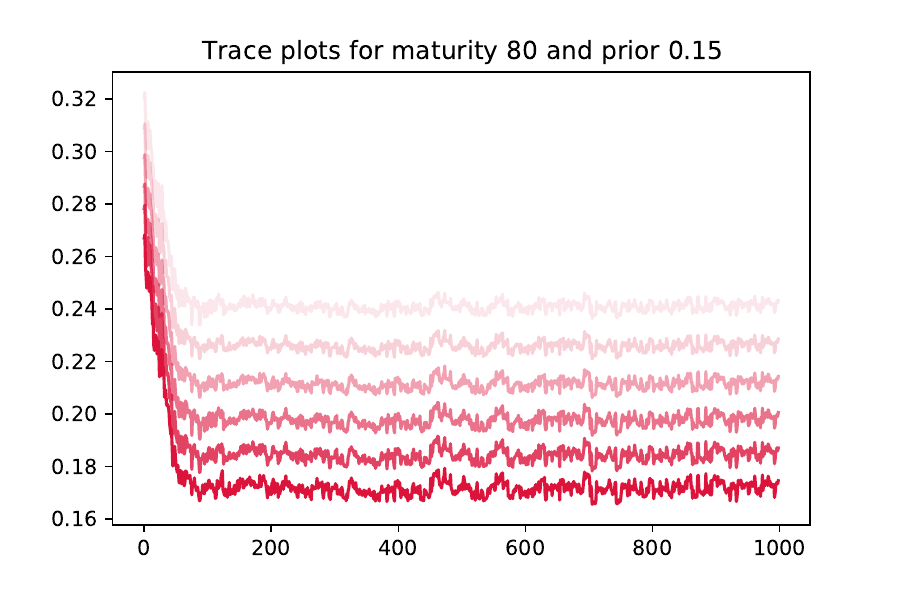}&
\includegraphics[scale=0.14]{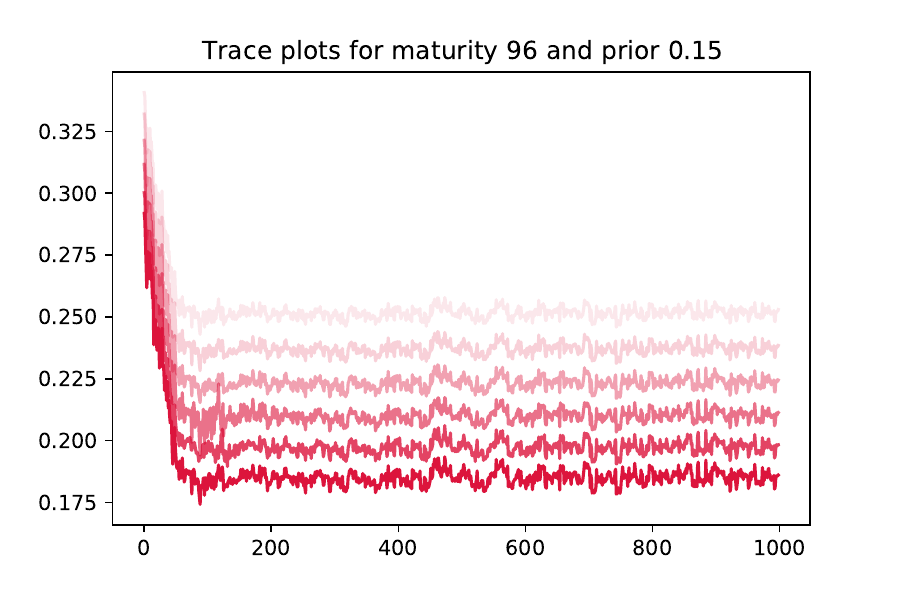}\\  [0.1ex] 
 \hline
$\sigma_{prior}=0.2$&\includegraphics[scale=0.14]{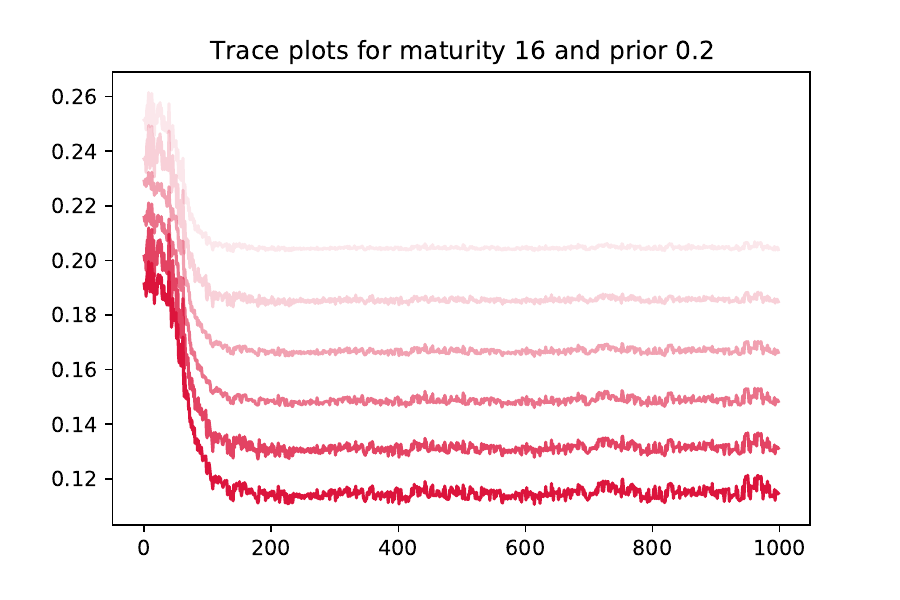}&
\includegraphics[scale=0.14]{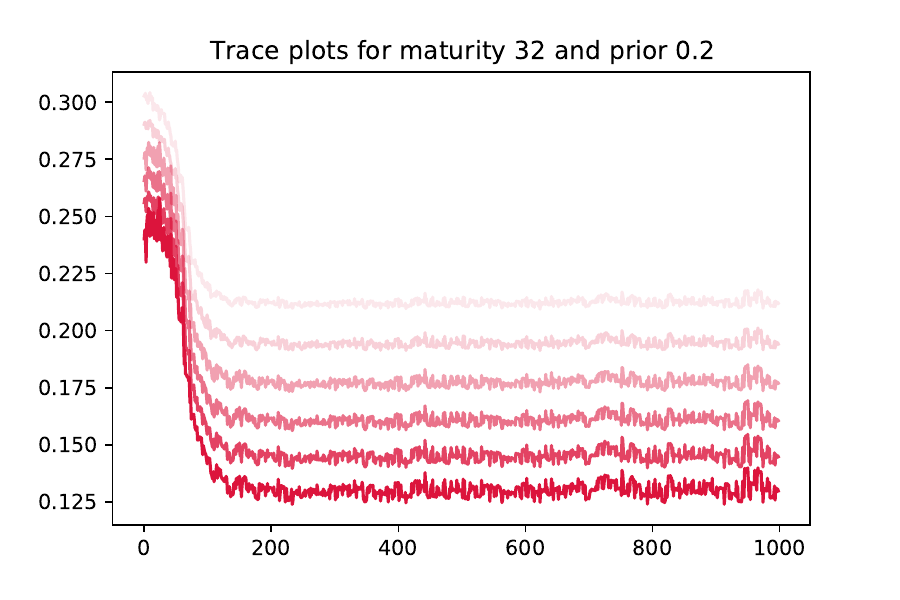}&
\includegraphics[scale=0.14]{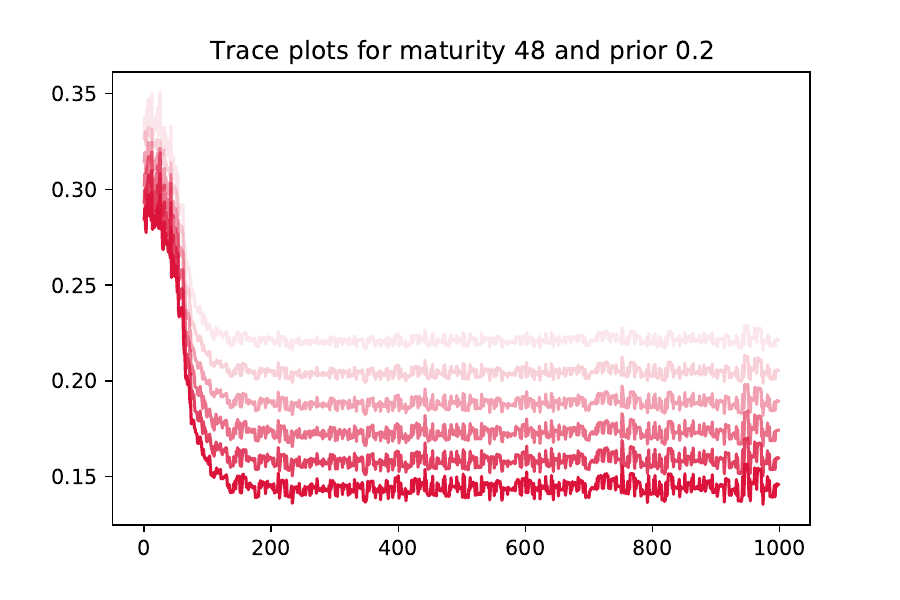}&
\includegraphics[scale=0.14]{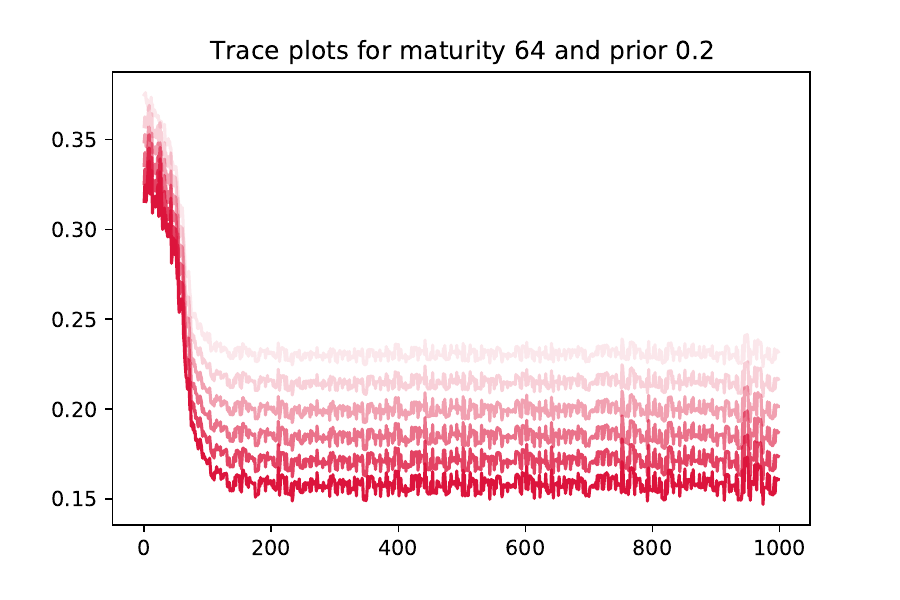}&
\includegraphics[scale=0.14]{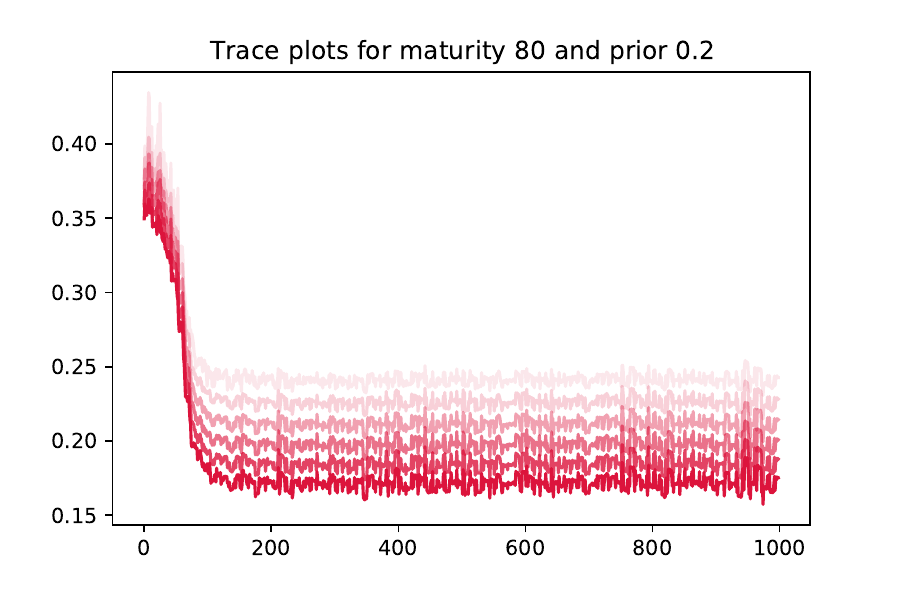}&
\includegraphics[scale=0.14]{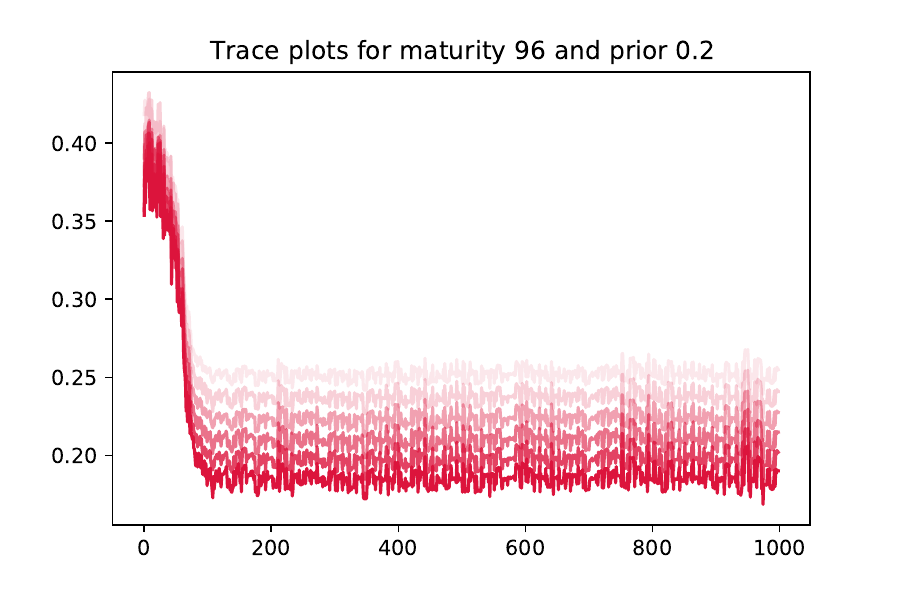}\\  [0.1ex] 
 \hline

$\sigma_{prior}=0.25$&\includegraphics[scale=0.14]{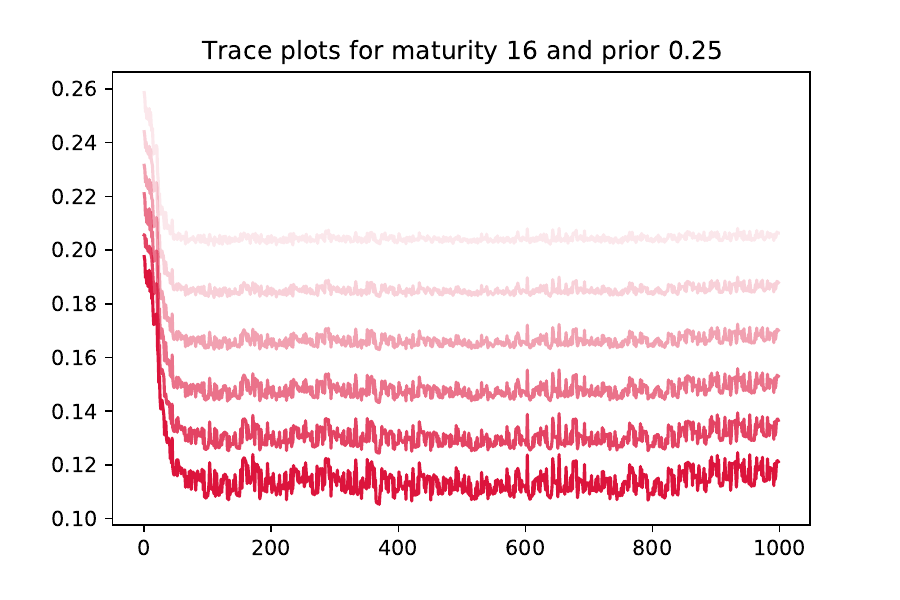}&
\includegraphics[scale=0.14]{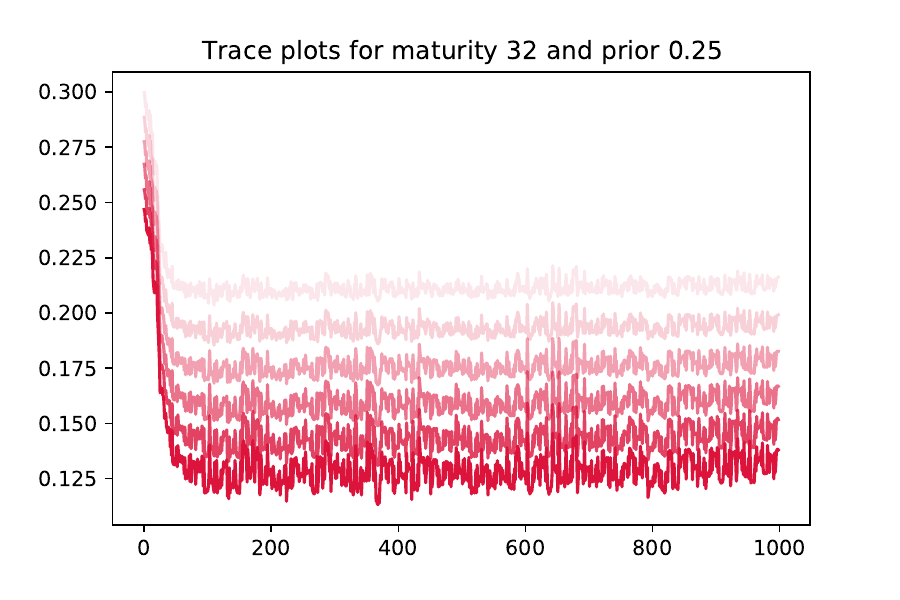}&
\includegraphics[scale=0.14]{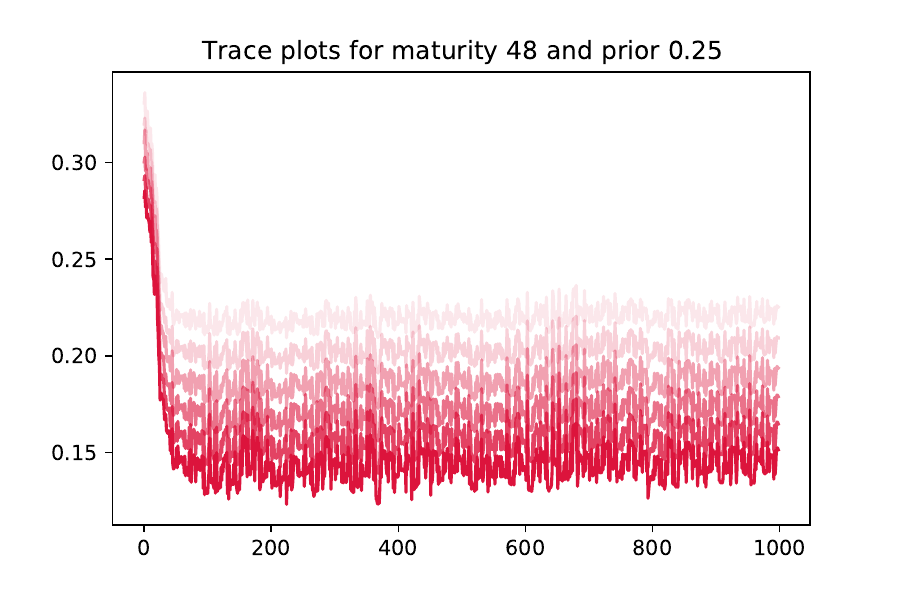}&
\includegraphics[scale=0.14]{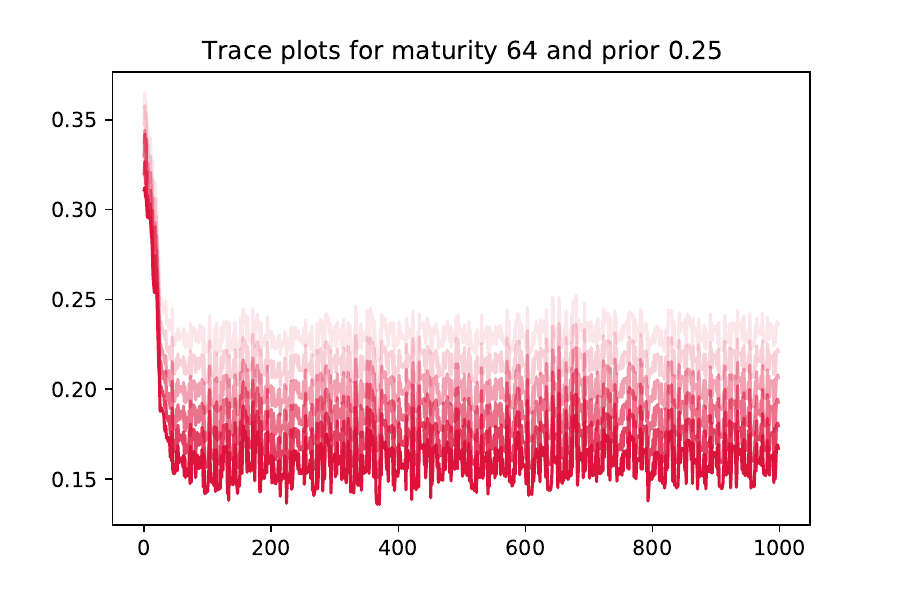}&
\includegraphics[scale=0.14]{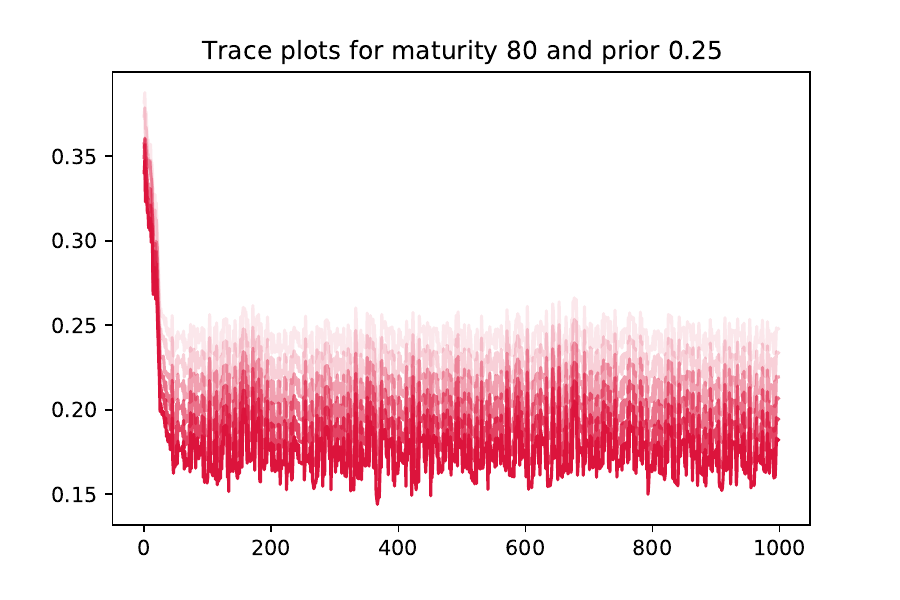}&
\includegraphics[scale=0.14]{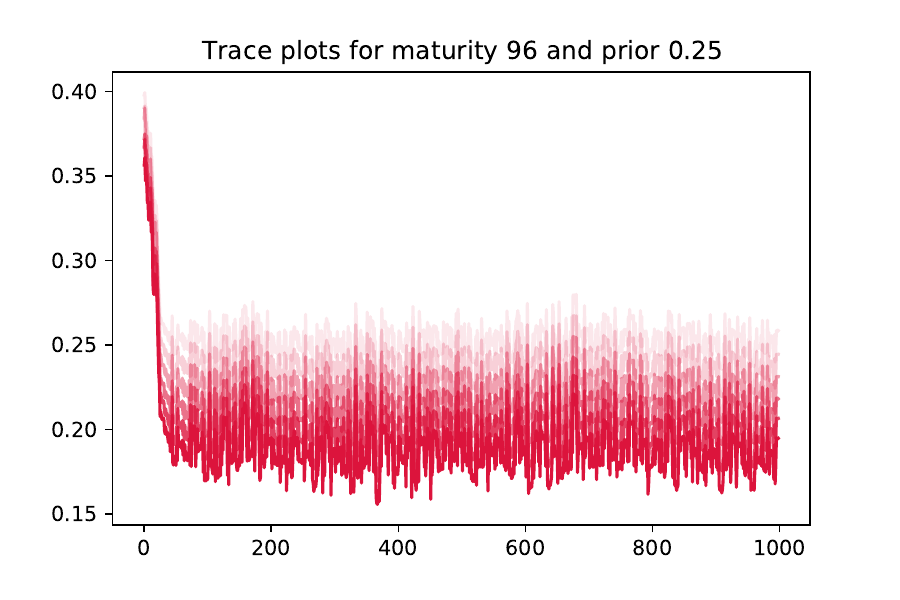}
\\  [0.1ex] 
 \hline
\end{tabular}
\end{table}

\begin{table}
\centering
  \small
  \caption{Sensitivity of calibrated implied volatilities, with respect to $\sigma_{prior}$.}
  \label{tb:table11}
\begin{tabular}{||c|c c c c c c||} \hline
$\sigma_{prior}=0.1$&
\includegraphics[scale=0.17]{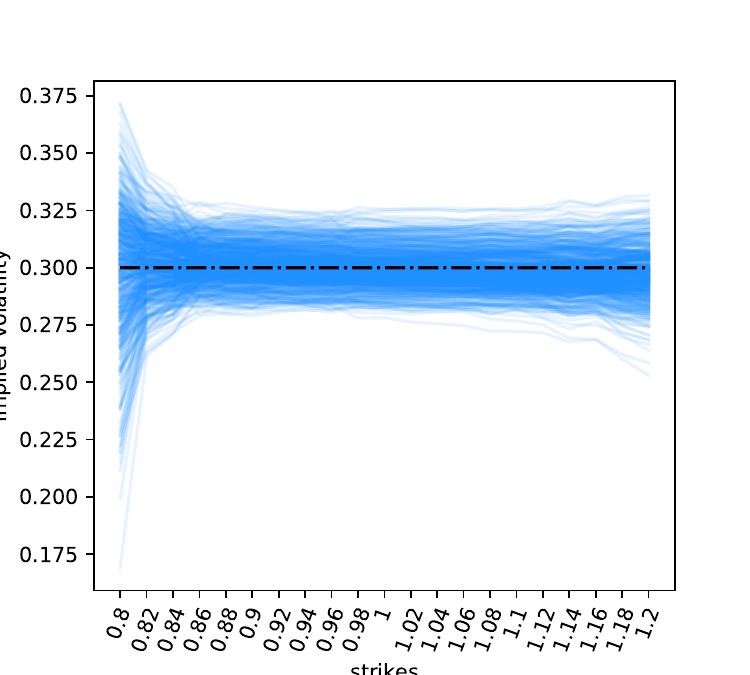}&
\includegraphics[scale=0.17]{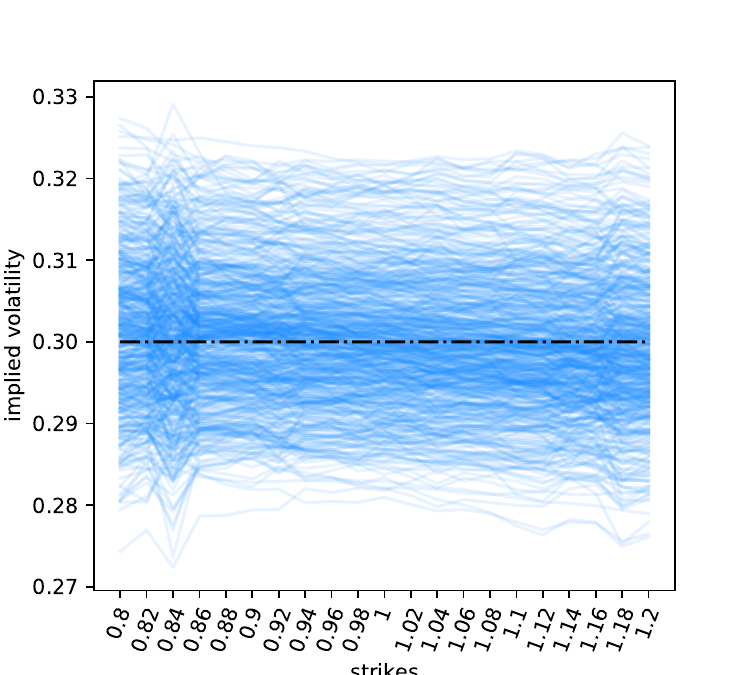}&
\includegraphics[scale=0.17]{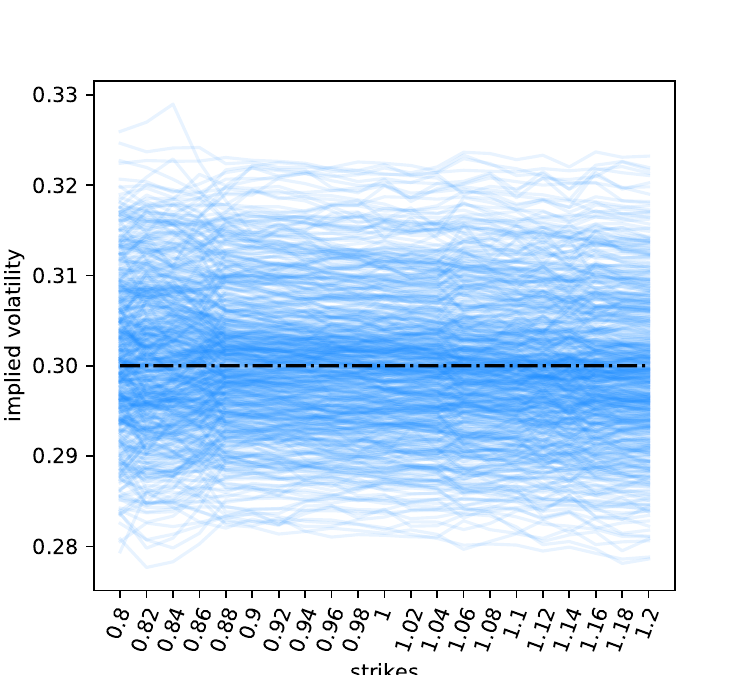}&
\includegraphics[scale=0.17]{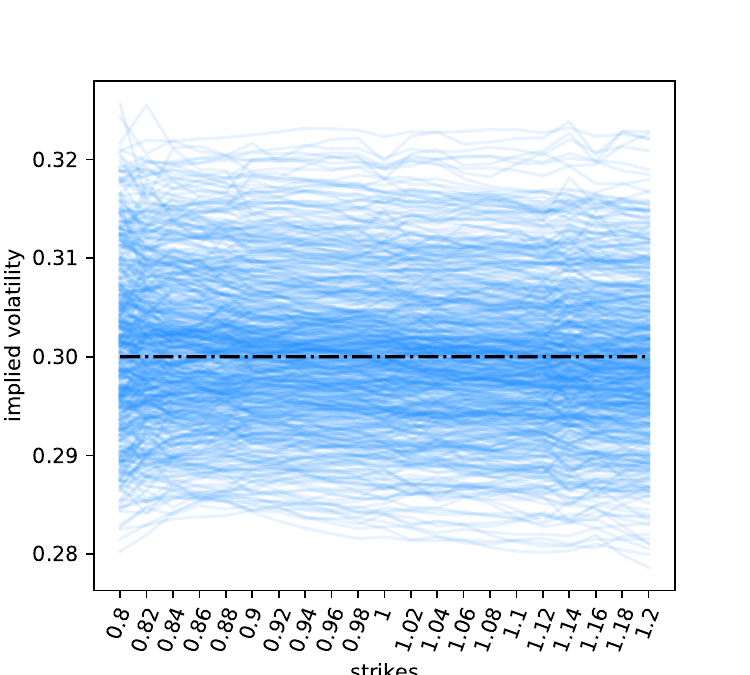}&
\includegraphics[scale=0.17]{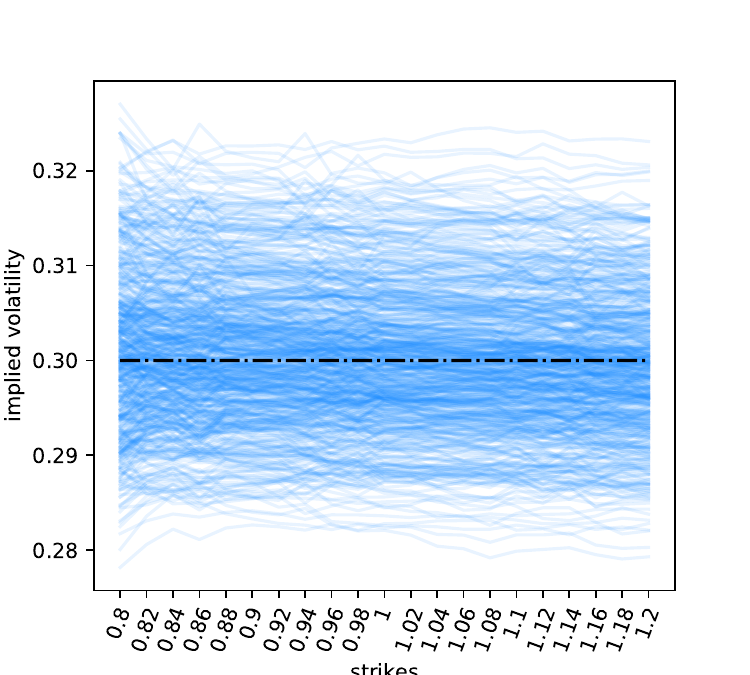}&
\includegraphics[scale=0.17]{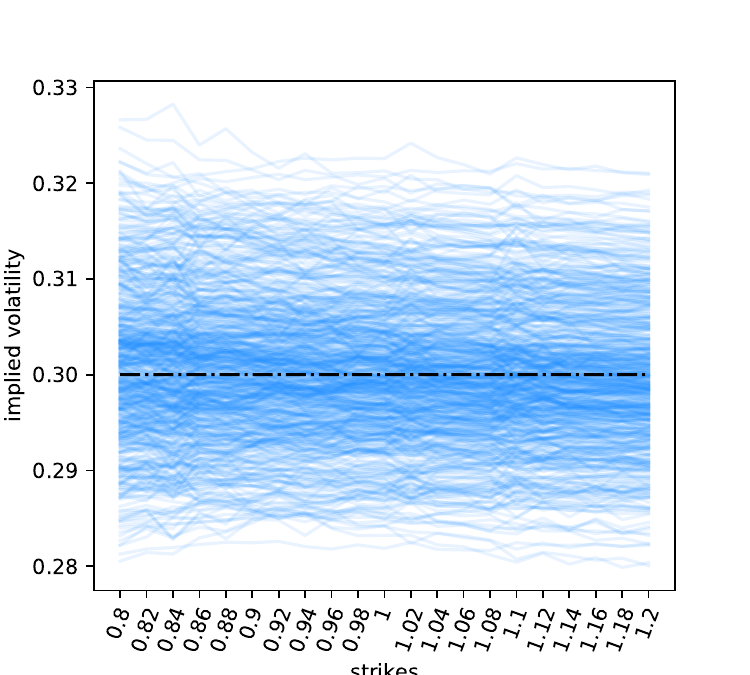}\\  [0.1ex] 
 \hline

$\sigma_{prior}=0.15$&\includegraphics[scale=0.17]{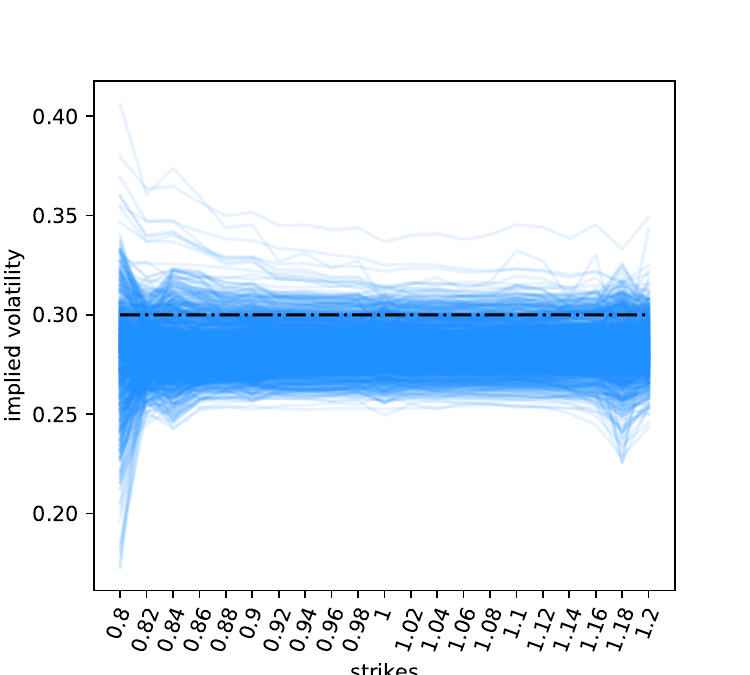}&
\includegraphics[scale=0.17]{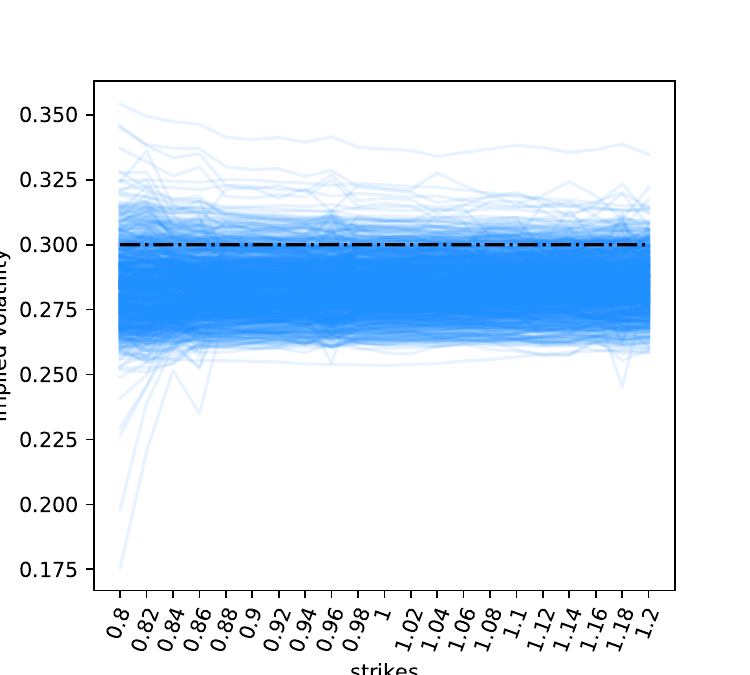}&
\includegraphics[scale=0.17]{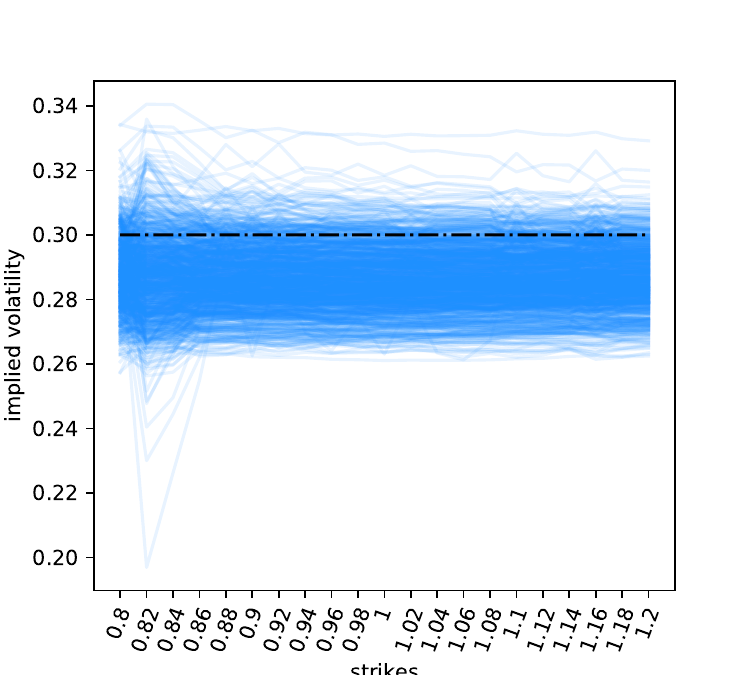}&
\includegraphics[scale=0.17]{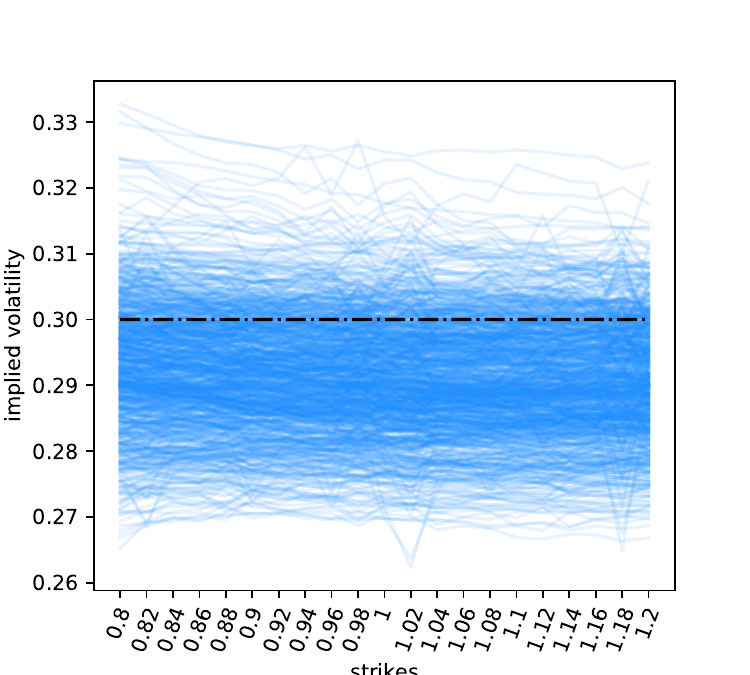}&
\includegraphics[scale=0.17]{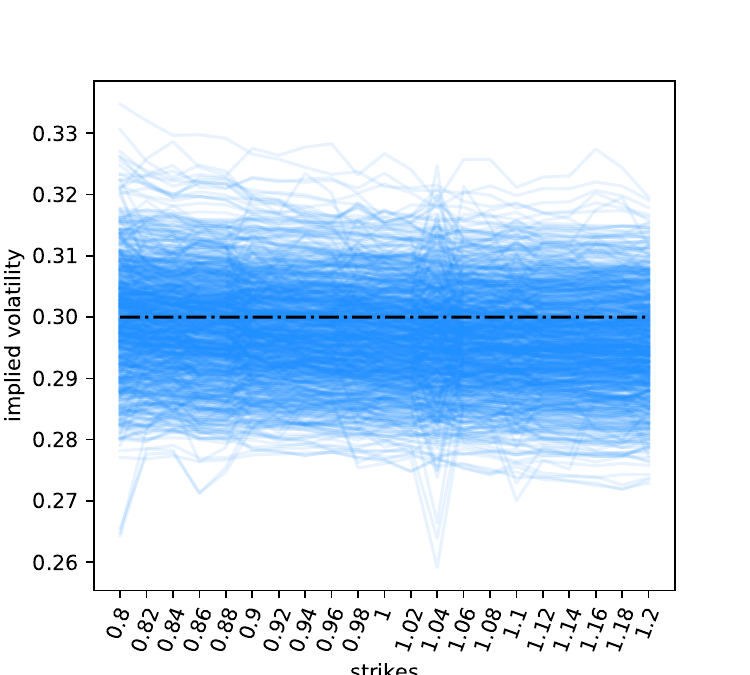}&
\includegraphics[scale=0.17]{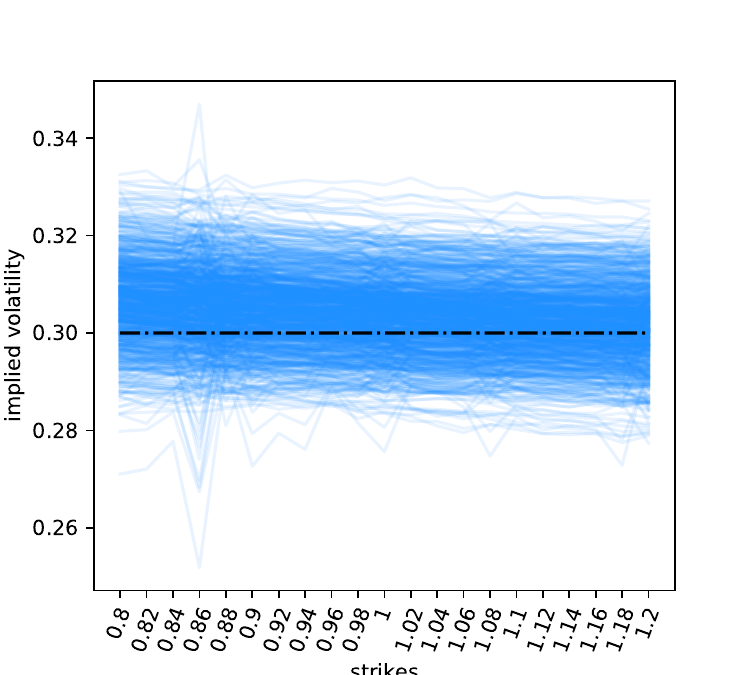}\\  [0.1ex] 
 \hline

$\sigma_{prior}=0.2$&\includegraphics[scale=0.17]{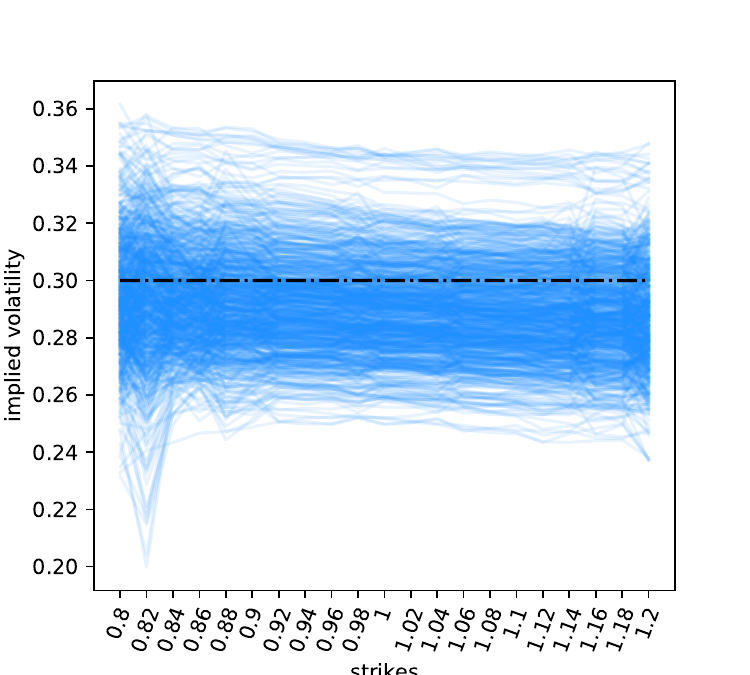}&
\includegraphics[scale=0.17]{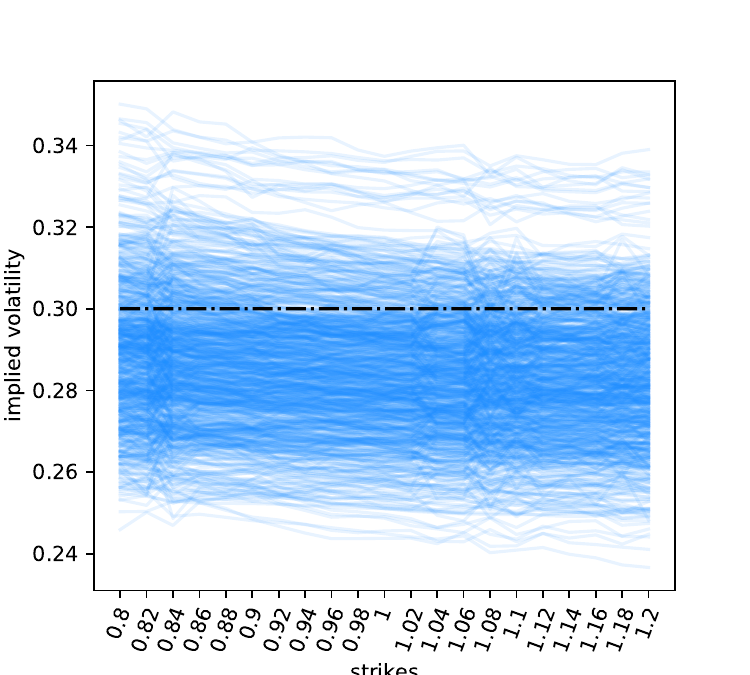}&
\includegraphics[scale=0.17]{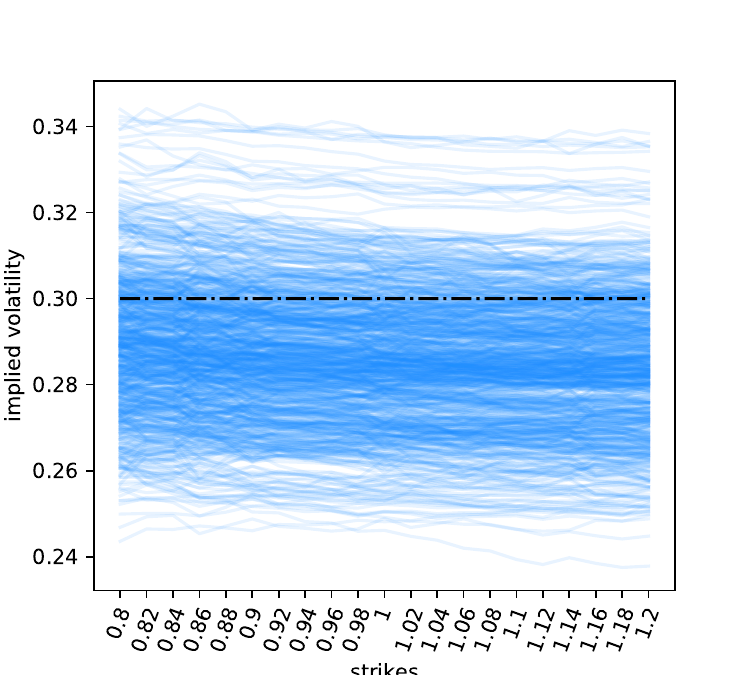}&
\includegraphics[scale=0.17]{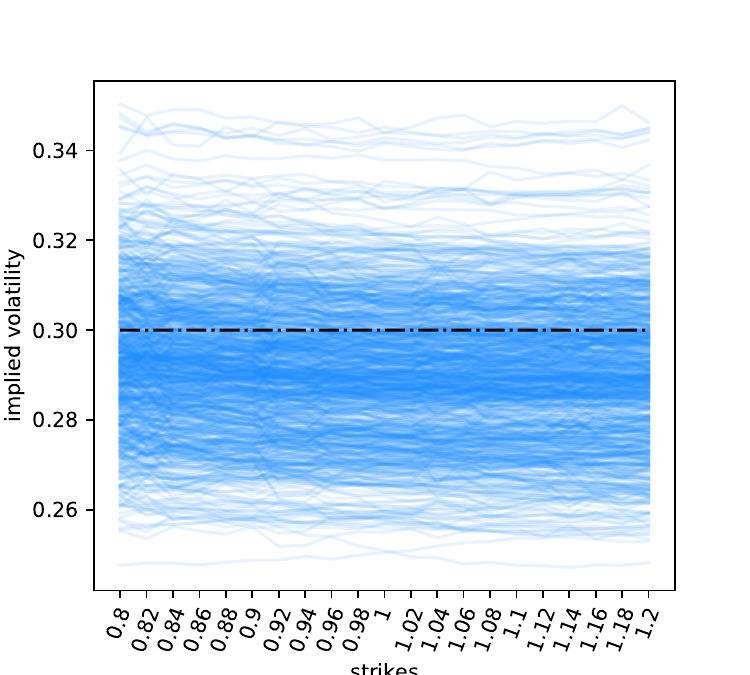}&
\includegraphics[scale=0.17]{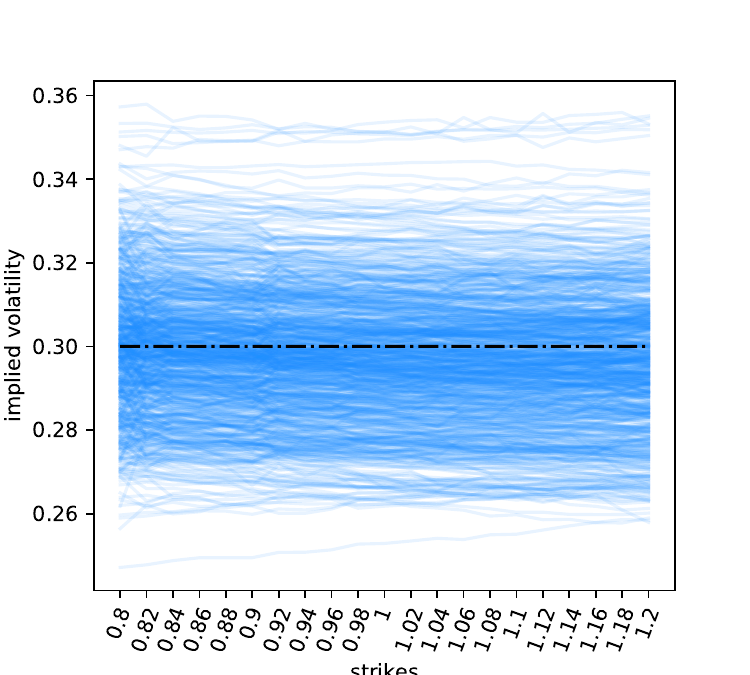}&
\includegraphics[scale=0.17]{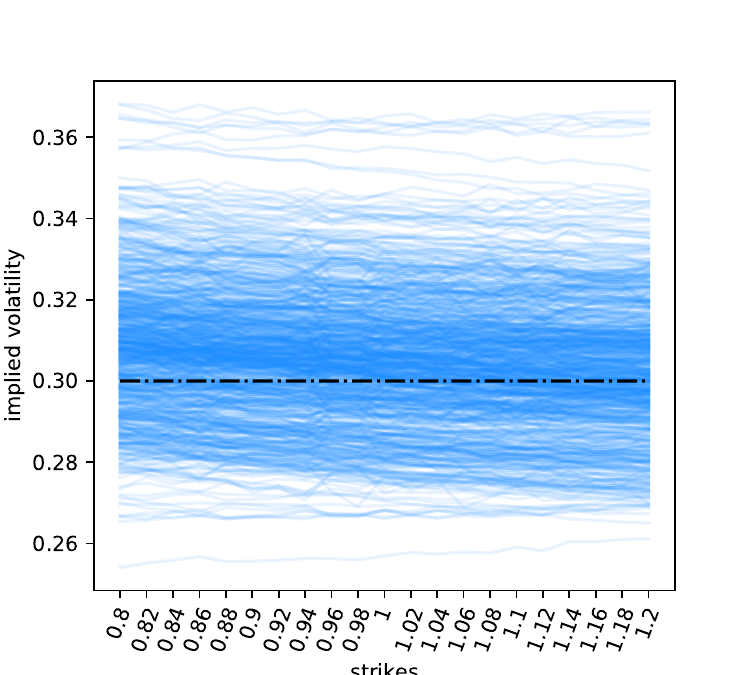}\\  [0.1ex] 
 \hline

$\sigma_{prior}=0.25$&\includegraphics[scale=0.17]{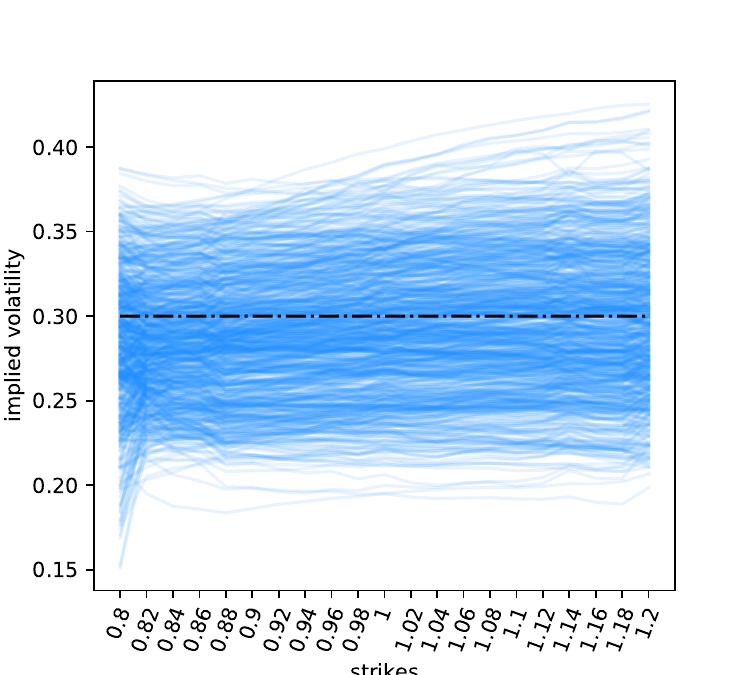}&
\includegraphics[scale=0.17]{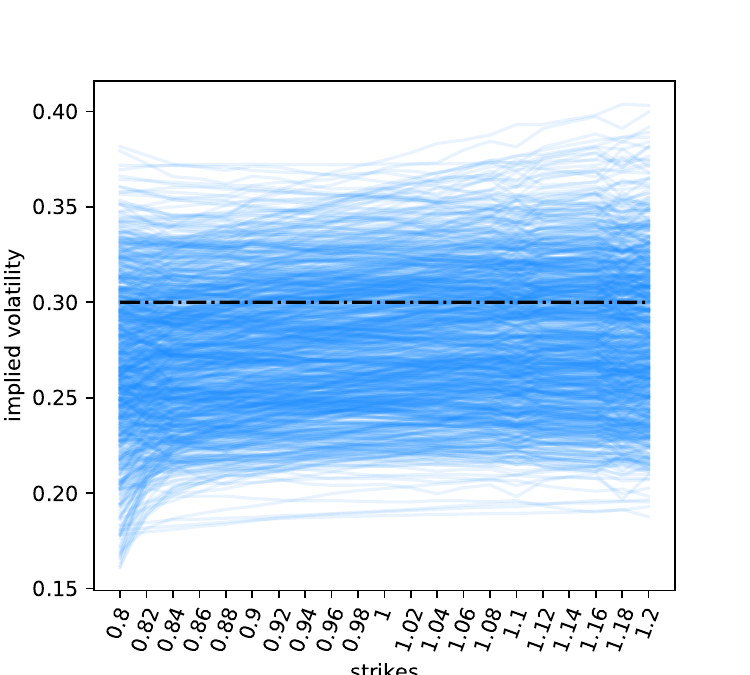}&
\includegraphics[scale=0.17]{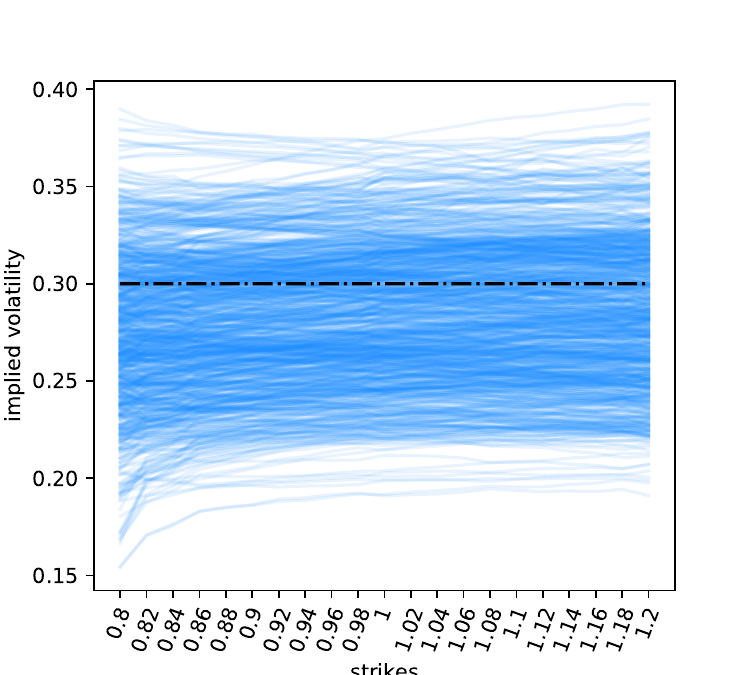}&
\includegraphics[scale=0.17]{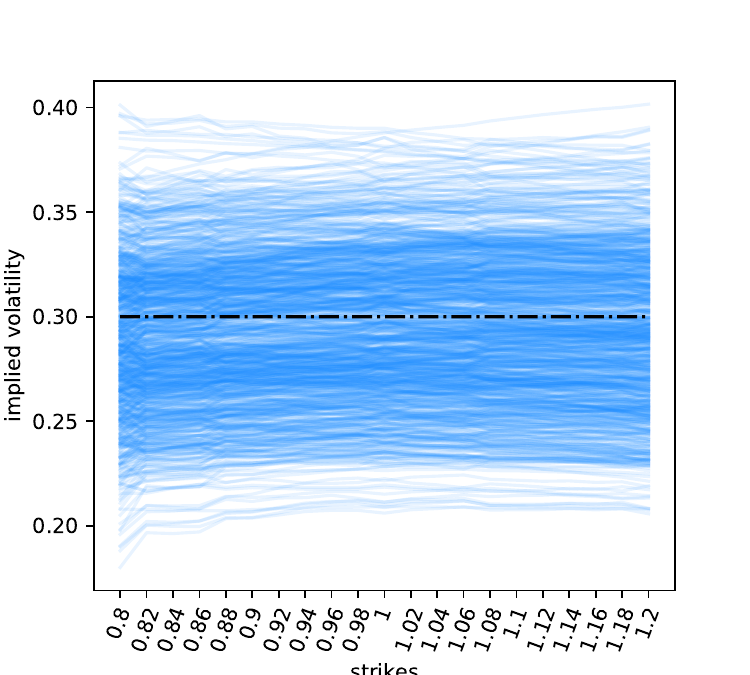}&
\includegraphics[scale=0.17]{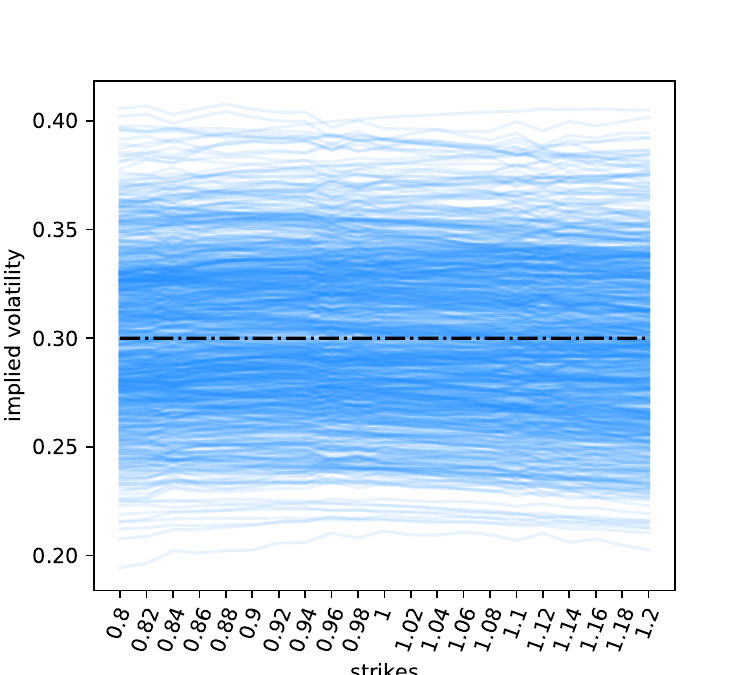}&
\includegraphics[scale=0.17]{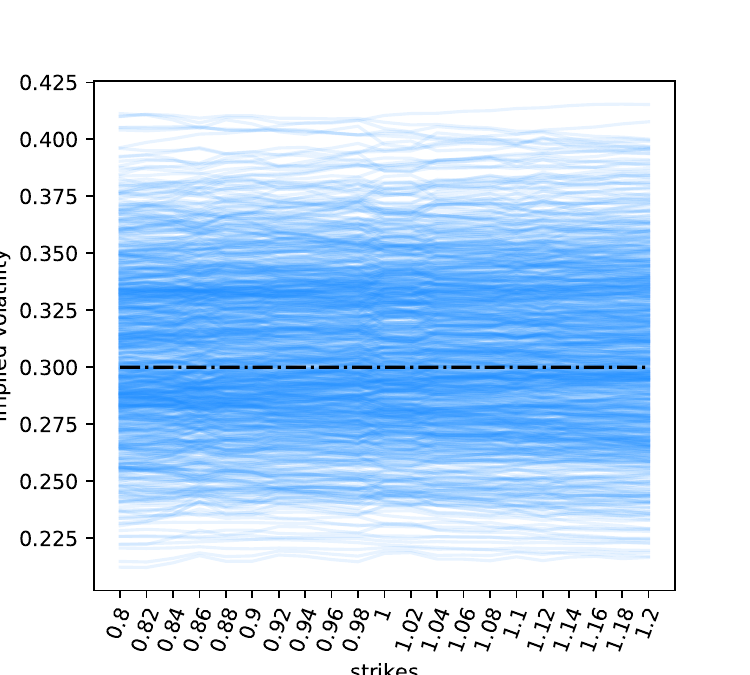}\\  [0.1ex] 
 \hline
 \hline
\end{tabular}
\end{table}

\begin{table}
  \centering
  \small
  \caption{Sensitivity of calibration to time series data, with respect to $\delta$.}
  \label{tb:table12}
\begin{tabular}{||c c ||} 
 \hline
   $\delta=1$&  $\delta=5$  \\ [0.1ex] 
 \hline\hline
 \includegraphics[scale=0.5]{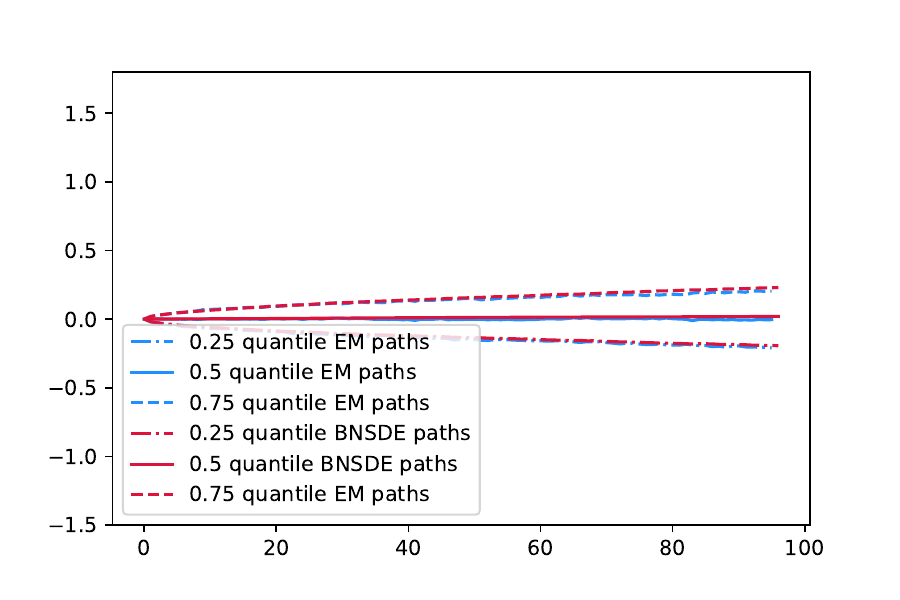}&
 \includegraphics[scale=0.5]{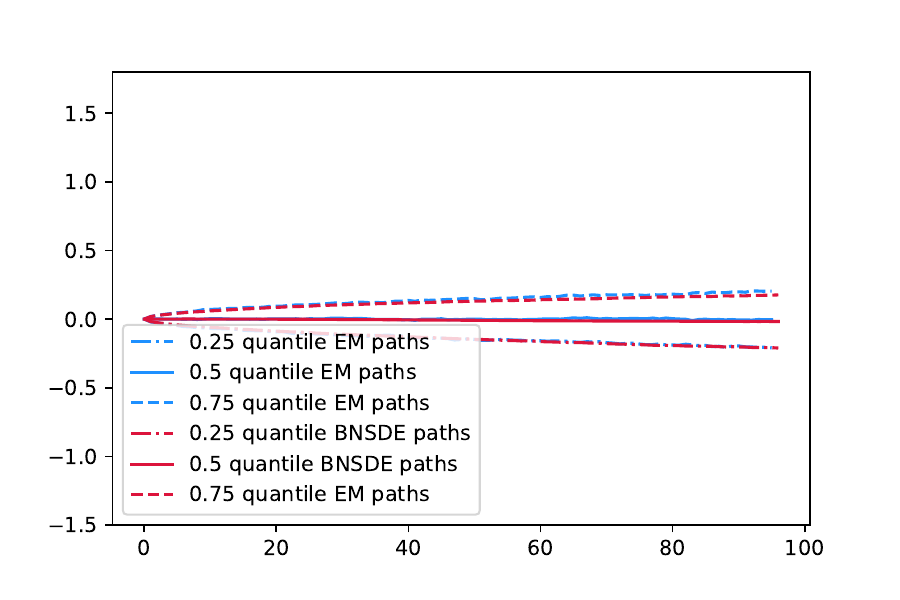} \\  [0.1ex] 
 \hline
 
  $\delta=10$ &  $\delta=38.44$   \\ [0.1ex] 
 \hline\hline
 \includegraphics[scale=0.5]{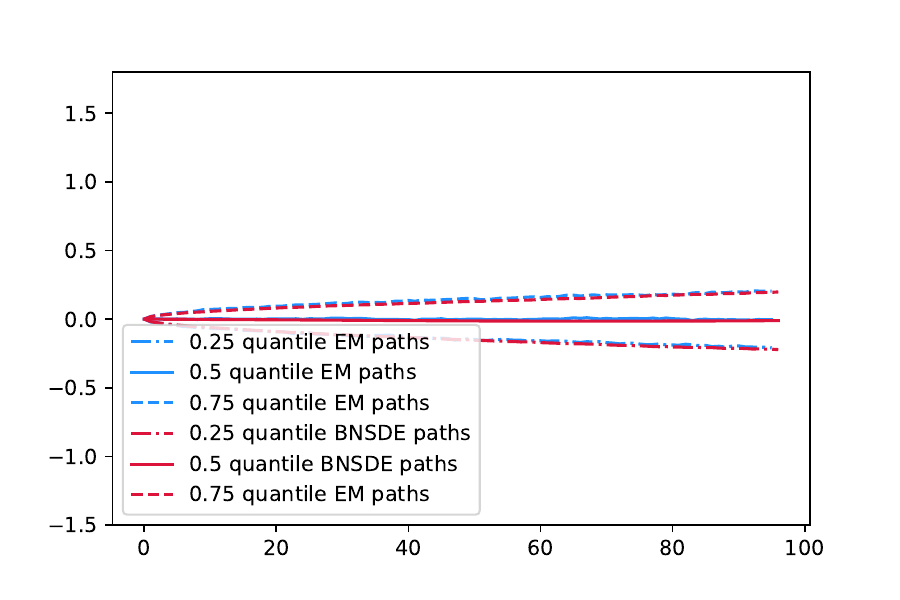}& \includegraphics[scale=0.5]{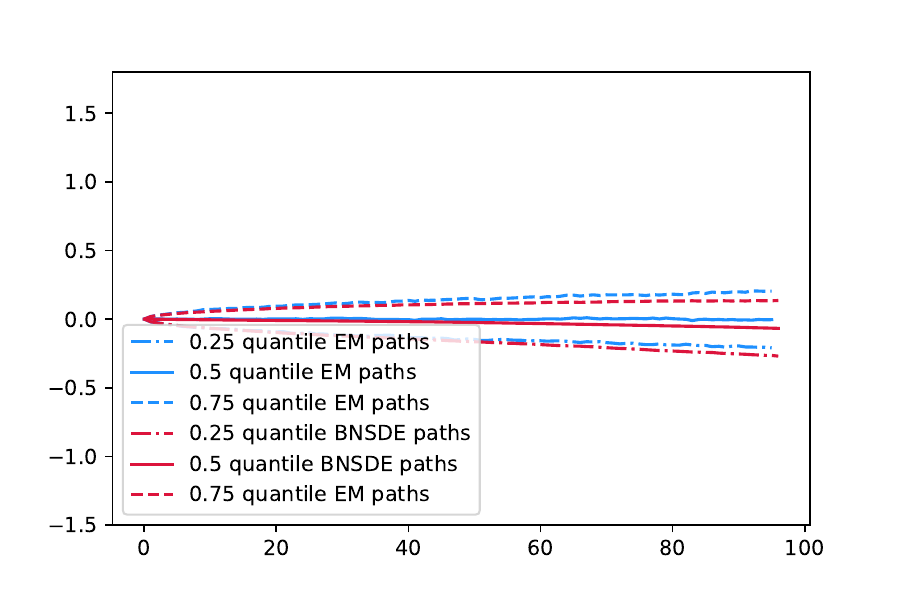}\\  [0.1ex] 
 \hline
\end{tabular}
\end{table}

\begin{table}
\centering
  \small
  \caption{Sensitivity of calibrated call option prices, with respect to $\delta$.}
  \label{tb:table13}
\begin{tabular}{||c c ||} 
 \hline
   $\delta=1$&   $\delta=5$  \\ [0.1ex] 
 \hline\hline
 \includegraphics[scale=0.6]{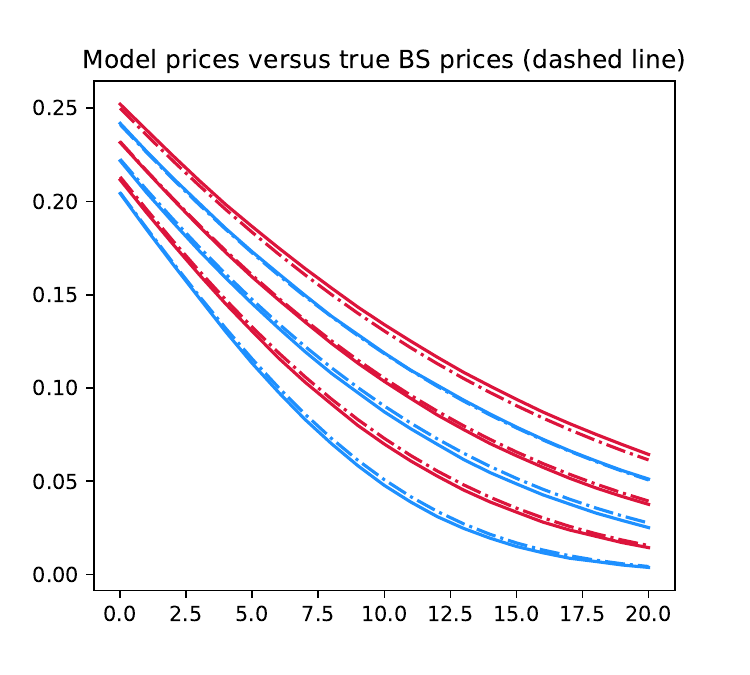}&
 \includegraphics[scale=0.6]{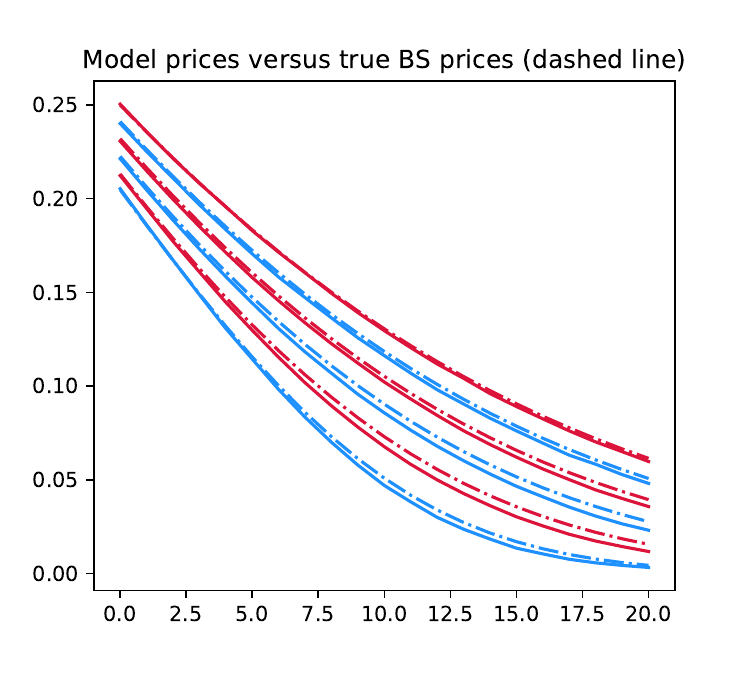} \\  [0.1ex] 
 \hline
 
  $\delta=10$&   $\delta=38.44$   \\ [0.1ex] 
 \hline\hline
 \includegraphics[scale=0.6]{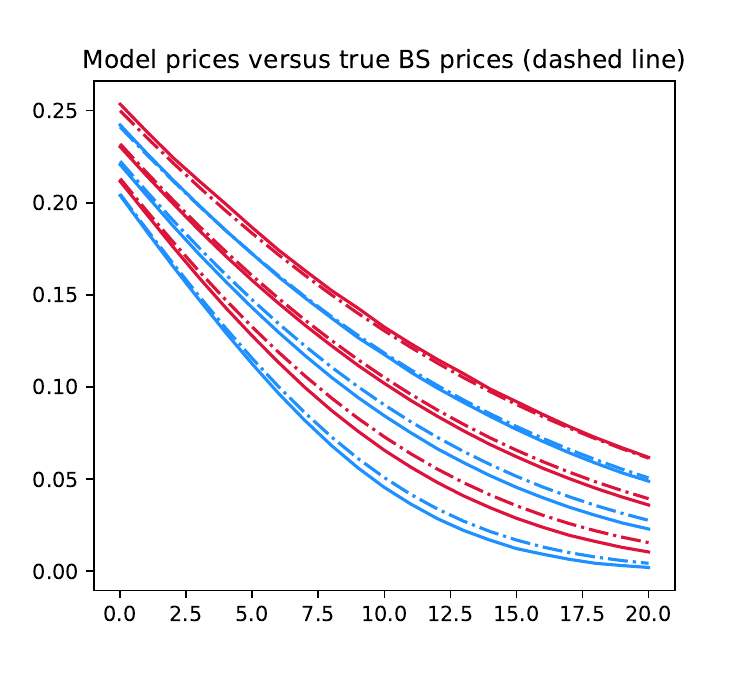}& \includegraphics[scale=0.6]{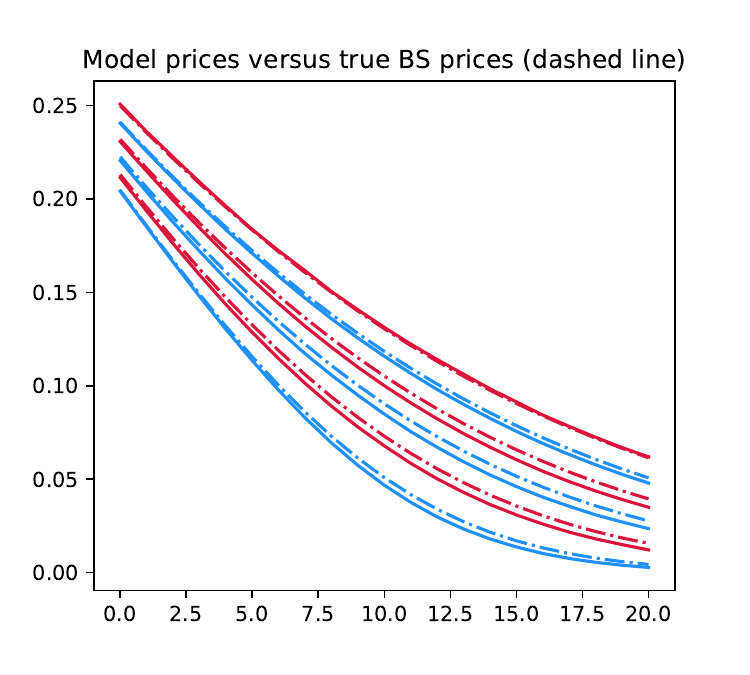}\\  [0.1ex] 
 \hline
\end{tabular}
\end{table}

\begin{table}
\centering
  \small
  \caption{Trace plots for calibration results with varying $\delta$.}
  \label{tb:table14}
\begin{tabular}{||c|c c c c c c||} \hline
$\delta=1$&
\includegraphics[scale=0.17]{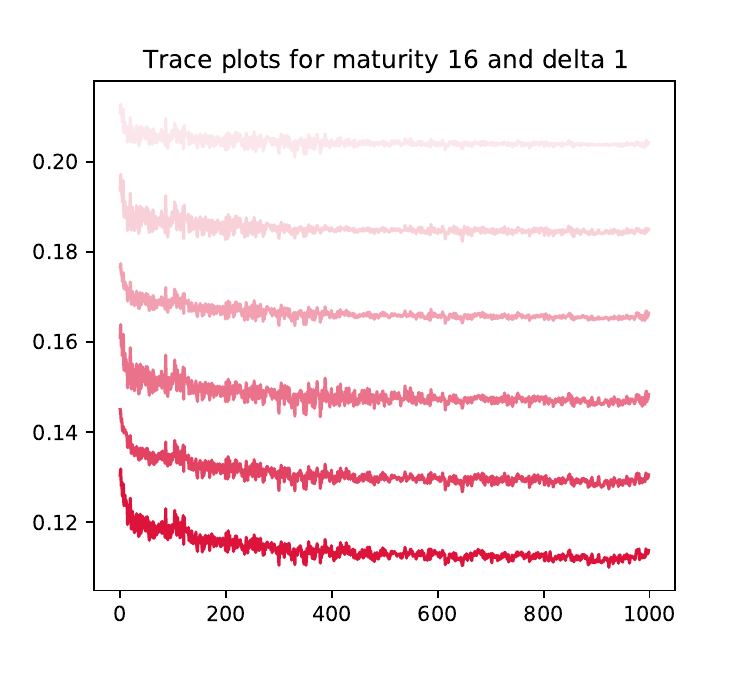}&
\includegraphics[scale=0.17]{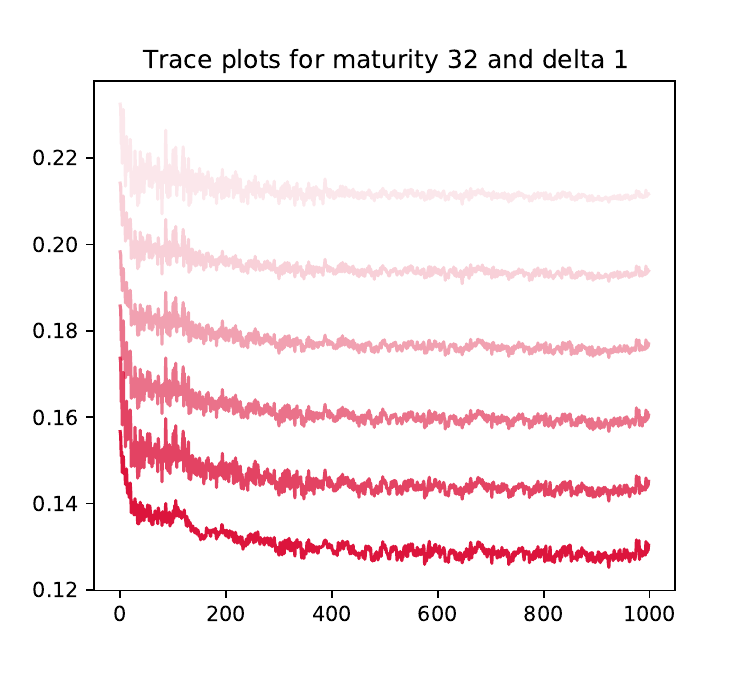}&
\includegraphics[scale=0.17]{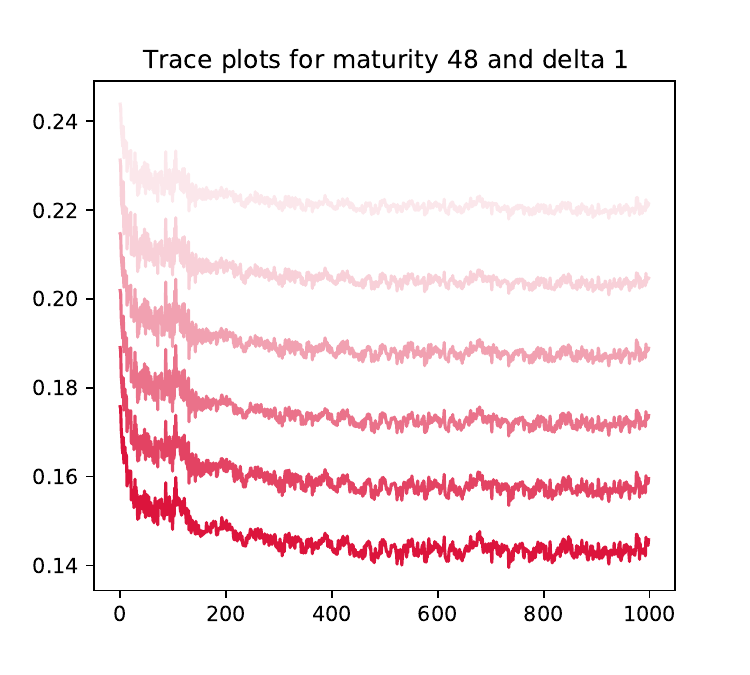}&
\includegraphics[scale=0.17]{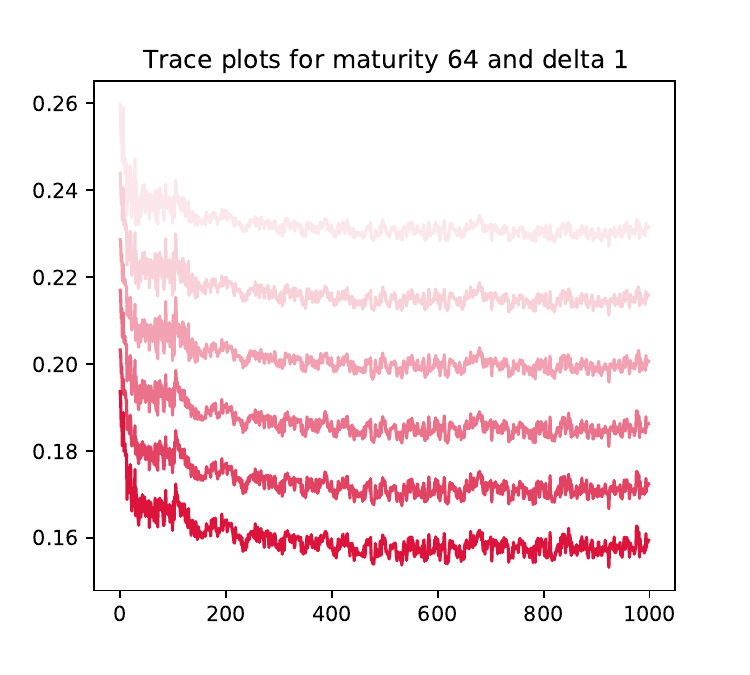}&
\includegraphics[scale=0.17]{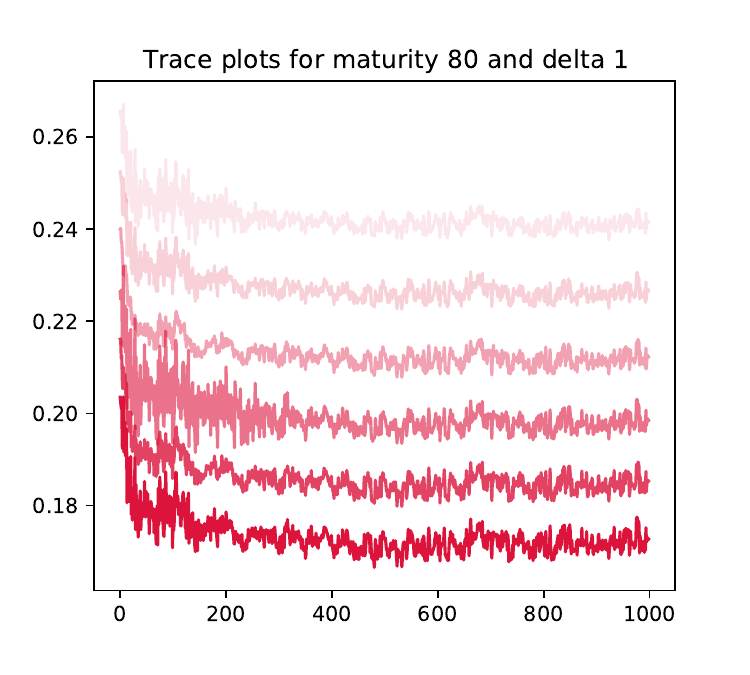}&
\includegraphics[scale=0.17]{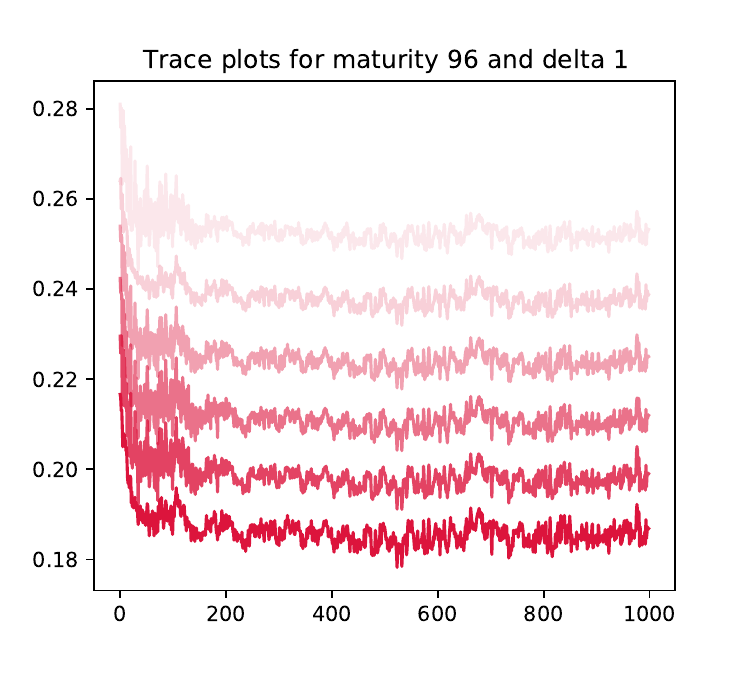}\\  [0.1ex] 
 \hline

$\delta=5$&\includegraphics[scale=0.17]{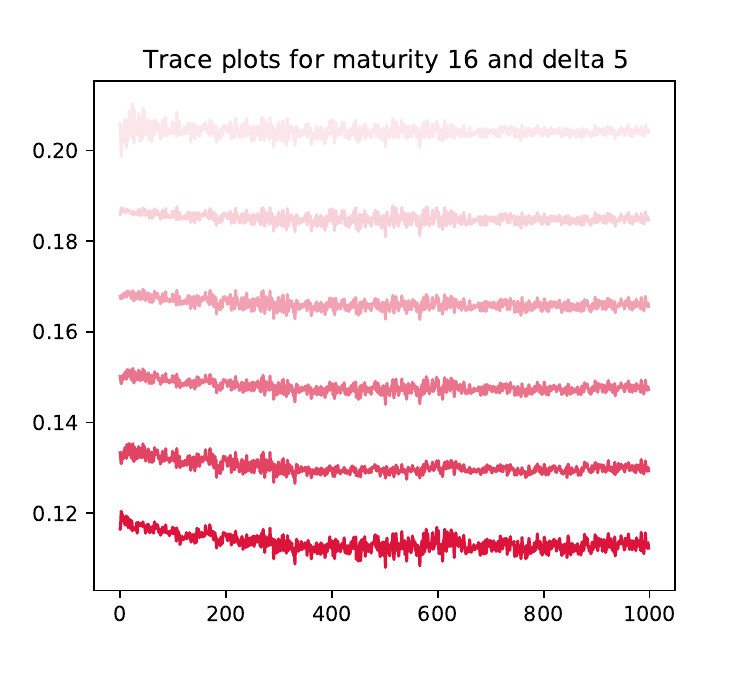}&
\includegraphics[scale=0.17]{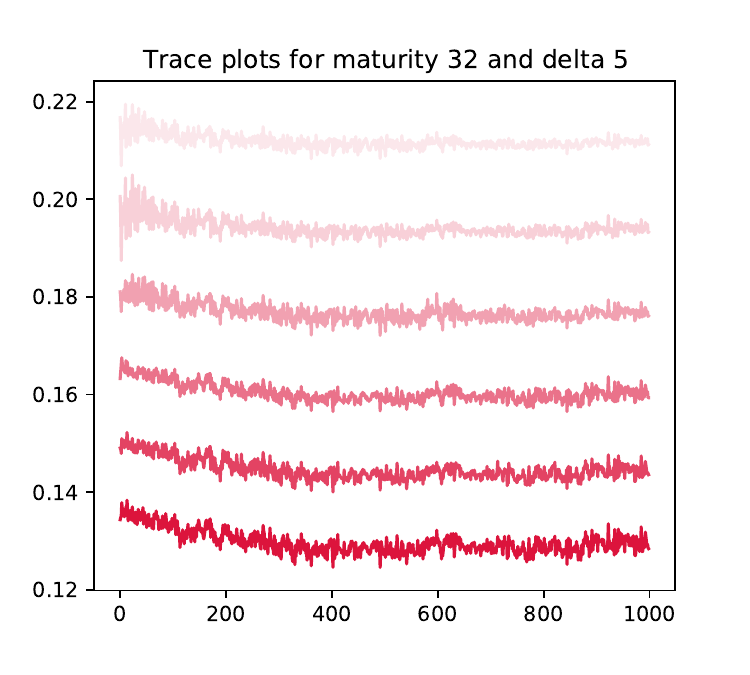}&
\includegraphics[scale=0.17]{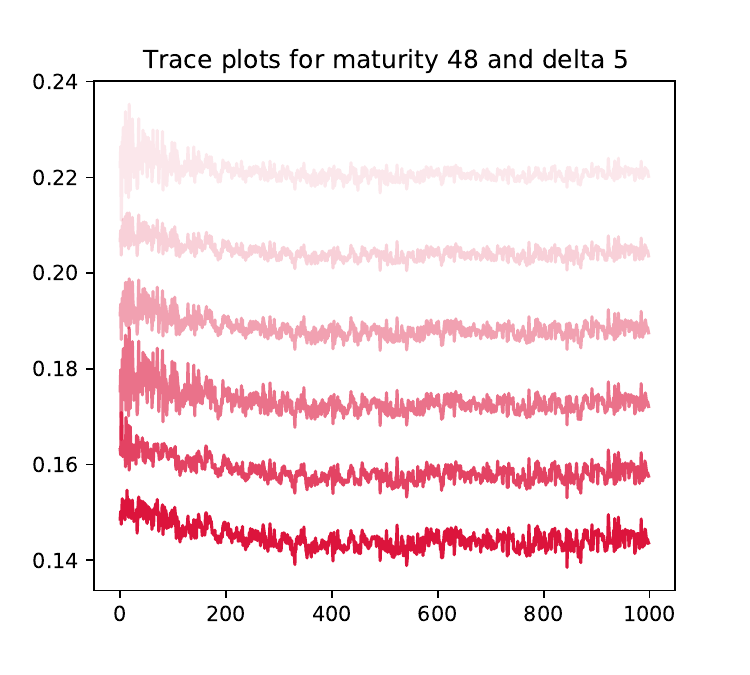}&
\includegraphics[scale=0.17]{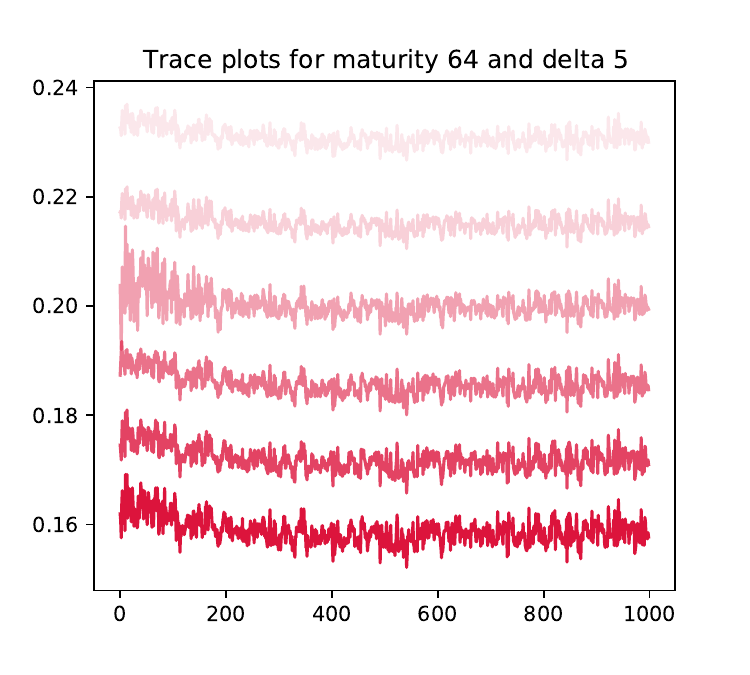}&
\includegraphics[scale=0.17]{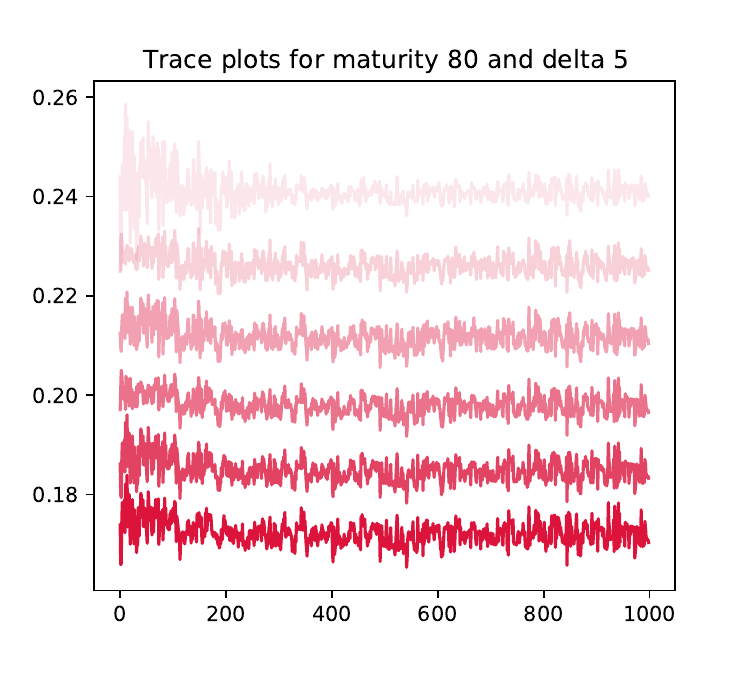}&
\includegraphics[scale=0.17]{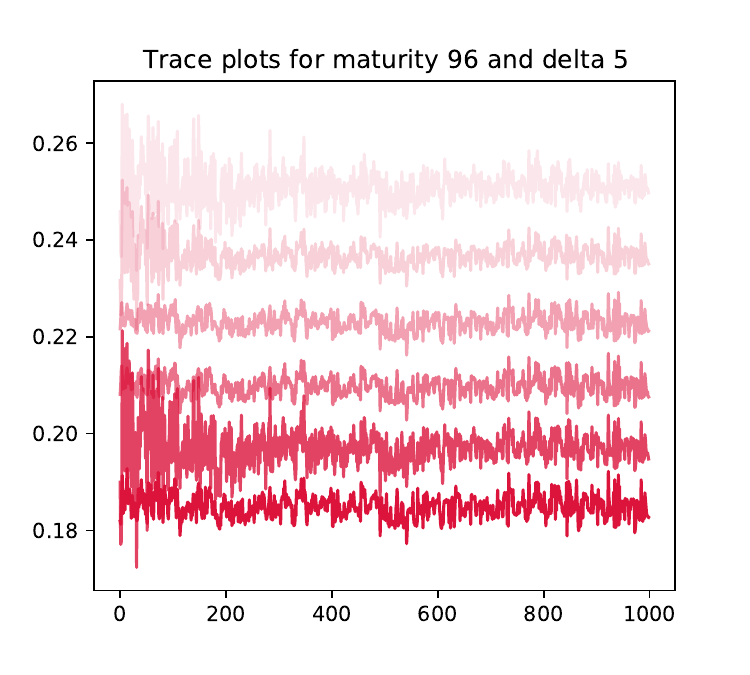}\\  [0.1ex] 
 \hline

$\delta=10$&\includegraphics[scale=0.17]{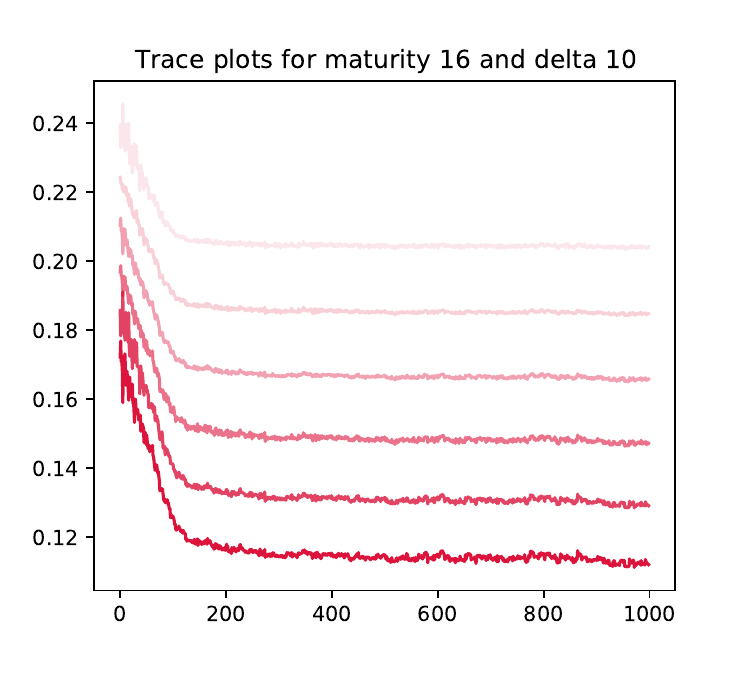}&
\includegraphics[scale=0.17]{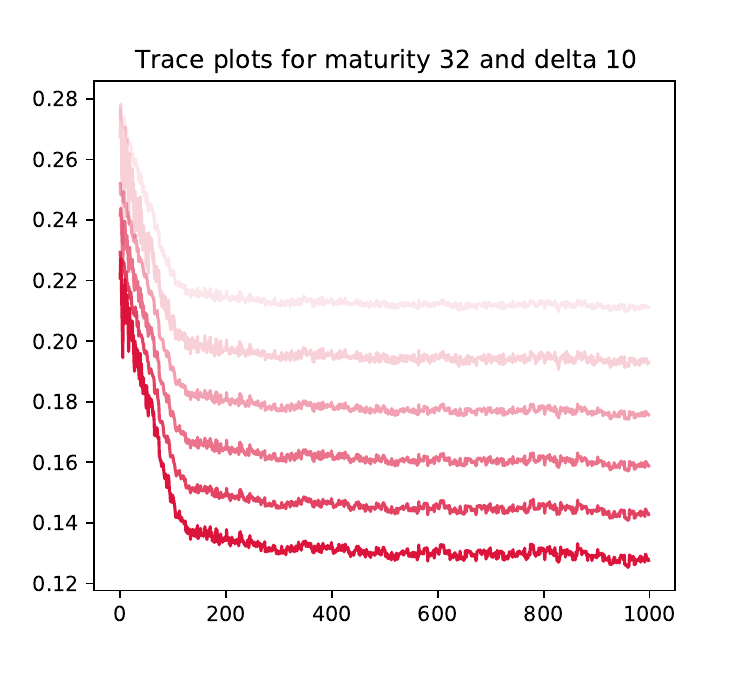}&
\includegraphics[scale=0.17]{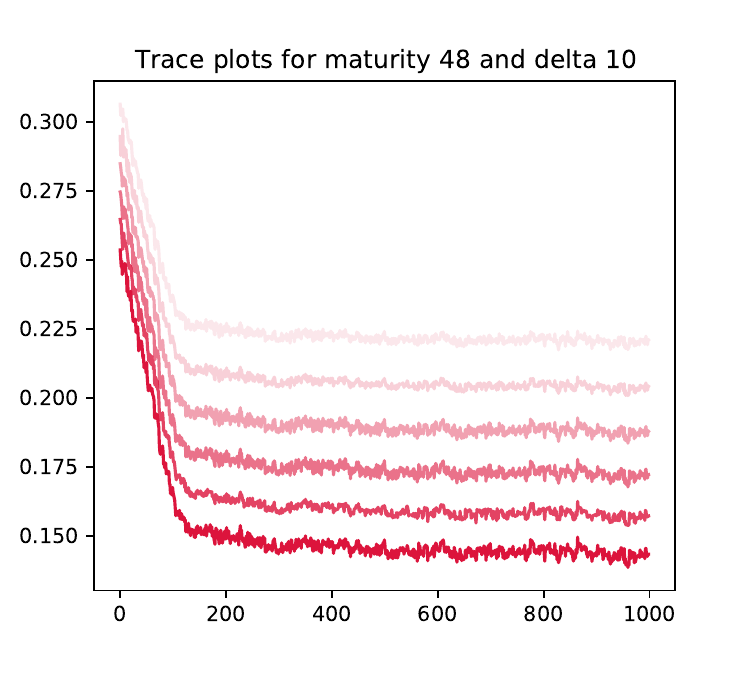}&
\includegraphics[scale=0.17]{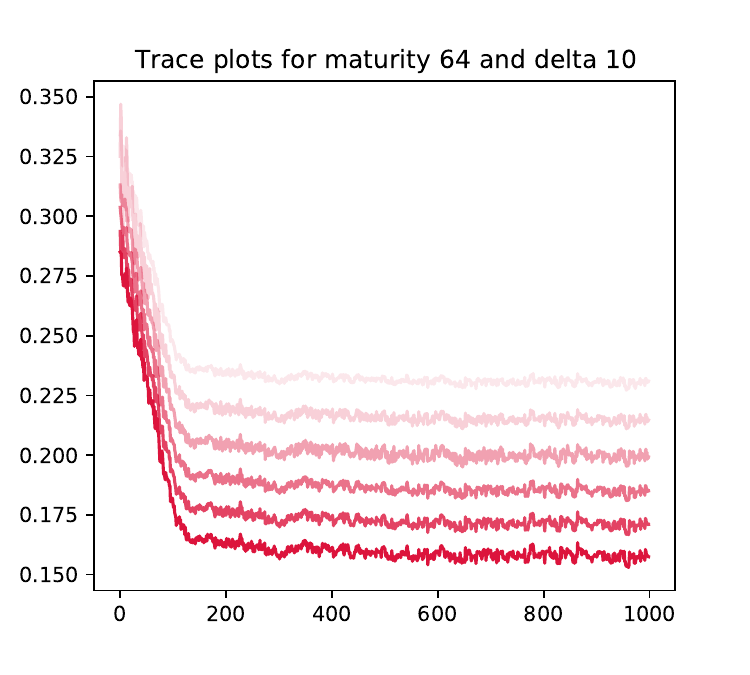}&
\includegraphics[scale=0.17]{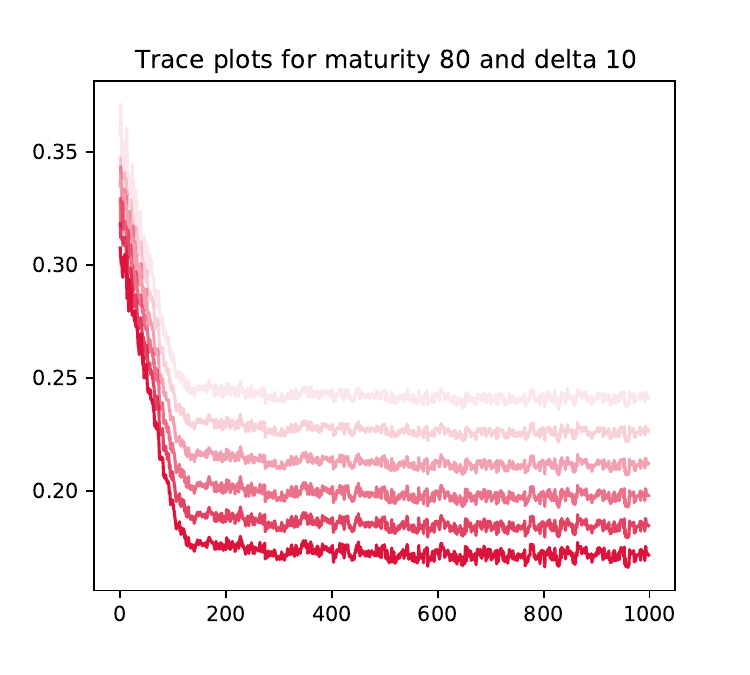}&
\includegraphics[scale=0.17]{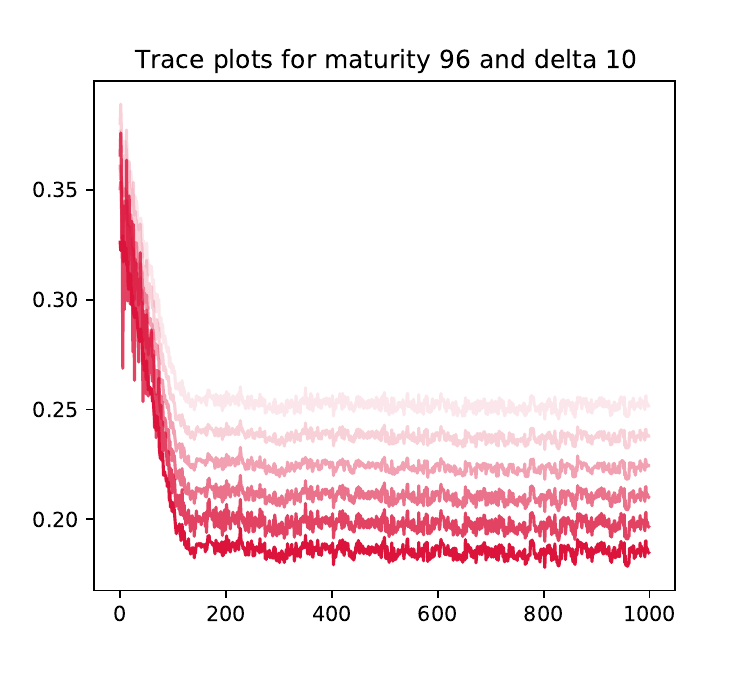}\\  [0.1ex] 
 \hline

$\delta=38.44$&\includegraphics[scale=0.17]{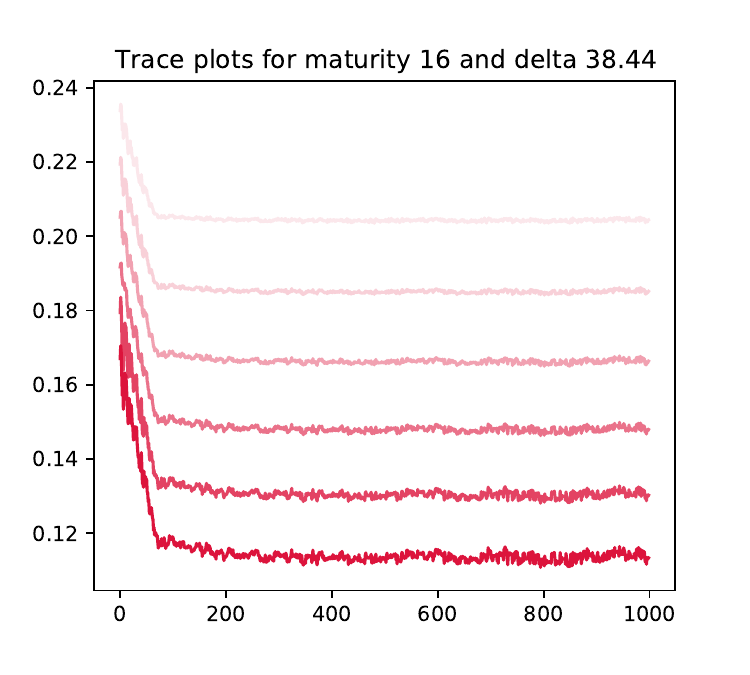}&
\includegraphics[scale=0.17]{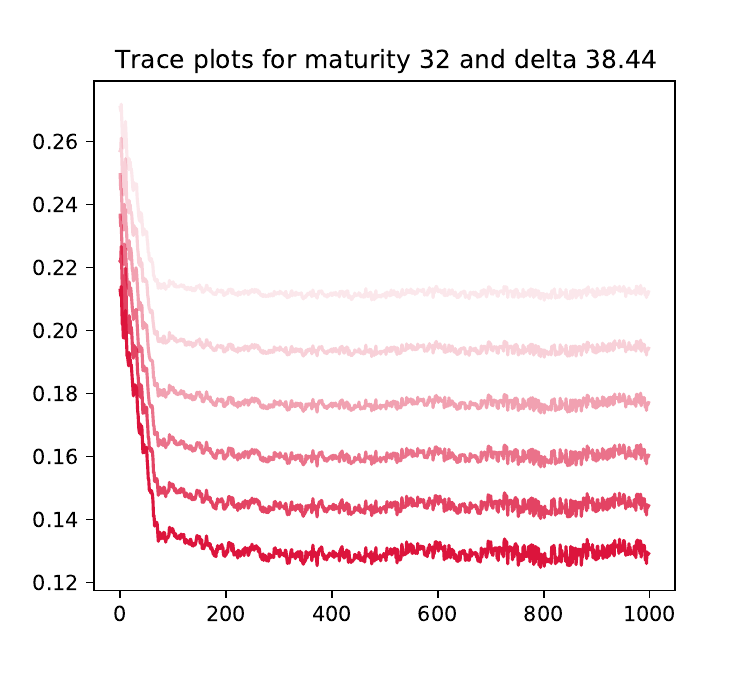}&
\includegraphics[scale=0.17]{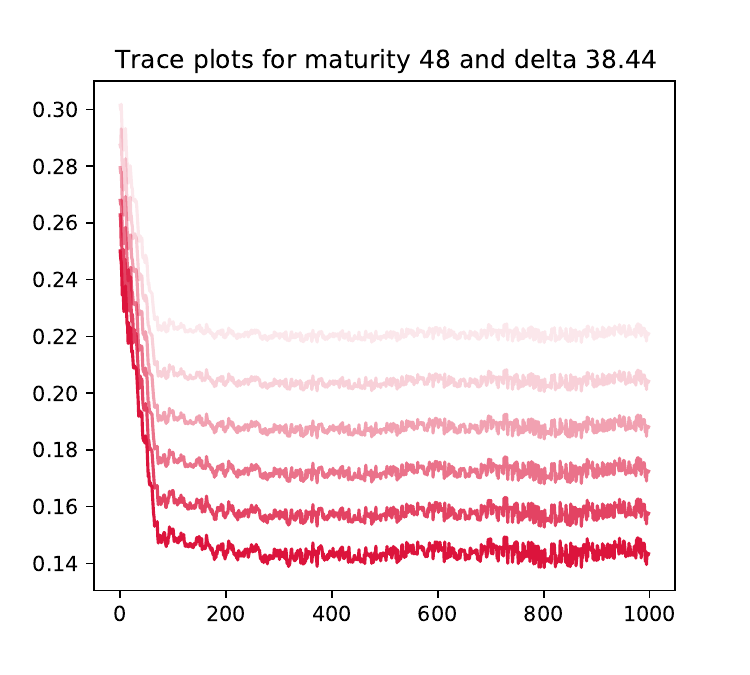}&
\includegraphics[scale=0.17]{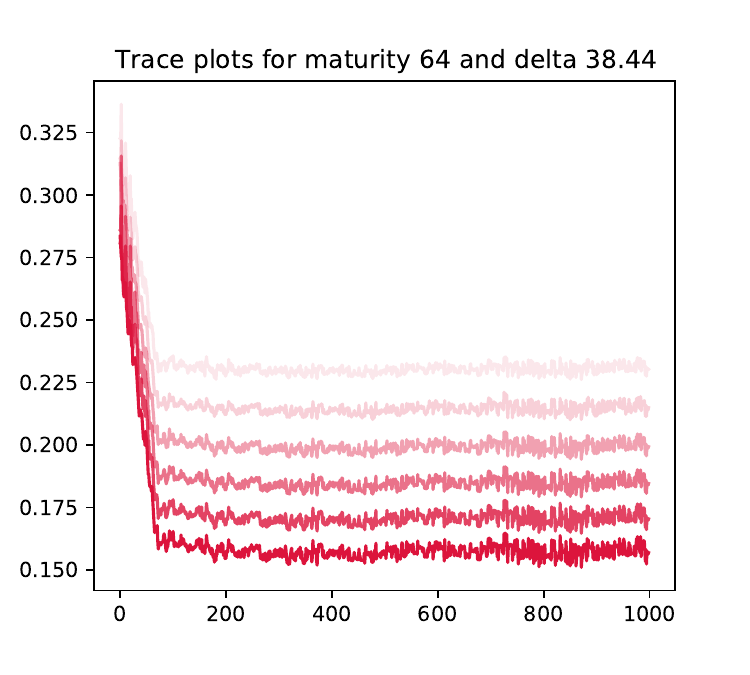}&
\includegraphics[scale=0.17]{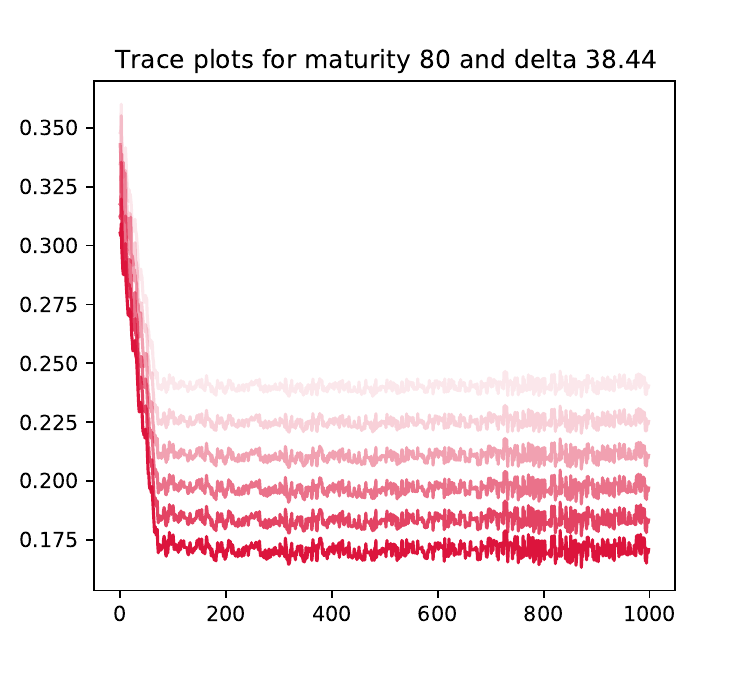}&
\includegraphics[scale=0.17]{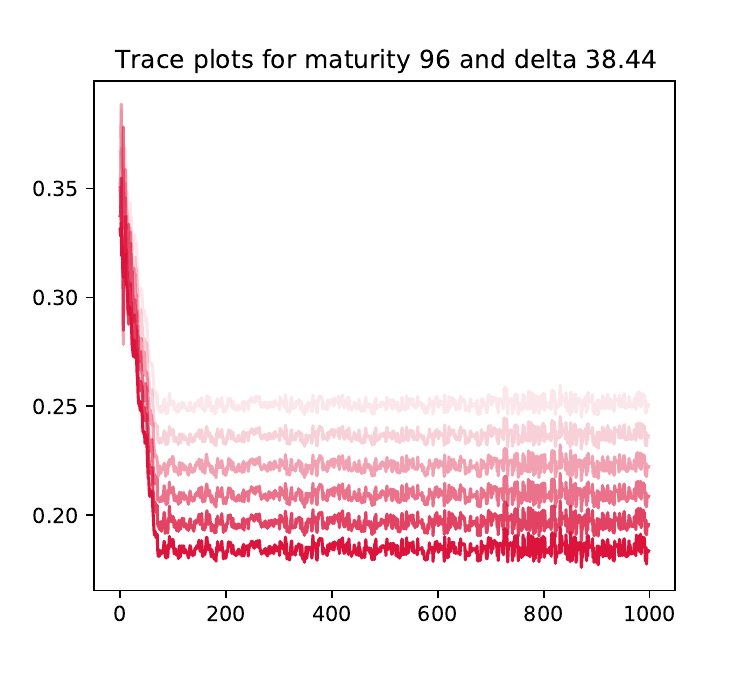}\\  [0.1ex] 
 \hline 
 \hline
\end{tabular}
\end{table}

\begin{table}
\centering
  \small
  \caption{Sensitivity of calibrated implied volatilities, with respect to $\delta$.}
  \label{tb:table15}
\begin{tabular}{||c|c c c c c c||} \hline
$\delta=1$&
\includegraphics[scale=0.17]{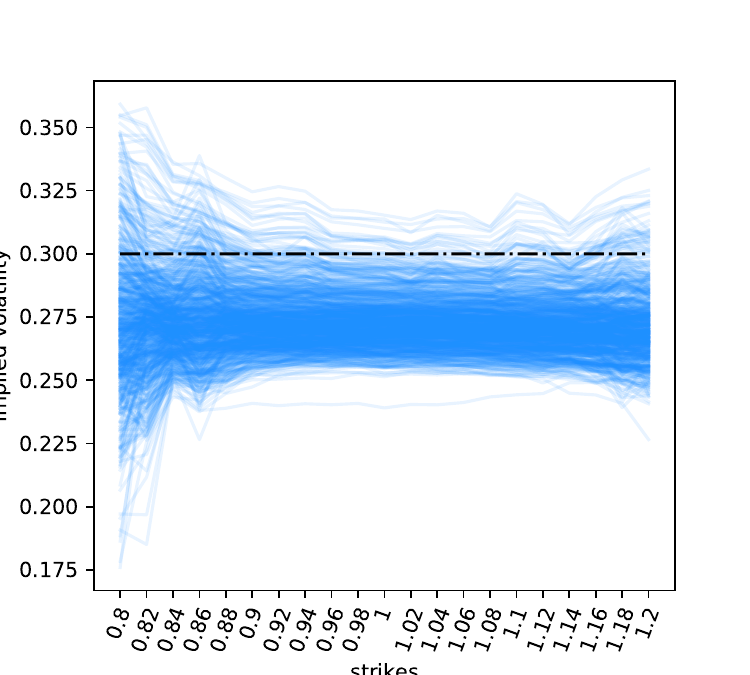}&
\includegraphics[scale=0.17]{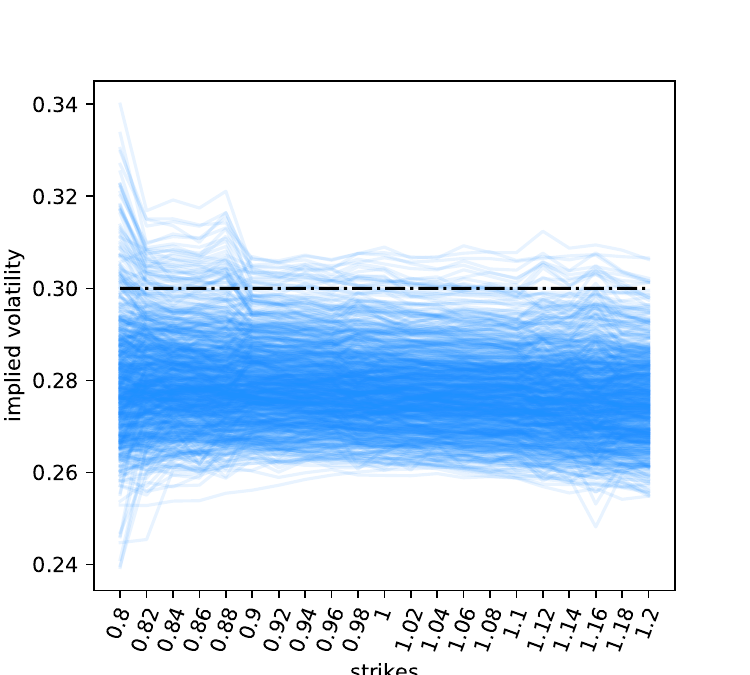}&
\includegraphics[scale=0.17]{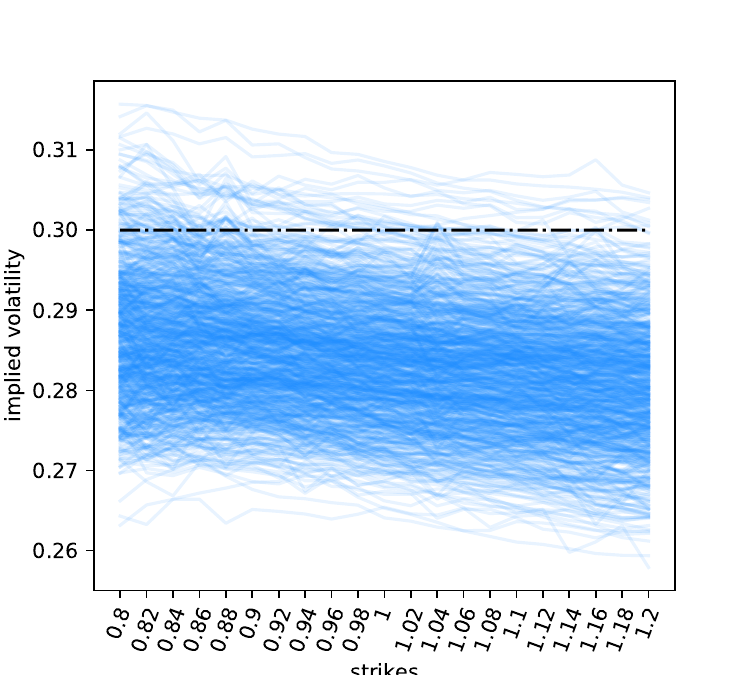}&
\includegraphics[scale=0.17]{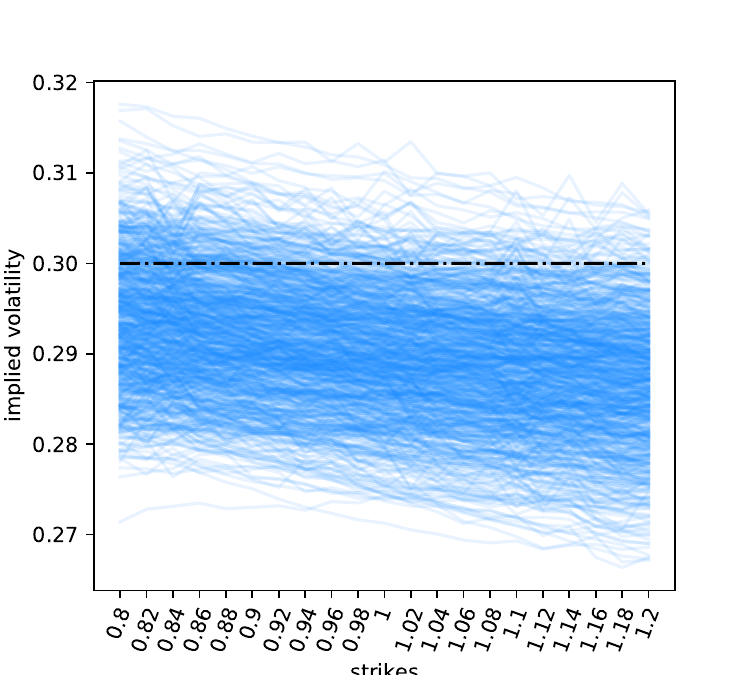}&
\includegraphics[scale=0.17]{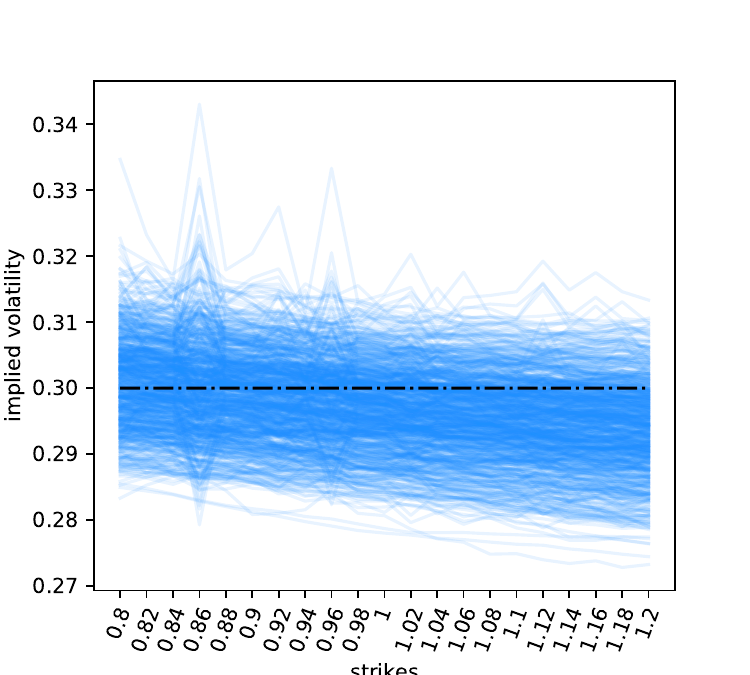}&
\includegraphics[scale=0.17]{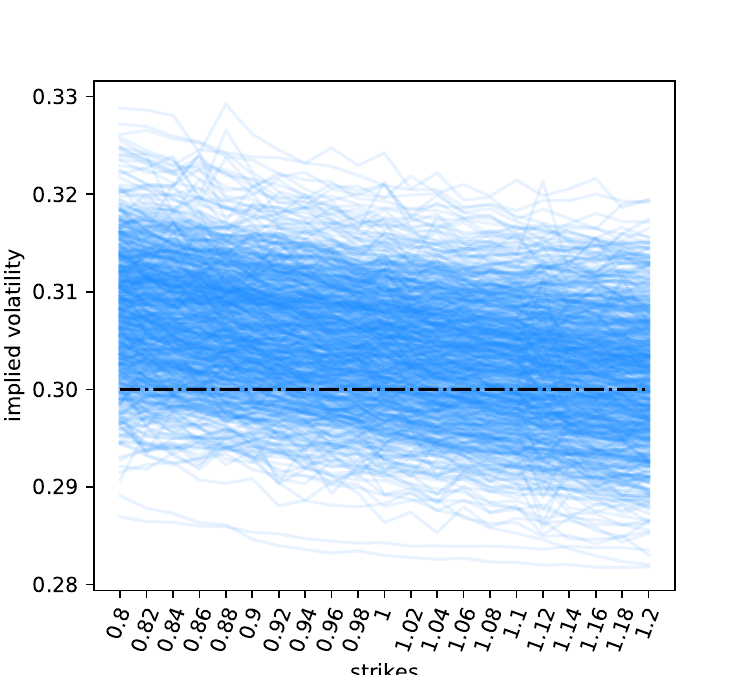}\\  [0.1ex] 
 \hline

$\delta=5$&\includegraphics[scale=0.17]{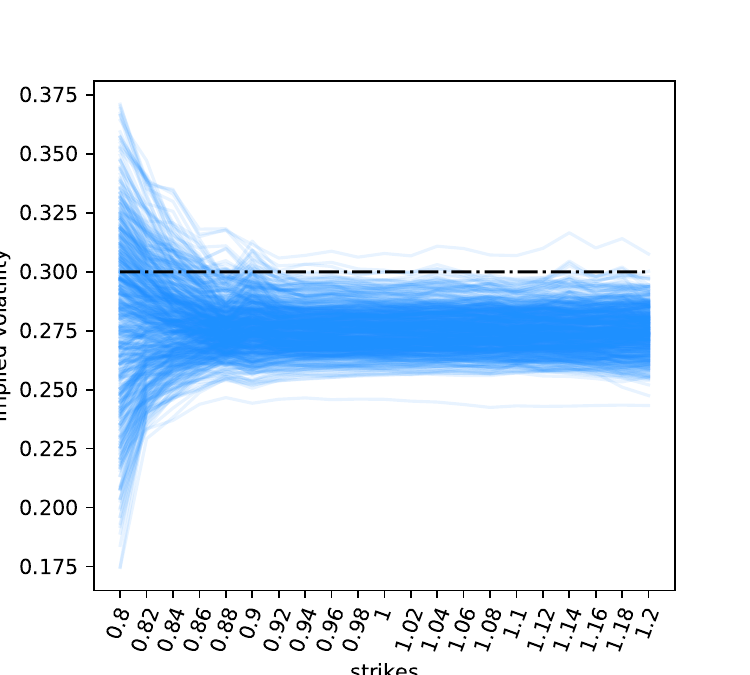}&
\includegraphics[scale=0.17]{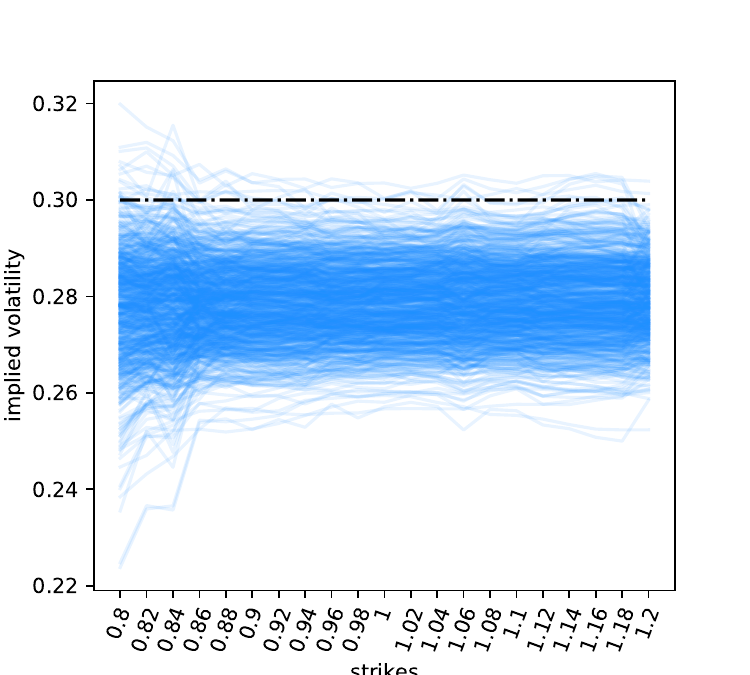}&
\includegraphics[scale=0.17]{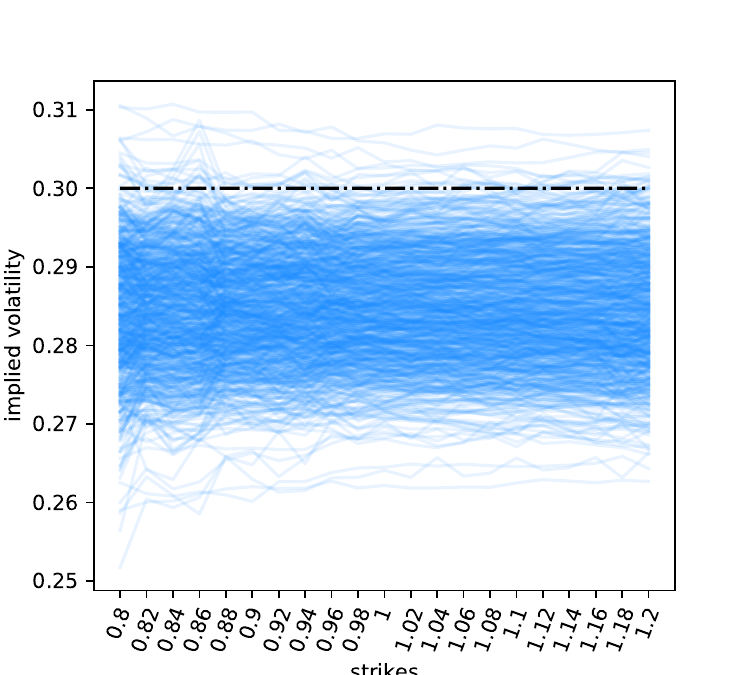}&
\includegraphics[scale=0.17]{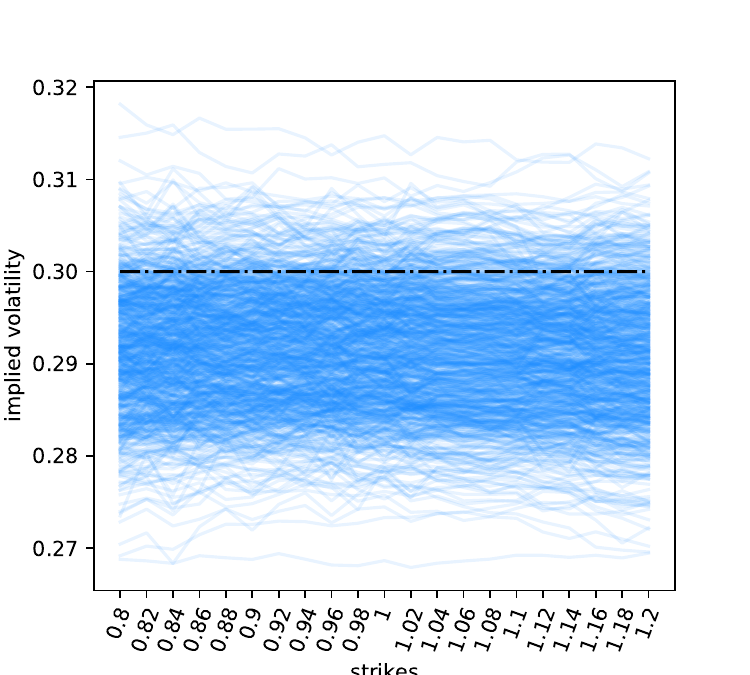}&
\includegraphics[scale=0.17]{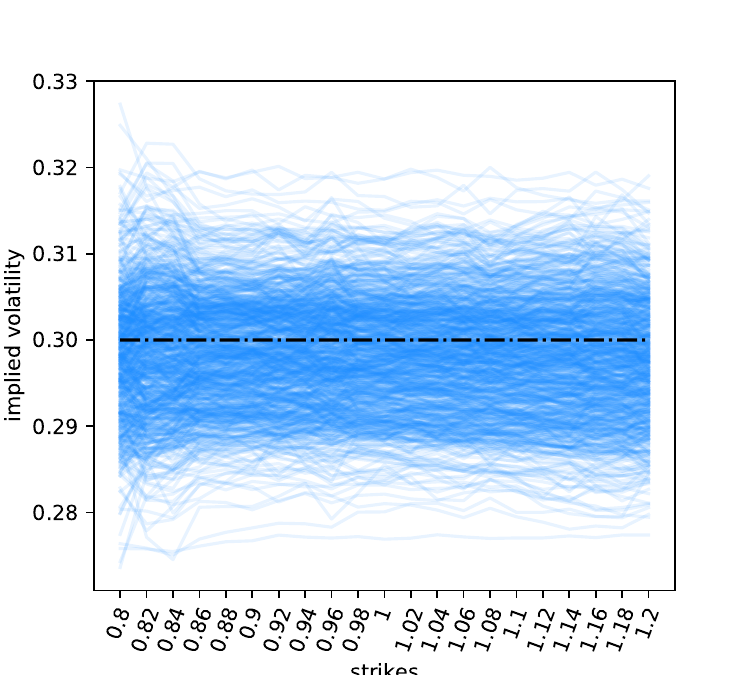}&
\includegraphics[scale=0.17]{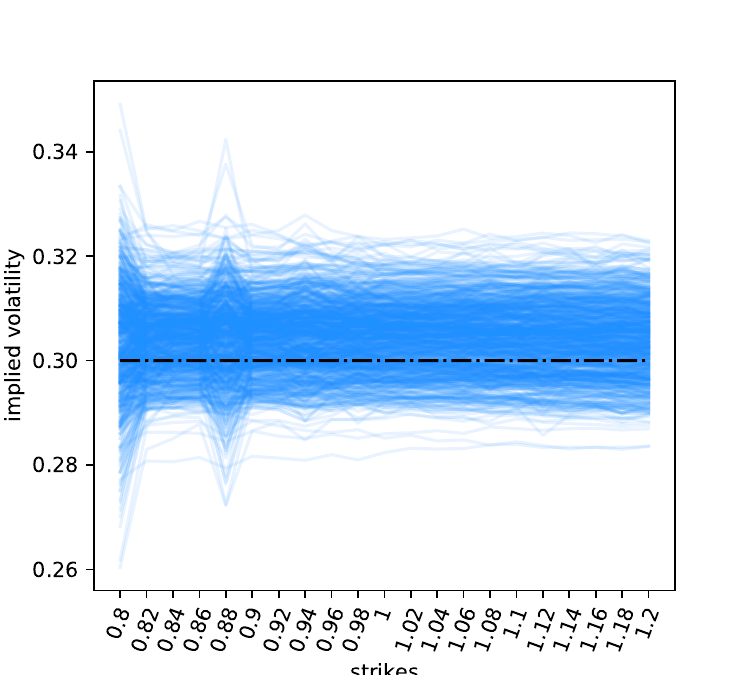}\\  [0.1ex] 
 \hline

$\delta=10$&\includegraphics[scale=0.17]{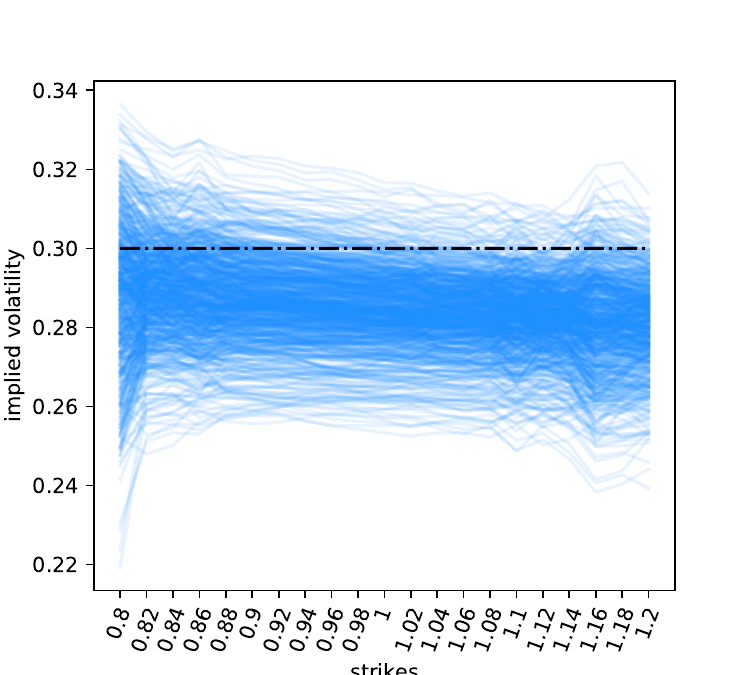}&
\includegraphics[scale=0.17]{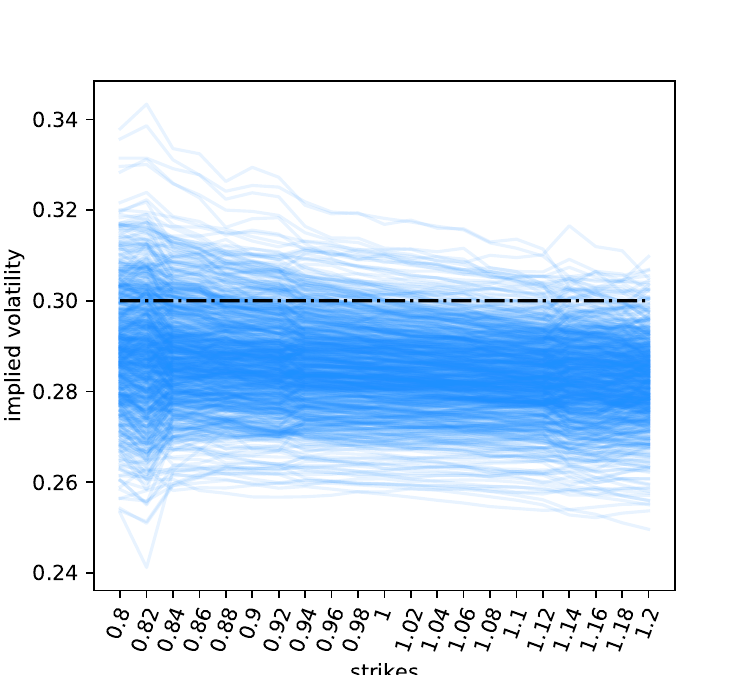}&
\includegraphics[scale=0.17]{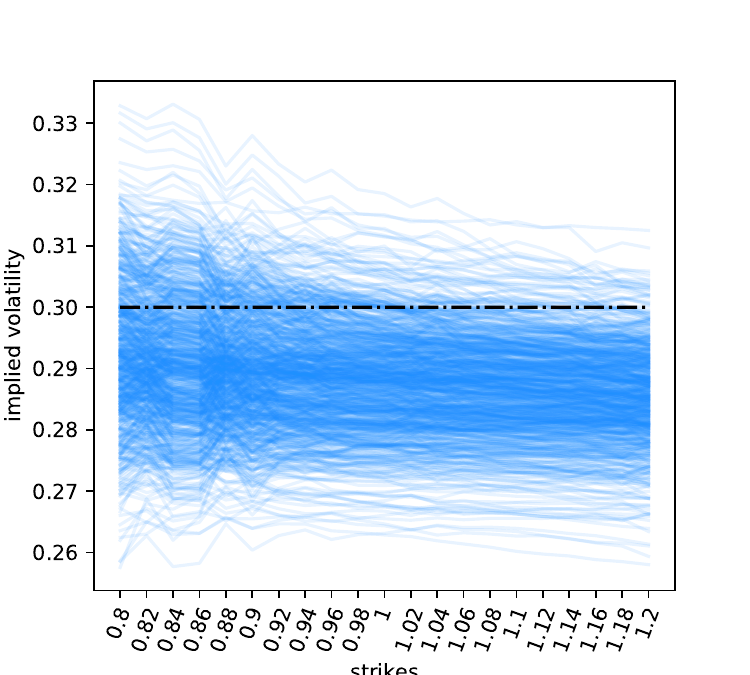}&
\includegraphics[scale=0.17]{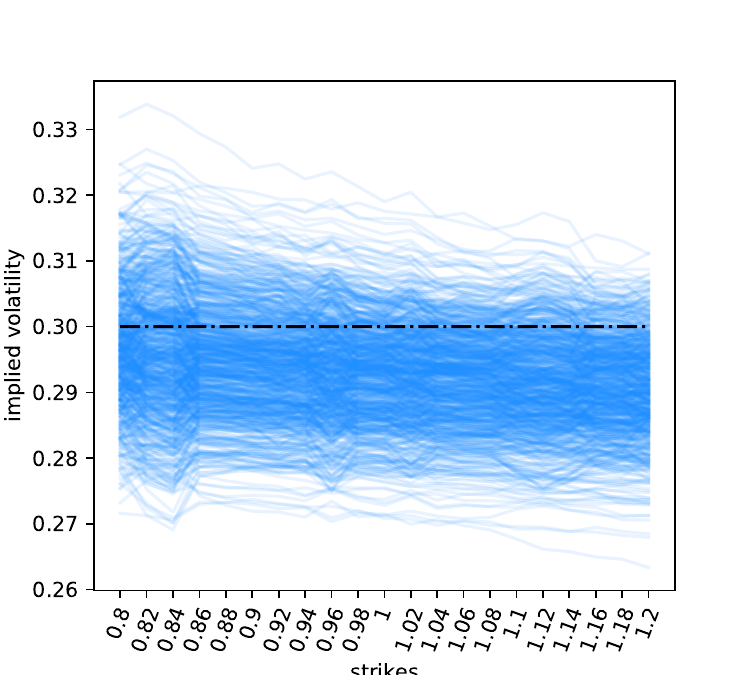}&
\includegraphics[scale=0.17]{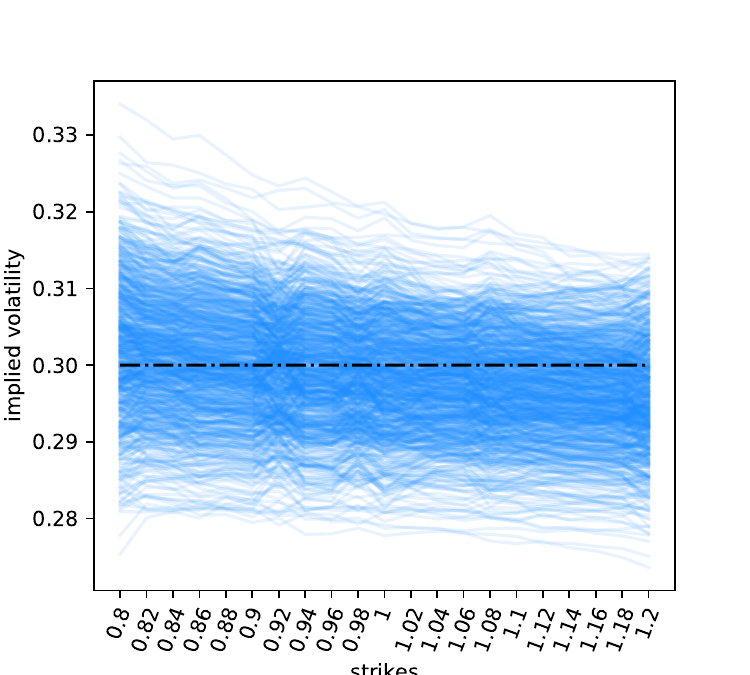}&
\includegraphics[scale=0.17]{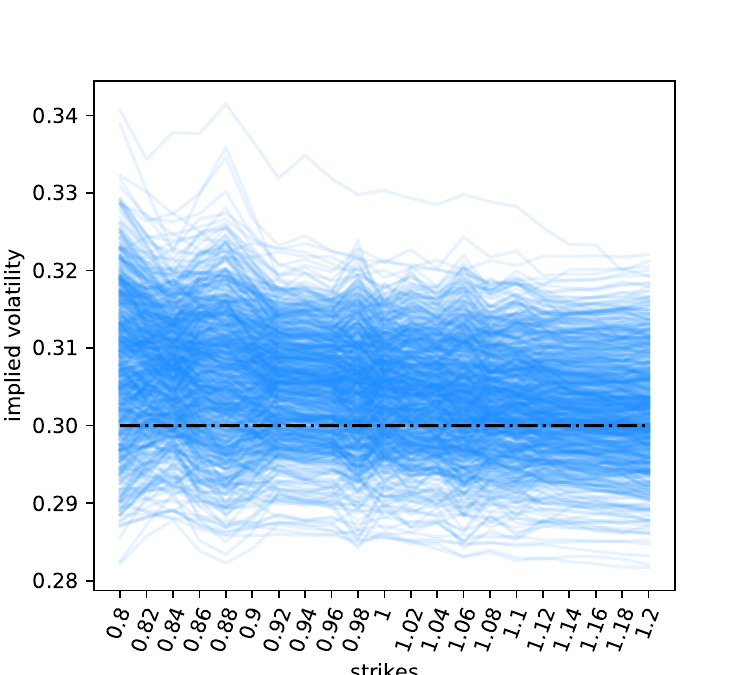}\\  [0.1ex] 
 \hline

$\delta=38.44$&\includegraphics[scale=0.17]{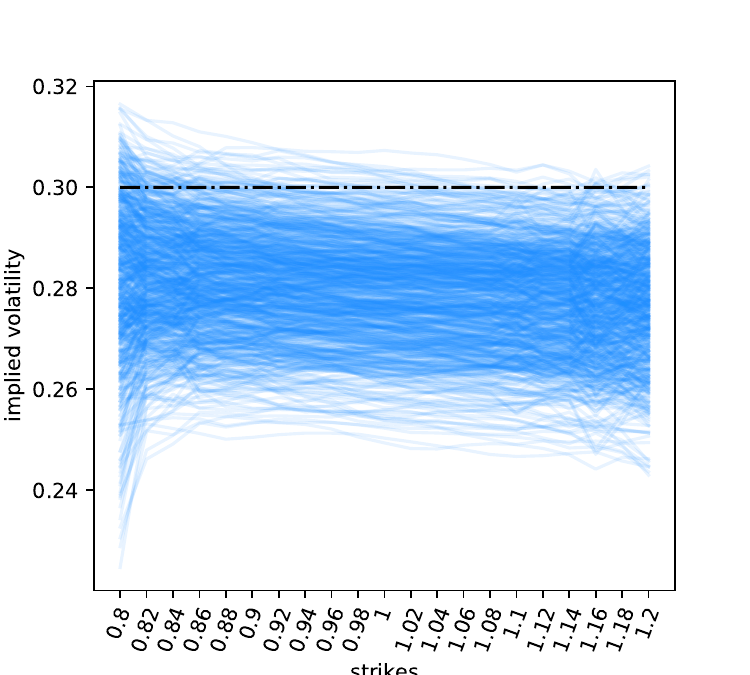}&
\includegraphics[scale=0.17]{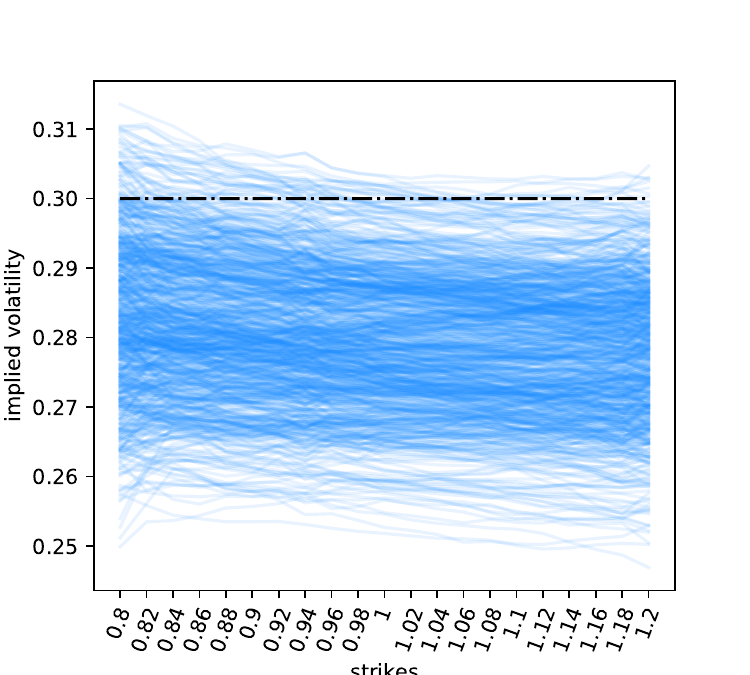}&
\includegraphics[scale=0.17]{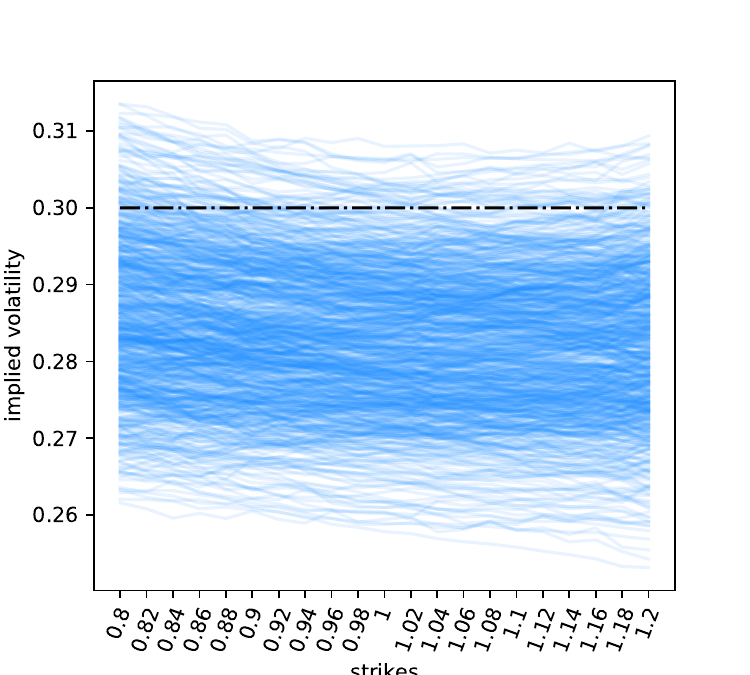}&
\includegraphics[scale=0.17]{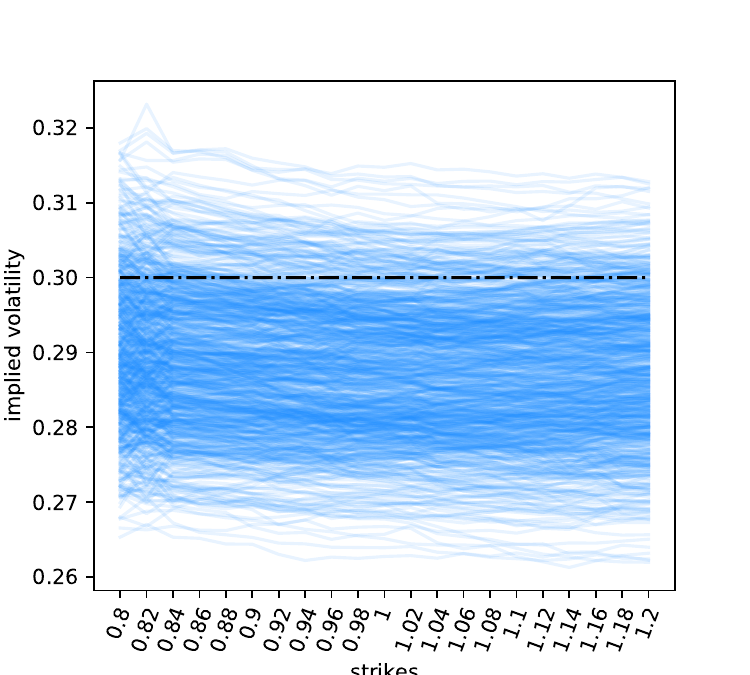}&
\includegraphics[scale=0.17]{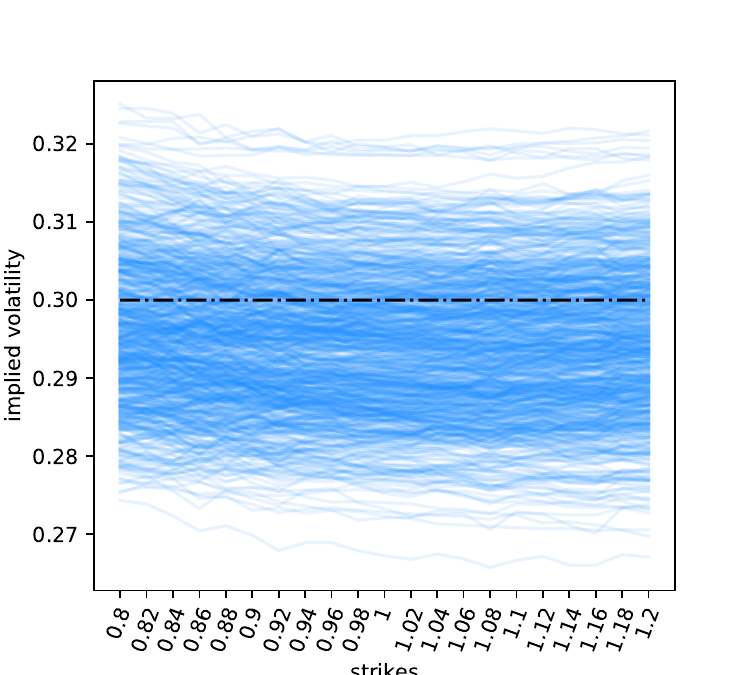}&
\includegraphics[scale=0.17]{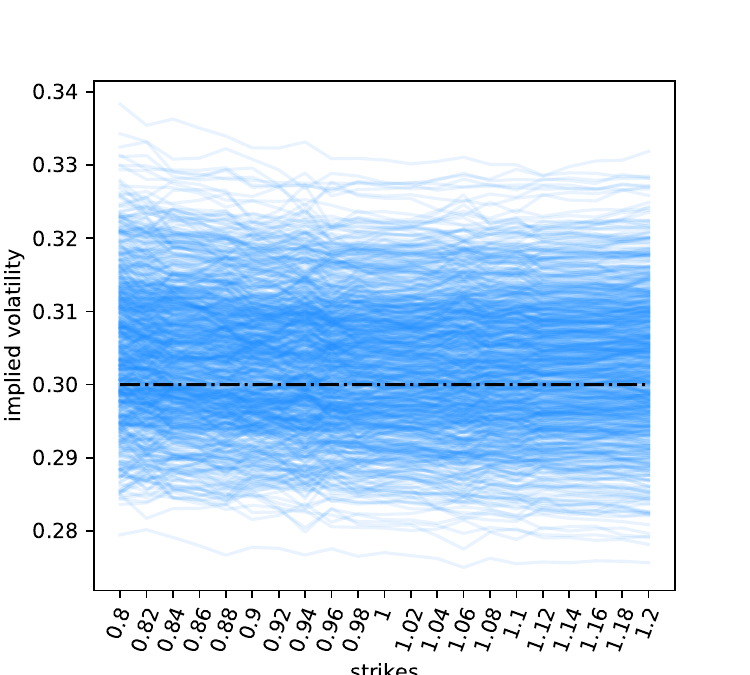}\\  [0.1ex] 
 \hline 
 \hline
\end{tabular}
\end{table}

\begin{table}[]
\centering
\begin{tabular}{l|llllllllll}
\multicolumn{1}{c}{Maturities}& \multicolumn{10}{c}{Strike (Units of S0)}                         \\ \hline
                               &0.85 &0.9 &0.95&1   &1.05&1.1 &1.15&1.2&1.3&1.4 \\ \hline
0.175                          &91.3 &62.8&35.2&12.9&2.1 &0.1 &0.0 &0.0&0.0&0.0 \\
0.425                          &96.3 &69.0&44.0&23.3&8.5 &2.3 &0.4 &0.2&0.0&0.0 \\
0.695                          &101.8&76.1&52.6&32.6&16.4&5.9 &1.9 &0.6&0.1&0.0 \\
0.940                          &106.8&82.2&59.9&39.9&23.8&11.3&4.7 &1.8&0.2&0.0
\end{tabular}
\caption{European call prices for S\&P 500 data set.}
  \label{tb:table16}
\end{table}
\clearpage
\begin{acknowledgements}
 Special thanks to Prof. Sylvia Fr\"uhwirth-Schnatter and Prof. Claudio Fontana for their helpful input and assistance with this project.
\end{acknowledgements}

\end{document}